# A Theory and Calculus for Reasoning about Sequential Behavior

FREDERICK FURTEK

Applied Combinatorics, Menlo Park, CA 94025 USA, fred@calculus.com

___

Basic results in combinatorial mathematics provide the foundation for a theory and calculus for reasoning about sequential behavior. A key concept of the theory is a generalization of Boolean implicant which deals with statements of the form:

*A sequence of Boolean expressions α is an implicant of*
*a set of sequences of Boolean expressions A*

This notion of a generalized implicant takes on special significance when each of the sequences in the set *A* describes a *disallowed* pattern of behavior. That's because a disallowed sequence of Boolean expressions represents a *logical/temporal dependency*, and because the implicants of a set of disallowed Boolean sequences *A* are themselves disallowed and represent precisely those dependencies that follow as a logical consequence from the dependencies represented by *A*. The main result of the theory is a necessary and sufficient condition for a sequence of Boolean expressions to be an implicant of a regular set of sequences of Boolean expressions. This result is the foundation for two new proof methods. *Sequential resolution* is a generalization of Boolean resolution which allows new logical/temporal dependencies to be inferred from existing dependencies. *Normalization* starts with a model (system) and a set of logical/temporal dependencies and determines which of those dependencies are satisfied by the model.

___

## 1. INTRODUCTION

Reasoning about sequential behavior is fundamental to the design of computing machinery. Hardware designers reason about sequential behavior in order to determine the input/output behavior of a *system* from the individual behaviors of the system's *components*. Software programmers reason about sequential behavior in order to determine the input/output behavior of a *program* from the individual behaviors of the program's *instructions*.

But, of course, the reasoning powers of designers and programmers are limited, and those limitations become apparent in the design of *discrete-time systems* with complex *logical/temporal dependencies*. The diversity and intricacy of those dependencies are hinted at in the following examples, where *P*, *Q* and *R* each represent a Boolean expression that *holds* (is *true*) in a subset of system states.

*If P, then Q in the next state*

*If P, then Q five states later*

*If P, then Q within five states*

*If P, then Q thereafter*

*If P, then Q until R*

*If P, then Q three states later and*
*every fourth state thereafter*

Not only must a designer/programmer deal with an initial set of such dependencies describing the components of a design or instructions of a program, the designer/programmer must also infer new dependencies in order to achieve the ultimate goal of determining how a system's outputs *depend* upon the system's inputs.

The present work is intended as a contribution towards ultimately replacing the error-prone mental models of designers and programmers with a mathematical framework for reasoning about sequential behavior *and* for reasoning about different mathematical domains – like integers and complex numbers. This contribution does not presume to solve the entire problem, but instead focuses on a theory and calculus for reasoning about sequential behavior. The theory has elements of Boolean logic and automata theory, but at its core are fundamental results in combinatorial mathematics. These results at the *combinatorics level* provide the foundation for results at the *logic level*, which, in turn, provide the foundation for the calculus. The calculus consists of two new proof methods, *sequential resolution* and *normalization*. Sequential resolution is a generalization of Boolean resolution which allows new logical/temporal dependencies to be inferred from existing dependencies. Normalization starts with a model (system) and a set of logical/temporal dependencies and determines which of those dependencies are satisfied by the model.

The following subsections provide an informal introduction to both the theory and calculus.

## 1.1 Allowed and Disallowed Behaviors

In order to make more precise the notion of a logical/temporal dependency, we must first understand the distinction between allowed and disallowed system behaviors.

A sequence of states of a discrete-time system *K* is an *allowed behavior of K* if and only if it is *possible* for *K* to traverse that sequence of consecutive states. A sequence of states of *K* is a *disallowed behavior of K* if and only if it is *impossible* for *K* to traverse that sequence of consecutive states. (*K* stands for generalized *K*ripke structure, the formal counterpart to our informal notion of discrete-time system. A *GKS* is defined in Section 4.1.)

Now suppose that $\omega$ is a sequence of states of System $K$ and that $\omega_{ss}$ is an arbitrary subsequence of $\omega$. (A *subsequence* of a sequence $\alpha$ is a sequence of consecutive elements appearing in $\alpha$.) Assume that $\omega$ is an allowed behavior of $K$. Because $\omega$ is allowed, we know that it is possible for the system to traverse $\omega$. But if it is possible for the system to traverse $\omega$, then it must be possible for the system to traverse $\omega_{ss}$ since in the process of traversing $\omega$, the system must traverse $\omega_{ss}$. $\omega_{ss}$ is therefore also an allowed behavior of $K$. So if $\omega$ is allowed, then so must be $\omega_{ss}$. And, of course, there is the contrapositive of this statement: If $\omega_{ss}$ is disallowed, then so must be $\omega$. These two equivalent properties are expressed as the following axiom.

AXIOM 1. (a) *Every subsequence of an allowed behavior of a discrete-time system (generalized Kripke structure) K is also an allowed behavior of K.* (b) *Every sequence of states of a discrete-time system (generalized Kripke structure) K having a disallowed behavior of K as a subsequence is also a disallowed behavior of K.*

From this single axiom – and a body of combinatorial mathematics – there follows a theory for reasoning about sequential behavior. That theory begins with the two closely related notions of *logical/temporal dependency* and *sequential constraint*.

## 1.2 Sequential Constraints

A logical/temporal dependency is a property of a discrete-time system (generalized Kripke structure) that *constrains*, or *reduces*, the set of allowed system behaviors. But a property that *reduces* the set of *allowed* behaviors must *expand* the set of *disallowed* behaviors. That means that a logical/temporal dependency can be identified with the set of behaviors disallowed (prohibited, forbidden) by that dependency. Furthermore, from Axiom 1(b) we know that prepending and appending arbitrary state sequences to a disallowed state sequence must yield another disallowed state sequence. A logical/temporal dependency, or set of dependencies, is therefore completely characterized by a set of disallowed state sequences if and only if that set contains all *minimal* state sequences prohibited by the dependency(ies) – that is, all state sequences that cannot be shortened at either end without yielding a state sequence that is not prohibited by the dependency.

In what follows, we use a *regular set of sequential constraints* to describe such a set of disallowed state sequences. (See Sipser [1] or Hopcroft [2] for definitions of *regular language* (*regular set*), *regular expression* and *finite state automaton*. Note especially the equivalence of regular expressions and finite-state automata in defining regular sets of sequences.)

A *sequential constraint* is defined with the aid of the *holds tightly* relation [3,4].

*Definition* 1.1. A sequence of Boolean expressions $\alpha$ *holds tightly* on a state sequence $\omega$ if and only if $\alpha$ and $\omega$ are the same length and each Boolean expression of $\alpha$ holds in the corresponding state of $\omega$. A set of sequences of Boolean expressions $A$ *holds tightly* on a state sequence $\omega$ if and only if there exists a sequence of Boolean expressions in $A$ that holds tightly on $\omega$.

*Definition* 1.2. A *sequential constraint* of a discrete-time system (generalized Kripke structure) $K$ is a finite sequence of Boolean expressions $\alpha$ such that all sequences of states of $K$ on which $\alpha$ holds tightly are disallowed behaviors of $K$.

A sequential constraint thus describes a disallowed pattern of behavior. Moreover, when we declare that a sequence of Boolean expressions $\alpha$ is a sequential constraint of a system $K$, we are declaring not only that the state sequences on which $\alpha$ holds tightly are disallowed behaviors of $K$ but that all state sequences containing a subsequence on which $\alpha$ holds tightly are disallowed behaviors of $K$.

To illustrate these ideas, we return to the six logical/temporal dependences listed above. They are represented in disallowed form by the following six regular expressions, with each regular expression defining a regular set of sequential constraints and each sequential constraint defining a set of disallowed state sequences via the holds-tightly relation.

$$\langle P, \neg Q \rangle$$

$$\langle P, true, true, true, true, \neg Q \rangle$$

$$\langle P, \neg Q, \neg Q, \neg Q, \neg Q, \neg Q \rangle$$

$$\langle P, true^*, \neg Q \rangle$$

$$\langle P, (\neg R)^*, (\neg R \wedge \neg Q) \rangle$$

$$\langle P, true, true, \langle true, true, true, true \rangle^*, \neg Q \rangle$$

Consider, for example, the logical/temporal dependency *If P, then Q five states later*. It is represented by the sequential constraint $\langle P, \text{true}, \text{true}, \text{true}, \text{true}, \neg Q \rangle$, which asserts that in a sequence of six consecutive states $\langle s_0, s_1, s_2, s_3, s_4, s_5 \rangle$, it is never the case that $P$ holds in $s_0$, *true* holds in each of $s_1$, $s_2$, $s_3$ and $s_4$ and $\neg Q$ holds in $s_5$. But since *true* holds (is *true*) in every state, the constraint actually asserts that it is never the case that $P$ holds in $s_0$ and $\neg Q$ holds in $s_5$. So if $P$ holds in $s_0$, then $Q$ must hold in $s_5$. Consider also the logical/temporal dependency *If P, then Q thereafter*. It is represented by the regular expression $\langle P, \text{true}^*, \neg Q \rangle$, which defines the infinite set of sequential constraints:

$$\langle P, \neg Q \rangle$$
$$\langle P, \text{true}, \neg Q \rangle$$
$$\langle P, \text{true}, \text{true}, \neg Q \rangle$$
$$\cdot$$
$$\cdot$$
$$\cdot$$

This set of sequential constraints asserts that for all positive integers $n$, a state in which $P$ holds cannot be followed $n$ states later by a state in which $\neg Q$ holds.

*Comment*: Ours is not the only approach to use regular expressions over a set of Boolean expressions to express logical/temporal dependencies. Both PSL, the industry-standard property specification language [4], and its predecessor, the temporal logic Sugar [3], have such constructs.

Consider now the following question, which is the central issue addressed by the present theory:

*How do we know whether a logical/temporal dependency follows as a logical consequence from a given set of logical/temporal dependencies?*

It is equivalent to the following question expressed in terms of sequential constraints:

*How do we know whether a sequence of Boolean expressions is a sequential constraint as a consequence of a given set of sequential constraints?*

A generalization of the notion of Boolean implicant provides the key to answering this question.

## 1.3 Implicants

Suppose that $\alpha$ is a finite sequence of Boolean expressions and that $A$ is a set of sequential constraints of a system $K$. Suppose further that for every state sequence $\omega$ on which $\alpha$ holds tightly, there exists a subsequence $\omega_{ss}$ of $\omega$ on which $A$ holds tightly. Because $A$ is a set of sequential constraints of $K$ and $A$ holds tightly on $\omega_{ss}$, $\omega_{ss}$ must be a disallowed behavior of $K$. From Axiom 1(b), we know that because $\omega_{ss}$ is disallowed, $\omega$ must also be disallowed. $\alpha$ is therefore a sequential constraint of System $K$.

Now suppose – in contrast to the preceding supposition – that there exists a state sequence $\omega$ on which $\alpha$ holds tightly such that there is *no* subsequence of $\omega$ on which $A$ holds tightly. Then there is no basis on which to conclude that $\omega$ is a disallowed behavior of System $K$, and there is therefore no basis on which to conclude that $\alpha$ is a sequential constraint of $K$. Thus,

*A finite sequence of Boolean expressions $\alpha$ is a*
*sequential constraint of a system K as a consequence*
*of a set A of sequential constraints of K*

*if and only if*

*For every sequence of states $\omega$ of K on which*
*$\alpha$ holds tightly, there exists a subsequence of $\omega$*
*on which A holds tightly*

So we see that the original question posed above – *How do we know whether a logical/temporal dependency follows as a logical consequence from a given set of logical/temporal dependencies?* – reduces to the existence of a specific relationship between a sequence of Boolean expressions $\alpha$ and a set of such sequences $A$.

Consider now the purely Boolean case in which $\alpha$ and all of the sequences in $A$ are of length 1 – that is, each sequence consists of a single Boolean expression. Let $\alpha_{BE}$ be the Boolean expression appearing in $\alpha$ and let $A_{BE}$ be the disjunction (*OR*) of the Boolean expressions appearing in $A$. The above relationship between $\alpha$ and $A$ can then be simplified to:

*The set of states in which $\alpha_{BE}$ holds is a subset of*
*the set of states in which $A_{BE}$ holds*

Now suppose that this property is valid not just for a particular discrete-time system (generalized Kripke structure) *K* but for all possible discrete-time systems (generalized Kripke structures) *K*. But that means that

$$\alpha_{BE} \text{ implies } A_{BE}$$

Moreover, when $\alpha_{BE}$ is a product of literals – a quite common case – the relationship between $\alpha_{BE}$ and $A_{BE}$ can be re-expressed using the terminology of Boolean algebra [5]:

$$\alpha_{BE} \text{ is an implicant of } A_{BE}$$

This observation for the purely Boolean case leads us to generalize the notion of Boolean implicant to the realm of sequential behavior as follows.

*Definition* 1.3. An *implicant* of a set of sequences of Boolean expressions *A* is a finite sequence of Boolean expressions $\alpha$ such that for all discrete-time systems (generalized Kripke structures) *K*, for all sequences of states $\omega$ of *K* on which $\alpha$ holds tightly, there exists a subsequence of $\omega$ on which *A* holds tightly.

It follows that

*For all systems K, a finite sequence of Boolean expressions $\alpha$ is a sequential constraint*
*of K as a consequence of a set A of sequential constraints of K*

*if and only if*

*$\alpha$ is an implicant of A*

## 1.4 An Example

To illustrate the preceding ideas, consider the following set of logical/temporal dependencies:

| | |
|---|---|
| *If P, then Q in the next state* | (1a) |
| *If R, then S in the next state* | (1b) |
| *If $(Q \wedge S)$, then T in the next state* | (1c) |

Let *A* be the set of sequential constraints corresponding to this set of dependencies. Thus

$$A = \{\langle P, \neg Q\rangle, \langle R, \neg S\rangle, \langle(Q \wedge S), \neg T\rangle\}$$

Now let
$$\alpha = \langle(P \wedge R), \text{ true}, \neg T\rangle$$

and let $\langle s_0, s_1, s_2 \rangle$ be an arbitrary state sequence on which $\alpha$ holds tightly. From the definition of *holds tightly*, we know that $(P \wedge R)$ holds in $s_0$ and that $\neg T$ holds in $s_2$. Now consider state $s_1$ and the truth values of $Q$ and $S$ in that state. There are four possibilities:

1. $Q$ and $S$ hold in $s_1$. Then $\langle (Q \wedge S), \neg T \rangle$ holds tightly on $\langle s_1, s_2 \rangle$.
2. $\neg Q$ and S hold in $s_1$. Then $\langle P, \neg Q \rangle$ holds tightly on $\langle s_0, s_1 \rangle$.
3. $Q$ and $\neg S$ hold in $s_1$. Then $\langle R, \neg S \rangle$ holds tightly on $\langle s_0, s_1 \rangle$.
4. $\neg Q$ and $\neg S$ hold in $s_1$. Then both $\langle P, \neg Q \rangle$ and $\langle R, \neg S \rangle$ hold tightly on $\langle s_0, s_1 \rangle$.

Notice that in all four cases, there exists a subsequence of $\langle s_0, s_1, s_2 \rangle$ on which $A$ holds tightly. $\alpha$ is therefore an implicant of $A$, which by our argument above means that $\alpha$ is a sequential constraint. It is equivalent to the logical/temporal dependency

$$\text{If } (P \wedge R), \text{ then } T \text{ two states later} \qquad (2)$$

So our reasoning using sequential constraints has shown that Statement 2 follows as a logical consequence from Statements 1a, 1b and 1c. In our reasoning, furthermore, we have deduced that $\alpha$ is an implicant of $A$ with no assumptions whatsoever about the underlying state space. So this result is applicable to all discrete-time systems (generalized Kripke structures) as required by the definition of an implicant. (Section 5.4 shows how to deduce $\alpha$ from $A$ using sequential resolution.)

## 1.5 Combinatorics and Logic

The central problem addressed by the present theory is determining the implicants of a regular set of sequences of Boolean expressions. The main result of the theory is a necessary and sufficient condition for a sequence of Boolean expressions to be an implicant of such a regular set. Arriving at this result entails the proof of theorems at two levels, the *combinatorics level* (discussed in Section 3) and the *logic level* (discussed in Section 4), with results at the combinatorics level providing the foundation for results at the logic level.

At both levels, a directed graph with labeled arcs serves as a finite state automaton which accepts a regular set of sequences – sequences of *sets* at the combinatorics level and sequences of *Boolean expressions* at the logic level. Also at both levels, there is the notion of an *implicant* of a regular set of sequences – or, equivalently, of a directed graph with labeled arcs defining such a set.

At the combinatorics level, the objects of study are:

- Sequences of sets
- *Set graphs*: Directed graphs in which each arc is labeled with a set
- *Links* of a set graph $G$: Ordered triples $\langle \textit{aft}, \alpha, \textit{fore} \rangle$ satisfying special properties, where *aft* and *fore* are each a set of sets of vertices of $G$ and $\alpha$ is a sequence of sets
- *Elaborations* of a set graph $G$: Set graphs in which each vertex is an ordered pair $\langle \textit{aft}, \textit{fore} \rangle$ satisfying special properties, where *aft* and *fore* are each a set of sets of vertices of $G$

The main result at the combinatorics level (Theorem 3.6) is a necessary and sufficient condition for a sequence of sets to be an implicant of a set graph:

*A sequence of sets $\alpha$ is an implicant of a set graph $G$*

*if and only if*

*a subsequence of $\alpha$ is accepted by an elaboration of $G$*

At the logic level, the objects of study are:

- Sequences of Boolean expressions
- *Boolean graphs*: Directed graphs in which each arc is labeled with a Boolean expression
- *Links* of a Boolean graph $G$: Ordered triples $\langle \textit{aft}, \alpha, \textit{fore} \rangle$ satisfying special properties, where *aft* and *fore* are each a set of sets of vertices of $G$ and $\alpha$ is a sequence of Boolean expressions
- *Elaborations* of a Boolean graph $G$: Boolean graphs in which each vertex is an ordered pair $\langle \textit{aft}, \textit{fore} \rangle$ satisfying special properties, where *aft* and *fore* are each a set of sets of vertices of $G$

The connection between these four constructs and their counterparts at the combinatorics level is provided by the function $L$ associated with a generalized Kripke structure $(S, B, L)$ over a set of atomic propositions $AP$ (see Sections 4.1 and 4.2). $L$ maps an atomic proposition into the set of states in which the proposition holds (is *true*), while extension of $L$ map: (a) a Boolean expression into the set of states in which the Boolean expression holds, (b) a sequence of Boolean expressions into a sequence of sets of states and (c) a Boolean graph into a set(-of-states) graph.

The main result at the logic level, and of the theory, (Theorem 4.10) is a necessary and sufficient condition for a sequence of Boolean expressions to be an implicant of a Boolean graph:

*A sequence of Boolean expressions α is*

*an implicant of a Boolean graph G*

*if and only if*

*a subsequence of α is accepted by an elaboration of G*

So the problem of determining whether a sequence of Boolean expressions $\alpha$ is an implicant of a regular set of sequences of Boolean expressions defined by a Boolean graph *G* reduces to the problem of constructing an elaboration of *G* that accepts a subsequence of $\alpha$. Sections 5 and 6 describe two different methods, *sequential resolution* and *normalization*, for constructing such elaborations.

## 1.6 Resolution and Normalization

Boolean resolution is a powerful inference rule in Boolean logic. It comes in two forms. The *disjunctive* form [6,7] – which is sometimes called *consensus* [8] – is applied to a sum of products, while the *conjunctive* form [9] is applied to a product of sums. *Sequential resolution*, a generalization of the disjunctive form, is applied to a succession of elaborations of a Boolean graph *G* starting with an *initial elaboration* that is isomorphic to *G*. Each resolution is performed on two equal-length paths in an elaboration, and yields a new path that is the same length as the two resolved paths. This *inferred path* is added to the existing elaboration to create a new elaboration which accepts an expanded set of sequences of Boolean expressions. These added sequences represent logical/temporal dependencies that are *inferred* from the dependencies associated with the previous elaboration.

*Normalization*, the second method for constructing elaborations, starts with two Boolean graphs: (1) a graph representing a set of *known* logical/temporal dependencies and (2) a graph representing a set of *conjectured* logical/temporal dependencies. The first graph typically represents a system (model), while the second represents properties that one conjectures about the behavior of the system. Normalization determines which of those conjectured properties are satisfied by the system. The process involves transforming the conjectured graph, using arcs from the system graph, into an elaboration of the system graph. The resulting *verified* graph satisfies two properties:

1. Each sequence of Boolean expressions accepted by the verified graph is (a) an implicant of the system graph and (b) a subsequence of a sequence accepted by the conjectured graph.

2. For each sequence of Boolean expressions $\alpha$ that is (a) an implicant of the system graph and (b) accepted by the conjectured graph, there exists a subsequence of $\alpha$ that is accepted by the verified graph

The process of normalization is thus able to extract from a set of conjectured logical/temporal dependencies those dependencies that follow from a set of known dependencies. This capability means that someone who is unsure about a system's exact behavior can make an overly broad conjecture about that behavior – a conjecture known to be *false* – in order to find a version of the conjecture that is *true*.

## 1.7 Related Work

The need for mathematical/formal techniques for verifying the behavior of digital systems has long been recognized in the research community. Three approaches to *formal verification* are most relevant here: (1) the early work on sequential constraints, (2) *model checking* [10,11] and (3) *theorem proving* [12,13].

The very earliest work with sequential constraints sought to provide mathematical foundations for secure computation [14,15]. Later work broadened the scope to specifying and verifying the behavior of distributed systems [16,17]. The present work is an expansion of the theory developed at that time [18,19].

*Model checking is an automatic verification technique for finite state concurrent systems. In this approach to verification, temporal logic specifications are checked by an exhaustive search of the state space of the concurrent system* [20]. There are similarities, but also significant differences, between model checking and the present approach.

1. In both approaches, finite state automata play a central role. In the case of model checking, a finite state automaton describes a system's state space – that is, the set of all *allowed* system state sequences. In the present approach, a finite state automaton describes a set of *disallowed* system state sequences – but not necessarily *all* disallowed state sequences. This last difference is significant. In model checking, all allowed system behaviors must be represented in the finite state automaton because to ignore any allowed behaviors would jeopardize the soundness of proofs. The methodology described here, however, relies on *deductive reasoning* (see next point), and therefore ignoring disallowed behaviors affects what is provable but does not affect the soundness of proofs (unless one is trying to prove the *absence* of certain sequential constraints. See Point 6.).

2. In model checking, verification entails an exhaustive search of a system's state space. In the present approach, verification is accomplished through deductive reasoning – entirely within the realm of logical/temporal dependencies – using either sequential resolution or normalization. No attempt is made to model a system's state-transition function (nor is such a function even assumed to exist), and no attempt is made to explore, traverse or enumerate a system's state space.

3. A basic assumption (axiom) of model checking is that a system state is *total* – that is, a system state completely determines, through the system's state-transition function, the set of all possible next-states. But there are situations where it is useful to reason about *partial states*, and in these circumstances it is a *sequence of partial states* that determines a system's possible next-states. For example, in analyzing a system's behavior we may want to distinguish between those state variables that are *hidden* and those that are *visible* (typically the input/output variables), and we may wish to reason about the behavior of just the visible state variables. In the present approach, the assumption that a state is total is replaced by a more basic assumption, Axiom 1. The increased generality afforded by this axiom means that we can derive and reason about the sequences of partial states that define a system's visible (*black box*) behavior.

4. Because model checking involves an exhaustive search of a system's state space, it must deal with the exponential growth in the size of that space. In fact, *the main challenge in model checking is dealing with the state space explosion problem* [20]. In the present approach, there is no state space explosion problem because when a new component or instruction is added to a system, the sequential constraints associated with that component or instruction are *added* to the set of sequential constraints defining the system. The regular expression or finite-state automaton for the set of sequential constraints defining a system thus grows *linearly*, not *exponentially*, with the size of a system. However, although a combinatorial explosion does not occur in the mere act of modeling a system, as it does in model checking, an explosion is still possible – although not inevitable – through repeated applications of sequential resolution or in the normalization process.

5. In model checking, there are two types of constructs: finite-state automata for describing systems and temporal logic specifications for describing logical/temporal properties. In the present approach, there is only one type of construct for describing both systems and properties: a regular set of sequential constraints equivalently defined by either a regular expression or finite state automaton.

6. The expressive power of the temporal logics commonly used in model checking and the expressive power of sequential constraints differ in two fundamental respects: (1) The temporal logics of model checking are able to reason not only about properties involving finite behaviors but also *infinite* behaviors. So, for example, they can express the property *If P*, *then eventually Q*. Sequential constraints cannot express these properties since each constraint is restricted to being finite (although a *set* of constraints may be infinite). (2) The temporal logics of model checking can express properties involving allowed (permitted, possible) patterns of behavior. So, for example, these temporal logics can express the property *If P*, *then Q is possible in the next state*. Sequential constraints cannot express such properties directly since sequential constraints describe disallowed behavior. These properties can only be expressed indirectly by the *absence* of sequential constraints. The property *If P*, *then Q is possible in the next state*, for example, is expressed by the absence of sequential constraints of the form $\langle R, Q \rangle$, where $R$ is a Boolean expression such that $P \wedge R$ is satisfiable.

*Theorem proving* employs higher-order logic, predefined theories and a variety of inference procedures to provide exceptionally powerful and expressive proof systems. As with model checking, there are similarities, but also significant differences, between theorem proving and the present approach.

1. Like theorem proving, the present approach supports *deductive reasoning* by which new properties are inferred from existing properties using *inference rules*. Sequential resolution is just such a rule, and although normalization does not fit the definition of an inference rule, embedded within the algorithm are *micro inferences* in which new links are inferred from existing links.

2. Unlike theorem proving, which is largely symbolic and requires considerable human guidance, the present approach is based on a body of combinatorial mathematics, and that mathematics supports algorithmic proof systems employing either normalization or sequential resolution.

3. The theory described here is essentially an extension of propositional logic to handle sequential behavior. Although this logic has been further extended with *uninterpreted* *functions* (to be described in a future paper), it will be necessary to incorporate techniques from theorem proving in order to achieve the power and expressiveness of theorem proving together with the algorithmic techniques of the present approach.

## 2. PRELIMINARIES

Sections 2.1, 2.2, 2.3 and 2.4 provide notations and terminology for the familiar concepts of *ordered pair*, *sequence*, *Cartesian product* and *directed graph with labeled arcs*, respectively. Sections 2.5 and 2.6 introduce some less-familiar concepts: *De Morgan algebras* and a particular class of such algebras in which the elements are *sets of sets*.

### 2.1 Ordered Pairs

For an *ordered pair* $\langle x, y \rangle$,

$$aft(\langle x, y \rangle) = x$$

$$fore(\langle x, y \rangle) = y$$

### 2.2 Sequences

A *sequence* is a finite ordered list of elements, written $\langle x_0, x_1, \ldots, x_{n-1} \rangle$. The set of all sequences over a set of elements $E$ is denoted $E^*$. A *subsequence* of sequence $\alpha$ is a sequence of consecutive elements appearing in $\alpha$. A sequence $\alpha_P$ is a *prefix* of sequence $\alpha$ if and only if $\alpha$ begins with the sequence $\alpha_P$. A sequence $\alpha_S$ is a *suffix* of sequence $\alpha$ if and only if $\alpha$ ends with the sequence $\alpha_S$. The *concatenation* of sequences $\alpha_1$ and $\alpha_2$ is denoted $\alpha_1 \bullet \alpha_2$. The *length* of sequence $\alpha$ is denoted $|\alpha|$. Thus,

$$\langle b, c, d \rangle \text{ is a subsequence of } \langle a, b, c, d, e \rangle$$

$$\langle a, b, c \rangle \text{ is a prefix of } \langle a, b, c, d, e \rangle$$

$$\langle c, d, e \rangle \text{ is a suffix of } \langle a, b, c, d, e \rangle$$

$$\langle a, b \rangle \bullet \langle c, d, e \rangle = \langle a, b, c, d, e \rangle$$

$$|\langle a, b, c, d, e \rangle| = 5$$

### 2.3 Cartesian Products

The *Cartesian product* (*cross product*) of sets $A$ and $B$, written $A \times B$, is the set of all ordered pairs $\langle a, b \rangle$ such that $a \in A$ and $b \in B$. If $\alpha = \langle \alpha_0, \alpha_1, \ldots, \alpha_{n-1} \rangle$ is a sequence of sets, then $\times \alpha$ denotes the set of all sequences $\langle x_0, x_1, \ldots, x_{n-1} \rangle$ such that $x_i \in \alpha_i$ for $0 \leq i < n$. Thus

$$\times \langle \{a, b, c\}, \{d\}, \{e, f\} \rangle = \{\langle a, d, e \rangle, \langle a, d, f \rangle, \langle b, d, e \rangle, \langle b, d, f \rangle, \langle c, d, e \rangle, \langle c, d, f \rangle\}$$

## 2.4 Directed Graphs with Labeled Arcs

Set graphs and Boolean graphs play a central role in the theory that follows, and although their structures differ, they are both *directed graphs with labeled arcs*. Associated with such a graph are a finite set of *vertices V*, a set of *labels L* and a finite set of *arcs* $A \subseteq V \times L \times V$. Each arc is therefore of the form $\langle v_i, l, v_j \rangle$, where $v_i$ and $v_j$ are vertices and $l$ is a label.

We adopt the following notation and terminology for a directed graph with labeled arcs $G$. For an arc $\langle v_i, l, v_j \rangle$ of $G$,

$$tail(\langle v_i, l, v_j \rangle) = v_i$$
$$label(\langle v_i, l, v_j \rangle) = l$$
$$head(\langle v_i, l, v_j \rangle) = v_j$$

A *path* in $G$ is a non-null sequence of arcs $\langle a_0, a_1, \ldots, a_{n-1} \rangle$ such that for every pair of successive arcs $a_i$ and $a_{i+1}$ in $\langle a_0, a_1, \ldots, a_{n-1} \rangle$, $head(a_i) = tail(a_{i+1})$. For a path $\langle a_0, a_1, \ldots, a_{n-1} \rangle$ in $G$,

$$tail(\langle a_0, a_1, \ldots, a_{n-1} \rangle) = tail(a_0)$$
$$label(\langle a_0, a_1, \ldots, a_{n-1} \rangle) = \langle label(a_0), label(a_1), \ldots, label(a_{n-1}) \rangle$$
$$head(\langle a_0, a_1, \ldots, a_{n-1} \rangle) = head(a_{n-1})$$

An *initial vertex* of $G$ is a vertex of $G$ with no incoming arcs – that is, a vertex $v$ for which there does not exist an arc $a$ of $G$ such that $head(a) = v$. A *terminal vertex* of $G$ is a vertex of $G$ with no outgoing arcs – that is, a vertex $v$ for which there does not exist an arc $a$ of $G$ such that $tail(a) = v$. An *interior vertex* of $G$ is a vertex of $G$ that is neither an initial vertex of $G$ nor a terminal vertex of $G$. The set of interior vertices of $G$ is denoted $IV(G)$. $G$ *accepts* a sequence of labels $\alpha$ if and only if there exists a path $\mu$ in $G$ such that the following three properties hold: (1) $tail(\mu)$ is an initial vertex of $G$, (2) $\alpha = label(\mu)$ and (3) $head(\mu)$ is a terminal vertex of $G$.

*Comment*: Restricting initial vertices to just those vertices with no incoming arcs and terminal vertices to just those vertices with no outgoing arcs does not limit the generality of directed graphs with labeled arcs in defining sets of *disallowed sequences*. That's because prepending or appending arbitrary sequences to a disallowed sequence of Boolean expressions or disallowed sequence of sets of states only yields another, *weaker*, disallowed sequence.

## 2.5 De Morgan Algebras

Boolean algebra is well-known in both logic and computer science, but there is another algebra, not so well-known, that satisfies most – but not all – of the familiar properties of a Boolean algebra. It is a *De Morgan algebra* [Białynicki-Birula & Rasiowa 1957; Białynicki-Birula 1957; Kalman 1958; Balbes & Dwinger 1974; Cignoli 1975; Reed 1979; Sankappanavar 1980; Figallo & Monteiro 1981].

*Definition* 2.1. A *De Morgan algebra* is a 4-tuple $(A, \wedge, \vee, \sim)$, where $A$ is a set of elements, $\wedge$ and $\vee$ are binary operations on $A$ and $\sim$ is a unary operation on $A$ such that for all $a, b \in A$, the following three axioms hold:

1. $(A, \wedge, \vee)$ forms a distributive lattice
2. $\sim(a \wedge b) = (\sim a) \vee (\sim b)$  and  $\sim(a \vee b) = (\sim a) \wedge (\sim b)$ (De Morgan's laws)
3. $\sim\sim a = a$ (involution)

Notably absent from these axioms is the law of the excluded middle: $(a \vee \sim a) = 1$ or, equivalently, $(a \wedge \sim a) = 0$. The absence of this law is what distinguishes a De Morgan algebra from a Boolean algebra.

*Definition* 2.2. Let $(A, \wedge, \vee)$ be a lattice. Then the partial order $\leq$ on $A$ is defined such that: $a \leq b$ if and only if $a = a \wedge b$ (or, equivalently, $b = a \vee b$).

PROPERTY 2.1. *If* $(A, \wedge, \vee, \sim)$ *is a De Morgan algebra, then for all* $a, b \in A$:
(a) $\sim a \leq b \Leftrightarrow \sim b \leq a$
(b) $\sim a \leq c$ and $\sim b \leq d \Rightarrow \sim(a \wedge b) \leq (c \vee d)$
(c) $\sim a \leq c$ and $\sim b \leq d \Rightarrow \sim(a \vee b) \leq (c \wedge d)$

## 2.6 Sets of Sets

Sets of sets of vertices – together with are two binary operations and a unary operation defined on them – play a key role in characterizing the implicants of both set graphs and Boolean graphs.

*Definition* 2.3. For a set $V$,

$$SoS(V) = \{sos \subseteq 2^V \mid \text{For all } set_i, set_j \in sos: (set_i \subseteq set_j) \Rightarrow (set_i = set_j)\}$$

The elements of *SoS*($V$) are thus those sets of subsets of $V$ whose member sets are pairwise incomparable with respect to set inclusion ($\subseteq$). For example,

$SoS(\{v_0, v_1, v_2\}) = \{$ $\{\}$,
$\{\{v_0, v_1, v_2\}\}$,
$\{\{v_0, v_1\}\}$,
$\{\{v_0, v_2\}\}$,
$\{\{v_1, v_2\}\}$,
$\{\{v_0, v_1\}, \{v_0, v_2\}\}$,
$\{\{v_0, v_1\}, \{v_1, v_2\}\}$,
$\{\{v_0, v_2\}, \{v_1, v_2\}\}$,
$\{\{v_0\}\}$,
$\{\{v_1\}\}$,
$\{\{v_2\}\}$,
$\{\{v_0, v_1\}, \{v_0, v_2\}, \{v_1, v_2\}\}$,
$\{\{v_0\}, \{v_1, v_2\}\}$,
$\{\{v_1\}, \{v_0, v_2\}\}$,
$\{\{v_2\}, \{v_0, v_1\}\}$,
$\{\{v_0\}, \{v_1\}\}$,
$\{\{v_0\}, \{v_2\}\}$,
$\{\{v_1\}, \{v_2\}\}$,
$\{\{v_0\}, \{v_1\}, \{v_2\}\}$,
$\{\{\}\}$ $\}$

Two binary operations, $\wedge$ and $\vee$, and a unary operation, $\sim$, are now defined on *SoS*($V$). All three operations make use of the $min_\subseteq$ function which selects those member sets of a set of sets that are minimal with respect to set inclusion.

*Definition* 2.4. For a set of sets *sos*,

$$min_\subseteq(sos) = \{set_j \in sos \mid \text{For all } set_i \in sos: (set_i \subseteq set_j) \Rightarrow (set_i = set_j)\}$$

PROPERTY 2.2 *If $V$ is a set and $sos \subseteq 2^V$, then $min_\subseteq(sos) \in SoS(V)$.*

*Definition* 2.5. For a finite set of elements $V$ and for $sos_i, sos_j \in SoS(V)$,

$$sos_i \vee sos_j = min_{\subseteq}(sos_i \cup sos_j)$$

$$sos_i \wedge sos_j = min_{\subseteq}(\{set_i \cup set_j \mid set_i \in sos_i \text{ and } set_j \in sos_j\})$$

$$\sim sos_i = min_{\subseteq}(\{set_j \subseteq V \mid \text{For all } set_i \in sos_i: set_i \cap set_j \neq \varnothing\})$$

$sos_i \vee sos_j$ is thus the union of $sos_i$ and $sos_j$ with all but the minimal sets (with respect to set inclusion) discarded. $sos_i \wedge sos_j$ is the set of pairwise unions of sets $set_i$ from $sos_i$ and sets $set_j$ from $sos_j$ with all but the minimal sets discarded. $\sim sos_i$ is the set of minimal sets $set_j$ such that for all $set_i$ in $sos_i$, the intersection of $set_i$ and $set_j$ is nonempty.

PROPERTY 2.3. *If V is a finite set, then $(SoS(V), \wedge, \vee, \sim)$ is a De Morgan algebra.*

PROPERTY 2.4. *If V is a finite set, then for the lattice $(SoS(V), \wedge, \vee)$ and $sos_i, sos_j \in SoS(V)$,*

$$sos_i \leq sos_j$$

*if and only if*

*For all $set_i \in sos_i$, there exists $set_j \in sos_j$ such that $set_j \subseteq set_i$*

PROPERTY 2.5. *If V is a finite set, then*

(a) $\{\} \in SoS(V)$ *and* $\{\{\}\} \in SoS(V)$

(b) $\sim\{\} = \{\{\}\}$ *and* $\sim\{\{\}\} = \{\}$

(c) *For all $sos_i \in SoS(V)$, $\{\} \leq sos_i \leq \{\{\}\}$*

(d) *For all $sos_i, sos_j \in SoS(V)$, $(sos_i \wedge sos_j = \{\}) \Leftrightarrow (sos_i = \{\} \text{ or } sos_j = \{\})$*

(e) *For all $sos_i, sos_j \in SoS(V)$, $(sos_i \vee sos_j = \{\{\}\}) \Leftrightarrow (sos_i = \{\{\}\} \text{ or } sos_j = \{\{\}\})$*

These ideas are illustrated in Figure 1 and Table 1 for $(SoS(\{v_0, v_1, v_2\}), \wedge, \vee, \sim)$. In the lattice of Figure 1, $sos_i \vee sos_j$ is the least upper bound (join) of $sos_i$ and $sos_j$, while $sos_i \wedge sos_j$ is the greatest lower bound (meet) of $sos_i$ and $sos_j$.

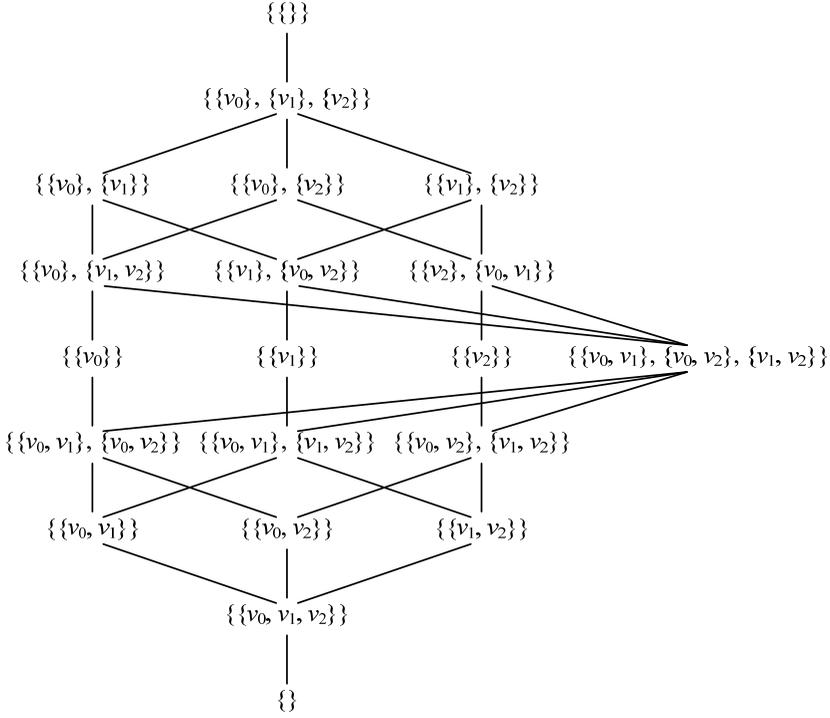

FIG. 1. Distributive Lattice for $(SoS(\{v_0, v_1, v_2\}), \wedge, \vee)$

TABLE 1. Inverses of Elements in $SoS(\{v_0, v_1, v_2\})$

| | | |
|---:|:---:|:---:|
| $\sim\{\}$ | = | $\{\{\}\}$ |
| $\sim\{\{v_0, v_1, v_2\}\}$ | = | $\{\{v_0\}, \{v_1\}, \{v_2\}\}$ |
| $\sim\{\{v_0, v_1\}\}$ | = | $\{\{v_0\}, \{v_1\}\}$ |
| $\sim\{\{v_0, v_2\}\}$ | = | $\{\{v_0\}, \{v_2\}\}$ |
| $\sim\{\{v_1, v_2\}\}$ | = | $\{\{v_1\}, \{v_2\}\}$ |
| $\sim\{\{v_0, v_1\}, \{v_0, v_2\}\}$ | = | $\{\{v_0\}, \{v_1, v_2\}\}$ |
| $\sim\{\{v_0, v_1\}, \{v_1, v_2\}\}$ | = | $\{\{v_1\}, \{v_0, v_2\}\}$ |
| $\sim\{\{v_0, v_2\}, \{v_1, v_2\}\}$ | = | $\{\{v_2\}, \{v_0, v_1\}\}$ |
| $\sim\{\{v_0\}\}$ | = | $\{\{v_0\}\}$ |
| $\sim\{\{v_1\}\}$ | = | $\{\{v_1\}\}$ |
| $\sim\{\{v_2\}\}$ | = | $\{\{v_2\}\}$ |
| $\sim\{\{v_0, v_1\}, \{v_0, v_2\}, \{v_1, v_2\}\}$ | = | $\{\{v_0, v_1\}, \{v_0, v_2\}, \{v_1, v_2\}\}$ |

*Comment*: The definition and properties of $(SoS(V), \wedge, \vee, \sim)$ can be more easily understood when each set of sets in $SoS(V)$ is interpreted as a reduced, negation-free Boolean sum of products. In this interpretation, $\{\}$ corresponds to *false*, $\{\{\}\}$ corresponds to *true*, $\leq$ corresponds to $\Rightarrow$ (implication) and the operations $\wedge$, $\vee$ and $\sim$ correspond, respectively, to conjunction (*AND*), disjunction (*OR*) and Boolean dual (the interchange of *true* and *false* and *AND* and *OR*). This interpretation is significant – aside from its pedagogical value – because it means that in algorithms based on the present theory, the two-level sets of sets in *SoS*(*V*) can be replaced by arbitrarily nested constructs, and techniques for Boolean minimization and equivalence can then be applied to these structures.

3. COMBINATORICS

The main results of this paper are in Section 4, where the objects of study are sequences of Boolean expressions and directed graphs in which each arc is labeled with a Boolean expression. This section provides the mathematical foundations for those results. Here, the objects of study are sequences of sets of elements and directed graphs in which each arc is labeled with a set of elements. At the logic level, each such set of elements will be interpreted as the set of states in which a Boolean expression holds (is *true*).

The theory at the combinatorics level proceeds as follows:

- Section 3.1 presents the *fundamental theorem* (Theorem 3.1), a basic result in combinatorial mathematics which provides a necessary and sufficient condition for a product (composite) relation to be total. This result is the foundation upon which the subsequent theory rests.

- Section 3.2 introduces *set graphs*, directed graphs in which each arc is labeled with a set of elements. A set graph plays the role of a finite state automaton and defines a regular set of sequences of sets. An *implicant* of a set of sequences of sets – or of a set graph defining a regular set of such sequences – is the combinatorial counterpart to the notion of sequential implicant defined above.

- Section 3.3 defines a *link* of a set graph $G$ as an ordered triple $\langle aft, \alpha, fore \rangle$ satisfying special properties, where *aft* and *fore* are each a set of sets of interior vertices of $G$ and $\alpha$ is a sequence of sets. The links of a set graph $G$ are the key to characterizing the implicants of $G$ since a sequence of sets $\alpha$ is an implicant of $G$ if and only if $\langle \{\{\}\}, \alpha, \{\{\}\} \rangle$ is a link of $G$ (Lemma 3.1). Theorem 3.2 provides a sufficient condition for two links to be *concatenated*: If $\langle aft_1, \alpha_1, fore_1 \rangle$ and $\langle aft_2, \alpha_2, fore_2 \rangle$ are links of set graph $G$ such that $\sim fore_1 \leq aft_2$, then $\langle aft_1, \alpha_1 \bullet \alpha_2, fore_2 \rangle$ is a link of $G$.

- Section 3.4 describes the special properties of those links ⟨aft, α, fore⟩ such that $|α| = 1$. These *links of length 1* are of interest because *elaborations* are defined in Section 3.6 solely in terms of such links and because the manipulations used in sequential resolution (described in Section 5) and the process of normalization (described in Section 6) involve only links of length 1. The *initial links of length 1* of a set graph $G$ are derived from the arcs of $G$ via Theorem 3.3. Additional links of length 1 of $G$ are derived from existing links of length 1 through the *micro inferences* described in Theorem 3.4.

- Section 3.5 defines a *forwards-* (*backwards-*) *maximal link* as a link ⟨aft, α, fore⟩ such that *fore* (*aft*) is the maximum set of sets of interior vertices of the set graph $G$ – with respect to the partial order $≤$ – such that ⟨aft, α, fore⟩ is a link of $G$. A key result involving such links (Theorem 3.5) allows us to construct for any implicant of $G$ an elaboration that accepts a subsequence of that implicant.

- Section 3.6 defines an *elaboration* of a set graph $G$ as another set graph $E$ in which each vertex is an ordered pair ⟨aft, fore⟩ satisfying special properties, where *aft* and *fore* are each a set of sets of interior vertices of $G$. The main result at the combinatorics level is Theorem 3.6 which states that a sequence of sets $α$ is an implicant of a set graph $G$ if and only if a subsequence of $α$ is accepted by an elaboration of $G$.

## 3.1 The Fundamental Theorem

Let $A$, $B$ and $C$ be sets with $B$ finite, and let $R_{AB} \subseteq A \times B$ and $R_{BC} \subseteq B \times C$ be binary relations. The *product* (*composite*) *relation* $R_{AB} \bullet R_{BC}$ is the relation

$$\{⟨a, c⟩ \in A \times C \mid \text{There exists } b \in B \text{ such that } aR_{AB}b \text{ and } bR_{BC}c\}$$

Question: Under what circumstances is $R_{AB} \bullet R_{BC}$ total – that is, under what circumstances does $R_{AB} \bullet R_{BC} = A \times C$? For example, the product of the relations in Figure l(a) is not total because there does not exist $b \in B$ such that $a_1 R_{AB} b$ and $b R_{BC} c_1$, but the product in Figure l(b) is total. The answer is provided by Theorem 3.1.

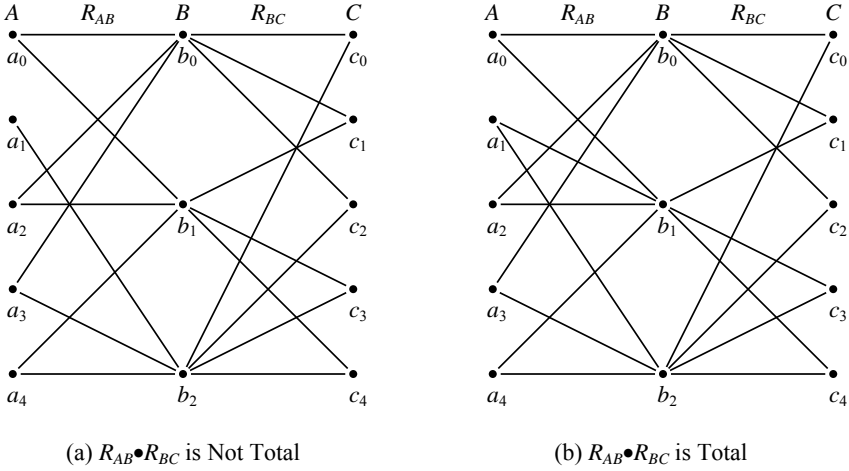

(a) $R_{AB} \bullet R_{BC}$ is Not Total    (b) $R_{AB} \bullet R_{BC}$ is Total

FIG. 2. Product Relations

THEOREM 3.1. (Furtek 1984). *Let A, B and C be sets with B finite, and let $R_{AB} \subseteq (A \times B)$ and $R_{BC} \subseteq (B \times C)$ be binary relations. Then*

$$R_{AB} \bullet R_{BC} = A \times C$$

*if and only if*

$$\sim min_{\subseteq}(\{B_i \subseteq B \mid R_{AB}^{-1}(B_i) = A\}) \leq min_{\subseteq}(\{B_j \subseteq B \mid R_{BC}(B_j) = C\})$$

So we see that $R_{AB} \bullet R_{BC}$ is total if and only if a certain relationship exists between two sets of subsets of *B*. From the definition of ~ and Property 2.4, we see that that relationship can be restated as follows: For each minimal subset $B_k$ of *B* that intersects each set in

$$min_{\subseteq}(\{B_i \subseteq B \mid R_{AB}^{-1}(B_i) = A\}),$$

there exists a subset of $B_k$ in

$$min_{\subseteq}(\{B_j \subseteq B \mid R_{BC}(B_j) = C\}).$$

But what is the significance of these two sets of subsets of *B*? We observe first that $\{B_i \subseteq B \mid R_{AB}^{-1}(B_i) = A\}$ is the set of subsets $B_i$ of *B* whose image under the inverse relation $R_{AB}^{-1}$ is *A*. In other words, for all $a \in A$, there exists $b \in B_i$ such that $bR_{AB}^{-1}a$ or, equivalently $aR_{AB}b$. Thus

$$min_{\subseteq}(\{B_i \subseteq B \mid R_{AB}^{-1}(B_i) = A\})$$

is the set of minimal subsets (with respect to set inclusion) $B_i$ of *B* such that for all $a \in A$, there exists $b \in B_i$ such that $aR_{AB}b$. Similarly,

$$min_{\subseteq}(\{B_j \subseteq B \mid R_{BC}(B_j) = C\})$$

is the set of minimal subsets (with respect to set inclusion) $B_j$ of $B$ such that for all $c \in C$, there exists $b \in B_j$ such that $bR_{BC}c$.

*Comment*: The asymmetry between $R_{AB}$ and $R_{BC}$ in the property

$$\sim min_{\subseteq}(\{B_i \subseteq B \mid R_{AB}^{-1}(B_i) = A\}) \leq min_{\subseteq}(\{B_j \subseteq B \mid R_{BC}(B_j) = C\})$$

may appear incongruous with the symmetry between $R_{AB}$ and $R_{BC}$ in the property $R_{AB} \bullet R_{BC} = A \times C$. But that asymmetry is only apparent since by Property 2.1(a) the above property is equivalent to:

$$\sim min_{\subseteq}(\{B_j \subseteq B \mid R_{BC}(B_j) = C\}) \leq min_{\subseteq}(\{B_i \subseteq B \mid R_{AB}^{-1}(B_i) = A\})$$

To illustrate the above ideas, consider again the two binary relations in Figure 2(a). Their product is not total, and the property in Theorem 3.1 is not satisfied (see the partial order in Figure 1):

$$min_{\subseteq}(\{B_i \subseteq B \mid R_{AB}^{-1}(B_i) = A\}) \;=\; \{\{b_0, b_2\}, \{b_1, b_2\}\}$$
$$\sim min_{\subseteq}(\{B_i \subseteq B \mid R_{AB}^{-1}(B_i) = A\}) \;=\; \{\{b_0, b_1\}, \{b_2\}\}$$
$$min_{\subseteq}(\{B_j \subseteq B \mid R_{BC}(B_j) = C\}) \;=\; \{\{b_0, b_1\}, \{b_0, b_2\}, \{b_1, b_2\}\}$$

Now consider the two binary relations Figure 2(b). Their product is total, and the property in Theorem 3.1 is satisfied:

$$min_{\subseteq}(\{B_i \subseteq B \mid R_{AB}^{-1}(B_i) = A\}) \;=\; \{\{b_0, b_1\}, \{b_0, b_2\}, \{b_1, b_2\}\}$$
$$\sim min_{\subseteq}(\{B_i \subseteq B \mid R_{AB}^{-1}(B_i) = A\}) \;=\; \{\{b_0, b_1\}, \{b_0, b_2\}, \{b_1, b_2\}\}$$
$$min_{\subseteq}(\{B_j \subseteq B \mid R_{BC}(B_j) = C\}) \;=\; \{\{b_0, b_1\}, \{b_0, b_2\}, \{b_1, b_2\}\}$$

## 3.2 Set Graphs

A set graph – the combinatorial counterpart to a Boolean graph – defines a regular set of sequences of sets as described in Section 2.3.

*Definition* 3.1. A *set graph* is a triple $(V, S, A)$, where

1. $V$ is a finite set of *vertices*
2. $S$ is a set of *elements*
3. $A \subseteq (V \times 2^S \times V)$ is a finite set of *labeled arcs*

The *labels* on the arcs of a set graph are thus subsets of $S$, which at the combinatorics level is just a set of arbitrary elements. At the logic level, $S$ will be interpreted as the set of states associated with a generalized Kripke structure.

*Example*: Figure 3 depicts a set graph in which the set of elements is $\{s_0, s_1, s_2, s_3, s_4, s_5, s_6, s_7, s_8, s_9, s_{10}, s_{11}, s_{12}, s_{13}, s_{14}, s_{15}\}$. (This set graph corresponds to the Boolean graph in Section 4.2.) The initial vertices of the graph are $v_0$, $v_2$, $v_5$, $v_8$, $v_{11}$, $v_{14}$ and $v_{16}$; the terminal vertices are $v_1$, $v_4$, $v_7$, $v_{10}$, $v_{13}$, $v_{15}$ and $v_{17}$; and the interior vertices are $v_3$, $v_6$, $v_9$ and $v_{12}$.

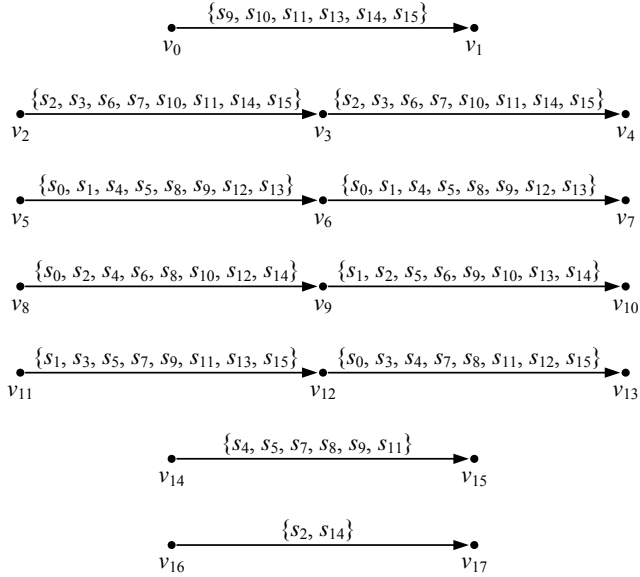

FIG. 3. A Set Graph

The sequences of sets accepted by the graph are:

$$\langle \{s_9, s_{10}, s_{11}, s_{13}, s_{14}, s_{15}\} \rangle$$
$$\langle \{s_2, s_3, s_6, s_7, s_{10}, s_{11}, s_{14}, s_{15}\}, \{s_2, s_3, s_6, s_7, s_{10}, s_{11}, s_{14}, s_{15}\} \rangle$$
$$\langle \{s_0, s_1, s_4, s_5, s_8, s_9, s_{12}, s_{13}\}, \{s_0, s_1, s_4, s_5, s_8, s_9, s_{12}, s_{13}\} \rangle$$
$$\langle \{s_0, s_2, s_4, s_6, s_8, s_{10}, s_{12}, s_{14}\}, \{s_1, s_2, s_5, s_6, s_9, s_{10}, s_{13}, s_{14}\} \rangle$$
$$\langle \{s_1, s_3, s_5, s_7, s_9, s_{11}, s_{13}, s_{15}\}, \{s_0, s_3, s_4, s_7, s_8, s_{11}, s_{12}, s_{15}\} \rangle$$
$$\langle \{s_4, s_5, s_7, s_8, s_9, s_{11}\} \rangle$$
$$\langle \{s_2, s_{14}\} \rangle$$

An *implicant* of a set of sequences of sets is the combinatorial counterpart to an implicant of a set of sequences of Boolean expressions (see Definition 1.3).

*Definition* 3.2. An *implicant* of a set of sequences of sets $A$ is a sequence of sets $\alpha$ such that for all $\omega \in \times\alpha$, there exists a subsequence $\omega'$ of $\omega$ and a sequence of sets $\alpha'$ in $A$ such that $\omega' \in \times \alpha'$. An *implicant* of a set graph $G = (V, S, A)$ is a sequence $\alpha$ of subsets of $S$ such that $\alpha$ is an implicant of the set of sequences of sets accepted by $G$.

## 3.3 Links

The *links* of a set graph are the key to characterizing the implicants of the set graph. (Recall that $IV(G)$ is the set of interior vertices of $G$ and that $SoS(V)$ is the set of sets of subsets of $V$ as defined in Section 2.6.)

*Definition* 3.3. A *link* of the set graph $G = (V, S, A)$ is a triple $\langle \mathit{aft}, \alpha, \mathit{fore} \rangle$, where *aft* and *fore* are elements of $SoS(IV(G))$ and $\alpha$ is a sequence of subsets of $S$, such that for all $\mathit{set}_a \in \mathit{aft}$, for all $\omega \in \times \alpha$, for all $\mathit{set}_f \in \mathit{fore}$, there exists a path $\mu$ in $G$ such that at least one of the following four properties holds:

1. (a) $\times \mathit{label}(\mu)$ contains a subsequence of $\omega$ and
   (b) $\mathit{tail}(\mu)$ is an initial vertex of $G$ and
   (c) $\mathit{head}(\mu)$ is a terminal vertex of $G$

2. (a) $\times \mathit{label}(\mu)$ contains a prefix of $\omega$ and
   (b) $\mathit{tail}(\mu) \in \mathit{set}_a$ and
   (c) $\mathit{head}(\mu)$ is a terminal vertex of $G$

3. (a) $\times \mathit{label}(\mu)$ contains a suffix of $\omega$ and
   (b) $\mathit{tail}(\mu)$ is an initial vertex of $G$ and
   (c) $\mathit{head}(\mu) \in \mathit{set}_f$

4. (a) $\times \mathit{label}(\mu)$ contains $\omega$ and
   (b) $\mathit{tail}(\mu) \in \mathit{set}_a$ and
   (c) $\mathit{head}(\mu) \in \mathit{set}_f$

*Example*: Let $G$ be the set graph in Figure 3. Now consider the triple $\langle \mathit{aft}, \alpha, \mathit{fore} \rangle$, where

$$\mathit{aft} = \{\{v_3, v_{12}\}\}$$
$$\alpha = \langle \{s_1, s_6, s_{12}\}, \{s_3\}, \{s_1, s_6\} \rangle$$
$$\mathit{fore} = \{\{v_6\}, \{v_9\}\}$$

Table 2 shows that for each combination of $\omega \in \times \alpha$, $\mathit{set}_a \in \mathit{aft}$ and $\mathit{set}_f \in \mathit{fore}$, there exists a path $\mu$ in $G$ such that at least one of the four properties in Definition 3.3 holds. Those elements in $\mathit{label}(\mu)$ forming a subsequence, prefix or suffix of $\omega$ are indicated in **red bold**. For those cases where Property 2 holds, the tail of $\mu$, which is in $\in \mathit{set}_a$, is indicated in **blue bold**. For those cases where Property 3 holds, the head of $\mu$, which is in $\in set_f$, is also indicated in **blue bold**.

$$\langle\{\{v_3,v_{12}\}\}, \langle\{s_1,s_6,s_{12}\},\{s_3\},\{s_1,s_6\}\rangle, \{\{v_6\},\{v_9\}\}\rangle$$

is therefore a link of *G*. (Note: This exhaustive enumeration of $\omega \in \times\alpha$, $set_a \in aft$ and $set_f \in fore$ is for illustrative purposes only. None of the techniques described below rely on such an enumeration.)

The connection between links and implicants is provided by Lemma 3.1.

LEMMA 3.1. *Let $G = (V, S, A)$ be a set graph and let $\alpha$ be a sequence of subsets of S. Then $\alpha$ is an implicant of G if and only if $\langle\{\{\}\}, \alpha, \{\{\}\}\rangle$ is a link of G.*

PROOF. Suppose that $\langle\{\{\}\}, \alpha, \{\{\}\}\rangle$ is a link of *G*. Then in Definition 3.3, $aft = \{\{\}\}$ and $fore = \{\{\}\}$. Thus for all $\omega \in \times\alpha$, for $set_a = \{\}$ and for $set_f = \{\}$, there exists a path $\mu$ in *G* such that at least one of Properties 1 – 4 in Definition 3.3 holds. But Properties 2 – 4 cannot hold since $set_a$ and $set_f$ are both empty. So Property 1 must hold. That means for all $\omega \in \times\alpha$, there exists a path $\mu$ in *G* such that (a) $\times label(\mu)$ contains a subsequence of $\omega$, (b) $tail(\mu)$ is an initial vertex of *G* and (c) $head(\mu)$ is a terminal vertex of *G*. There thus exists a subsequence $\omega'$ of $\omega$ and a sequence of sets $\alpha'$ (namely, $label(\mu)$) accepted by *G* such that $\omega' \in \times\alpha'$. $\alpha$ is therefore an implicant of *G*. A reverse argument shows that if $\alpha$ is an implicant of *G*, then $\langle\{\{\}\}, \alpha, \{\{\}\}\rangle$ is a link of *G*.

The next property states, in effect, that *weakening* any, or all, of the components of a link yields another link.

PROPERTY 3.1. *Let $\langle aft_1, \alpha_1, fore_1\rangle$ be a link of the set graph $G = (V, S, A)$, let $aft_2$ and $fore_2$ be elements of SoS(IV(G)) and let $\alpha_2$ be a sequence of subsets of S such that $|\alpha_2| = |\alpha_1|$. If each of the following three properties holds*

1. $aft_2 \leq aft_1$
2. $\alpha_2(i) \subseteq \alpha_1(i)$ *for* $0 \leq i < |\alpha_1|$
3. $fore_2 \leq fore_1$

*then $\langle aft_2, \alpha_2, fore_2\rangle$ is a link of G.*

As an illustration of this property, consider the set graph in Figure 3 and the triple

$$\langle\{\{v_3,v_6,v_{12}\}\}, \langle\{s_1,s_{12}\},\{s_3\},\{s_1,s_6\}\rangle, \{\{v_9\}\}\rangle$$

It is a weaker version of

$$\langle\{\{v_3,v_{12}\}\}, \langle\{s_1,s_6,s_{12}\},\{s_3\},\{s_1,s_6\}\rangle, \{\{v_6\},\{v_9\}\}\rangle$$

TABLE 2. Properties Satisfied by $\langle\{\{v_3,v_{12}\}\}, \langle\{s_1,s_6,s_{12}\},\{s_3\},\{s_1,s_6\}\rangle, \{\{v_6\},\{v_9\}\}\rangle$

| $\omega \in \times\alpha$ | $set_a \in$ aft | $set_f \in$ fore | Path $\mu$ in Set Graph $G$ | Prop. |
|---|---|---|---|---|
| $\langle s_1,s_3,s_1\rangle$ | $\{v_3,v_{12}\}$ | $\{v_6\}$ | $\langle\langle v_{11},\{s_1,s_3,s_5,s_7,s_9,s_{11},s_{13},s_{15}\},v_{12}\rangle,\langle v_{12},\{s_0,s_3,s_4,s_7,s_8,s_{11},s_{12},s_{15}\},v_{13}\rangle\rangle$ | 1 |
| | | | $\langle\langle v_5,\{s_0,s_1,s_4,s_5,s_8,s_9,s_{12},s_{13}\},v_6\rangle\rangle$ | 3 |
| | | $\{v_9\}$ | $\langle\langle v_{11},\{s_1,s_3,s_5,s_7,s_9,s_{11},s_{13},s_{15}\},v_{12}\rangle,\langle v_{12},\{s_0,s_3,s_4,s_7,s_8,s_{11},s_{12},s_{15}\},v_{13}\rangle\rangle$ | 1 |
| $\langle s_1,s_3,s_6\rangle$ | $\{v_3,v_{12}\}$ | $\{v_6\}$ | $\langle\langle v_2,\{s_2,s_3,s_6,s_7,s_{10},s_{11},s_{14},s_{15}\},v_3\rangle,\langle v_3,\{s_2,s_3,s_6,s_7,s_{10},s_{11},s_{14},s_{15}\},v_4\rangle\rangle$ | 1 |
| | | | $\langle\langle v_{11},\{s_1,s_3,s_5,s_7,s_9,s_{11},s_{13},s_{15}\},v_{12}\rangle,\langle v_{12},\{s_0,s_3,s_4,s_7,s_8,s_{11},s_{12},s_{15}\},v_{13}\rangle\rangle$ | 1 |
| | | $\{v_9\}$ | $\langle\langle v_2,\{s_2,s_3,s_6,s_7,s_{10},s_{11},s_{14},s_{15}\},v_3\rangle,\langle v_3,\{s_2,s_3,s_6,s_7,s_{10},s_{11},s_{14},s_{15}\},v_4\rangle\rangle$ | 1 |
| | | | $\langle\langle v_{11},\{s_1,s_3,s_5,s_7,s_9,s_{11},s_{13},s_{15}\},v_{12}\rangle,\langle v_{12},\{s_0,s_3,s_4,s_7,s_8,s_{11},s_{12},s_{15}\},v_{13}\rangle\rangle$ | 1 |
| | | | $\langle\langle v_8,\{s_0,s_2,s_4,s_6,s_8,s_{10},s_{12},s_{14}\},v_9\rangle\rangle$ | 3 |
| $\langle s_6,s_3,s_1\rangle$ | $\{v_3,v_{12}\}$ | $\{v_6\}$ | $\langle\langle v_2,\{s_2,s_3,s_6,s_7,s_{10},s_{11},s_{14},s_{15}\},v_3\rangle,\langle v_3,\{s_2,s_3,s_6,s_7,s_{10},s_{11},s_{14},s_{15}\},v_4\rangle\rangle$ | 1 |
| | | | $\langle\langle v_3,\{s_2,s_3,s_6,s_7,s_{10},s_{11},s_{14},s_{15}\},v_4\rangle\rangle$ | 2 |
| | | | $\langle\langle v_5,\{s_0,s_1,s_4,s_5,s_8,s_9,s_{12},s_{13}\},v_6\rangle\rangle$ | 3 |
| | | $\{v_9\}$ | $\langle\langle v_2,\{s_2,s_3,s_6,s_7,s_{10},s_{11},s_{14},s_{15}\},v_3\rangle,\langle v_3,\{s_2,s_3,s_6,s_7,s_{10},s_{11},s_{14},s_{15}\},v_4\rangle\rangle$ | 1 |
| | | | $\langle\langle v_3,\{s_2,s_3,s_6,s_7,s_{10},s_{11},s_{14},s_{15}\},v_4\rangle\rangle$ | 2 |
| $\langle s_6,s_3,s_6\rangle$ | $\{v_3,v_{12}\}$ | $\{v_6\}$ | $\langle\langle v_2,\{s_2,s_3,s_6,s_7,s_{10},s_{11},s_{14},s_{15}\},v_3\rangle,\langle v_3,\{s_2,s_3,s_6,s_7,s_{10},s_{11},s_{14},s_{15}\},v_4\rangle\rangle$ | 1 |
| | | | $\langle\langle v_3,\{s_2,s_3,s_6,s_7,s_{10},s_{11},s_{14},s_{15}\},v_4\rangle\rangle$ | 2 |
| | | $\{v_9\}$ | $\langle\langle v_2,\{s_2,s_3,s_6,s_7,s_{10},s_{11},s_{14},s_{15}\},v_3\rangle,\langle v_3,\{s_2,s_3,s_6,s_7,s_{10},s_{11},s_{14},s_{15}\},v_4\rangle\rangle$ | 1 |
| | | | $\langle\langle v_3,\{s_2,s_3,s_6,s_7,s_{10},s_{11},s_{14},s_{15}\},v_4\rangle\rangle$ | 2 |
| | | | $\langle\langle v_8,\{s_0,s_2,s_4,s_6,s_8,s_{10},s_{12},s_{14}\},v_9\rangle\rangle$ | 3 |
| $\langle s_{12},s_3,s_1\rangle$ | $\{v_3,v_{12}\}$ | $\{v_6\}$ | $\langle\langle v_{12},\{s_0,s_3,s_4,s_7,s_8,s_{11},s_{12},s_{15}\},v_{13}\rangle\rangle$ | 2 |
| | | | $\langle\langle v_5,\{s_0,s_1,s_4,s_5,s_8,s_9,s_{12},s_{13}\},v_6\rangle\rangle$ | 3 |
| | | $\{v_9\}$ | $\langle\langle v_{12},\{s_0,s_3,s_4,s_7,s_8,s_{11},s_{12},s_{15}\},v_{13}\rangle\rangle$ | 2 |
| $\langle s_{12},s_3,s_6\rangle$ | $\{v_3,v_{12}\}$ | $\{v_6\}$ | $\langle\langle v_2,\{s_2,s_3,s_6,s_7,s_{10},s_{11},s_{14},s_{15}\},v_3\rangle,\langle v_3,\{s_2,s_3,s_6,s_7,s_{10},s_{11},s_{14},s_{15}\},v_4\rangle\rangle$ | 1 |
| | | | $\langle\langle v_{12},\{s_0,s_3,s_4,s_7,s_8,s_{11},s_{12},s_{15}\},v_{13}\rangle\rangle$ | 2 |
| | | $\{v_9\}$ | $\langle\langle v_2,\{s_2,s_3,s_6,s_7,s_{10},s_{11},s_{14},s_{15}\},v_3\rangle,\langle v_3,\{s_2,s_3,s_6,s_7,s_{10},s_{11},s_{14},s_{15}\},v_4\rangle\rangle$ | 1 |
| | | | $\langle\langle v_{12},\{s_0,s_3,s_4,s_7,s_8,s_{11},s_{12},s_{15}\},v_{13}\rangle\rangle$ | 2 |
| | | | $\langle\langle v_8,\{s_0,s_2,s_4,s_6,s_8,s_{10},s_{12},s_{14}\},v_9\rangle\rangle$ | 3 |

which we've already seen is a link of the set graph in Figure 3. It follows from Property 3.1 that $\langle\{\{v_3,v_6,v_{12}\}\}, \langle\{s_1,s_{12}\},\{s_3\},\{s_1,s_6\}\rangle, \{\{v_9\}\}\rangle$ is also a link of the set graph in Figure 3.

The next result, together with Theorem 3.5 below, are crucial. They both depend directly on the Fundamental Theorem and are the two supporting pillars for Theorem 3.6, the main result at the combinatorics level.

THEOREM 3.2. *If $\langle aft_1, \alpha_1, fore_1\rangle$ and $\langle aft_2, \alpha_2, fore_2\rangle$ are links of the set graph G such that $\sim fore_1 \leq aft_2$, then $\langle aft_1, \alpha_1 \bullet \alpha_2, fore_2\rangle$ is a link of G.*

PROOF. See Appendix A.

*Example*: Let $G$ be the set graph in Figure 3 and let

$$\begin{align}
aft_1 &= \{\{v_3, v_{12}\}\} \\
\alpha_1 &= \langle\{s_1, s_6, s_{12}\}\rangle \\
fore_1 &= \{\{v_6\}, \{v_{12}\}\} \\
aft_2 &= \{\{v_6, v_{12}\}\} \\
\alpha_2 &= \langle\{s_3\}\rangle \\
fore_2 &= \{\{v_3\}, \{v_9\}\} \\
aft_3 &= \{\{v_3, v_9\}\} \\
\alpha_3 &= \langle\{s_1, s_6\}\rangle \\
fore_3 &= \{\{v_6\}, \{v_9\}\}
\end{align}$$

$\langle aft_1, \alpha_1, fore_1\rangle$, $\langle aft_2, \alpha_2, fore_2\rangle$ and $\langle aft_3, \alpha_3, fore_3\rangle$ are links of $G$ as can be verified by exhaustively enumerating all $\omega \in \times\alpha$, $set_a \in aft$ and $set_f \in fore$, as was done in Table 2, for each of these three new cases. Since $\sim fore_1 = \{\{v_6, v_{12}\}\} = aft_2$, it follows from Theorem 3.2 that $\langle aft_1, \alpha_1 \bullet \alpha_2, fore_2\rangle$ is a link of $G$. Furthermore, since $\sim fore_2 = \{\{v_3, v_9\}\}$ = $aft_3$, it follows – again from Theorem 3.2 – that $\langle aft_1, \alpha_1 \bullet \alpha_2 \bullet \alpha_3, fore_3\rangle$ is a link of $G$. Now notice an important property of links $\langle aft_1, \alpha_1, fore_1\rangle$, $\langle aft_2, \alpha_2, fore_2\rangle$ and $\langle aft_3, \alpha_3, fore_3\rangle$. They are all *links of length 1* – that is, $|\alpha_1| = |\alpha_2| = |\alpha_3| = 1$. Notice also that $\langle aft_1, \alpha_1 \bullet \alpha_2 \bullet \alpha_3, fore_3\rangle$ is the same link considered in Table 2. So we have established that $\langle aft_1, \alpha_1 \bullet \alpha_2 \bullet \alpha_3, fore_3\rangle$ is a link of $G$ without having to exhaustively enumerate all $\omega \in \times(\alpha_1 \bullet \alpha_2 \bullet \alpha_3)$, $set_a \in aft_1$ and $set_f \in fore_3$ as was done in Table 2. We instead relied on the *concatenation* of links of length 1, a technique that is key to constructing links – and ultimately implicants – of a set graph.

## 3.4 Links of Length 1

Although the general notion of a link in Definition 3.3 is needed to prove that a sequence of sets is an implicant of a set graph if and only if it is accepted by an elaboration of that set graph, the actual definition of an elaboration is in terms of links of length 1. Moreover, the manipulations used in sequential resolution (described in Section 5) and the process of normalization (described in Section 6) also involve only links of length 1. For these reasons, we provide a separate definition for this special class of links. The definition is much simpler than for the general case.

*Definition* 3.4. A *link of length 1* of a set graph $G = (V, S, A)$ is a triple $\langle \mathit{aft}, D, \mathit{fore} \rangle$, where *aft* and *fore* are elements of $SoS(IV(G))$ and $D$ is a subset of $S$, such that for all $\mathit{set}_a \in \mathit{aft}$, for all elements $e \in D$, for all $\mathit{set}_f \in \mathit{fore}$, there exists an arc $a \in A$ such that each of the following properties holds:

1. $\mathit{tail}(a)$ is an initial vertex of $G$ or is in $\mathit{set}_a$
2. $e \in \mathit{label}(a)$
3. $\mathit{head}(a)$ is a terminal vertex of $G$ or is in $\mathit{set}_f$

The question now arises: Where do links of length 1 come from? The answer is twofold: (1) the *initial links of length 1* of set graph $G$ are derived from the arcs of $G$ via Theorem 3.3; (2) additional links of length 1 of $G$ are derived from existing links of length 1 through *micro inferences* as described in Theorem 3.4.

THEOREM 3.3. *Let $\langle v_i, D, v_j \rangle$ be an arc of set graph G.*

(a) *If $v_i$ is an initial vertex of G and $v_j$ is a terminal vertex of G, then $\langle \{\{\}\}, D, \{\{\}\} \rangle$ is a link of length 1 of G*

(b) *If $v_i$ is an initial vertex of G and $v_j$ is an interior vertex of G, then $\langle \{\{\}\}, D, \{\{v_j\}\} \rangle$ is a link of length 1 of G*

(c) *If $v_i$ is an interior vertex of G and $v_j$ is a terminal vertex of G, then $\langle \{\{v_i\}\}, D, \{\{\}\} \rangle$ is a link of length 1 of G*

(d) *If $v_i$ and $v_j$ are interior vertices of G, then $\langle \{\{v_i\}\}, D, \{\{v_j\}\} \rangle$ is a link of length 1 of G*

PROOF. (a) If $v_i$ is an initial vertex of $G$ and $v_j$ is a terminal vertex of $G$, then for all $\mathit{set}_a \in \{\{\}\}$, for all $\mathit{set}_f \in \{\{\}\}$ and for all $e \in D$, there exists arc $a$ of $G$ – namely, $\langle v_i, D, v_j \rangle$ – such that: (1) $\mathit{tail}(a)$ is an initial vertex of $G$ or is in $\mathit{set}_a$, (2) $e \in \mathit{label}(a)$ and (3)

*head*(*a*) is a terminal vertex of *G* or is in *set*$_f$. It follows from Definition 3.4 that $\langle\{\{\}\}, D,$ $\{\{\}\}\rangle$ is a link of length 1 of *G*. Similar arguments apply to (b), (c) and (d).

The application of this theorem to the set graph of Figure 3 is illustrated in Table 3. Column (a) lists the arcs of the set graph, while Column (b) lists for each arc the corresponding link of length 1.

TABLE 3. Links of Length 1 Derived From the Arcs in Figure 3

| (a) Arc | (b) Link of Length 1 |
|---|---|
| $\langle v_0, \{s_9, s_{10}, s_{11}, s_{13}, s_{14}, s_{15}\}, v_1\rangle$ | $\langle\{\{\}\}, \{s_9, s_{10}, s_{11}, s_{13}, s_{14}, s_{15}\}, \{\{\}\}\rangle$ |
| $\langle v_2, \{s_2, s_3, s_6, s_7, s_{10}, s_{11}, s_{14}, s_{15}\}, v_3\rangle$ | $\langle\{\{\}\}, \{s_2, s_3, s_6, s_7, s_{10}, s_{11}, s_{14}, s_{15}\}, \{\{v_3\}\}\rangle$ |
| $\langle v_3, \{s_2, s_3, s_6, s_7, s_{10}, s_{11}, s_{14}, s_{15}\}, v_4\rangle$ | $\langle\{\{v_3\}\}, \{s_2, s_3, s_6, s_7, s_{10}, s_{11}, s_{14}, s_{15}\}, \{\{\}\}\rangle$ |
| $\langle v_5, \{s_0, s_1, s_4, s_5, s_8, s_9, s_{12}, s_{13}\}, v_6\rangle$ | $\langle\{\{\}\}, \{s_0, s_1, s_4, s_5, s_8, s_9, s_{12}, s_{13}\}, \{\{v_6\}\}\rangle$ |
| $\langle v_6, \{s_0, s_1, s_4, s_5, s_8, s_9, s_{12}, s_{13}\}, v_7\rangle$ | $\langle\{\{v_6\}\}, \{s_0, s_1, s_4, s_5, s_8, s_9, s_{12}, s_{13}\}, \{\{\}\}\rangle$ |
| $\langle v_8, \{s_0, s_2, s_4, s_6, s_8, s_{10}, s_{12}, s_{14}\}, v_9\rangle$ | $\langle\{\{\}\}, \{s_0, s_2, s_4, s_6, s_8, s_{10}, s_{12}, s_{14}\}, \{\{v_9\}\}\rangle$ |
| $\langle v_9, \{s_1, s_2, s_5, s_6, s_9, s_{10}, s_{13}, s_{14}\}, v_{10}\rangle$ | $\langle\{\{v_9\}\}, \{s_1, s_2, s_5, s_6, s_9, s_{10}, s_{13}, s_{14}\}, \{\{\}\}\rangle$ |
| $\langle v_{11}, \{s_1, s_3, s_5, s_7, s_9, s_{11}, s_{13}, s_{15}\}, v_{12}\rangle$ | $\langle\{\{\}\}, \{s_1, s_3, s_5, s_7, s_9, s_{11}, s_{13}, s_{15}\}, \{\{v_{12}\}\}\rangle$ |
| $\langle v_{12}, \{s_0, s_3, s_4, s_7, s_8, s_{11}, s_{12}, s_{15}\}, v_{13}\rangle$ | $\langle\{\{v_{12}\}\}, \{s_0, s_3, s_4, s_7, s_8, s_{11}, s_{12}, s_{15}\}, \{\{\}\}\rangle$ |
| $\langle v_{14}, \{s_4, s_5, s_7, s_8, s_9, s_{11}\}, v_{15}\rangle$ | $\langle\{\{\}\}, \{s_4, s_5, s_7, s_8, s_9, s_{11}\}, \{\{\}\}\rangle$ |
| $\langle v_{16}, \{s_2, s_{14}\}, v_{17}\rangle$ | $\langle\{\{\}\}, \{s_2, s_{14}\}, \{\{\}\}\rangle$ |

THEOREM 3.4. *If* $\langle aft_1, D_1, fore_1\rangle$ *and* $\langle aft_2, D_2, fore_2\rangle$ *are links of length* 1 *of the set graph G, then each of the following are links of length 1 of G*:

(a) $\langle aft_1 \vee aft_2, D_1 \cap D_2, fore_1 \wedge fore_2\rangle$

(b) $\langle aft_1 \wedge aft_2, D_1 \cup D_2, fore_1 \wedge fore_2\rangle$

(c) $\langle aft_1 \wedge aft_2, D_1 \cap D_2, fore_1 \vee fore_2\rangle$

PROOF. (a) Suppose that $set_a \in (aft_1 \vee aft_2)$, $e \in (D_1 \cap D_2)$ and $set_f \in (fore_1 \wedge fore_2)$. From Definition 2.5, we know that either $set_a \in aft_1$ or $set_a \in aft_2$ and that there exist $set_{f1} \in fore_1$ and $set_{f2} \in fore_2$ such that $set_f = set_{f1} \cup set_{f2}$. We also know that $e \in D_1$ and $e \in D_2$. So either: (1) $set_a \in aft_1$, $e \in D_1$ and there exists $set_{f1} \in fore_1$ such that $set_{f1} \subseteq set_f$ or (2) $set_a \in aft_2$, $e \in D_2$ and there exists $set_{f2} \in fore_2$ such that $set_{f2} \subseteq set_f$. Since $\langle aft_1, a_1, fore_1\rangle$ is a link of length 1 of *G*, there must exist an arc *a* of *G* such that *tail*(*a*) is an initial vertex of *G* or is in $set_a$, $e \in label(a)$ and *head*(*a*) is a terminal vertex of *G* or is in $set_{f1}$.

But if *head*(*a*) is in $set_{f1}$, it must also be in $set_f$ since $set_{f1} \subseteq set_f$. It follows for Case 1 that $\langle aft_1 \vee aft_2, D_1 \cap D_2, fore_1 \wedge fore_2 \rangle$ is a link of length 1 of *G*. A similar argument holds for Case 2. Hence $\langle aft_1 \vee aft_2, D_1 \cap D_2, fore_1 \wedge fore_2 \rangle$ is a link of length 1 of *G*. Similar proofs apply to $\langle aft_1 \wedge aft_2, D_1 \cup D_2, fore_1 \wedge fore_2 \rangle$ and $\langle aft_1 \wedge aft_2, D_1 \cap D_2, fore_1 \vee fore_2 \rangle$.

This theorem is illustrated by applying each of the three forms of micro inference to links of length 1 from Table 3. The following is an example of a micro inference according to Theorem 3.4(a):

$$\langle \{\{v_6\}\}, \{s_0, s_1, s_4, s_5, s_8, s_9, s_{12}, s_{13}\}, \{\{\}\} \rangle$$
$$\langle \{\{v_9\}\}, \{s_1, s_2, s_5, s_6, s_9, s_{10}, s_{13}, s_{14}\}, \{\{\}\} \rangle$$
$$\Downarrow$$
$$\langle \{\{v_6\}, \{v_9\}\}, \{s_1, s_5, s_9, s_{13}\}, \{\{\}\} \rangle$$

The following is an example of a micro inference according to Theorem 3.4(b):

$$\langle \{\{\}\}, \{s_2, s_3, s_6, s_7, s_{10}, s_{11}, s_{14}, s_{15}\}, \{\{v_3\}\} \rangle$$
$$\langle \{\{v_6\}\}, \{s_0, s_1, s_4, s_5, s_8, s_9, s_{12}, s_{13}\}, \{\{\}\} \rangle$$
$$\Downarrow$$
$$\langle \{\{v_6\}\}, \{s_0, s_1, s_2, s_3, s_4, s_5, s_6, s_7, s_8, s_9, s_{10}, s_{11}, s_{12}, s_{13}, s_{14}, s_{15}\}, \{\{v_3\}\} \rangle$$

The following is an example of a micro inference according to Theorem 3.4(c):

$$\langle \{\{\}\}, \{s_2, s_3, s_6, s_7, s_{10}, s_{11}, s_{14}, s_{15}\}, \{\{v_3\}\} \rangle$$
$$\langle \{\{\}\}, \{s_1, s_3, s_5, s_7, s_9, s_{11}, s_{13}, s_{15}\}, \{\{v_{12}\}\} \rangle$$
$$\Downarrow$$
$$\langle \{\{\}\}, \{s_3, s_7, s_{11}, s_{15}\}, \{\{v_3\}, \{v_{12}\}\} \rangle$$

### 3.5 Maximal Links

Theorem 3.6 below, the main result at the combinatorics level, states that a sequence of sets is an implicant of a set graph if and only if it is accepted by an elaboration of that set graph. Theorem 3.2 above is sufficient to prove the *if* part of the theorem. To prove the *only if* part, we need the concept of a *maximal link*, which allows us to construct an elaboration of set graph *G* from an implicant of *G*. (Maximal links are also used in the normalization process described in Section 6.)

*Definition* 3.5. Let $G = (V, S, A)$ be a set graph, let *aft* and *fore* be elements of *SoS(IV(G))* and let $\alpha$ be a sequence of subsets of $S$. Then

$$max^+(G, \mathit{aft}, \alpha) = min_\subseteq(\{U \subseteq IV(G) \mid \langle \mathit{aft}, \alpha, \{U\}\rangle \text{ is a link of } G\})$$

$$max^-(G, \mathit{fore}, \alpha) = min_\subseteq(\{U \subseteq IV(G) \mid \langle \{U\}, \alpha, \mathit{fore}\rangle \text{ is a link of } G\})$$

PROPERTY 3.2. *If $G = (V, S, A)$ is a set graph, aft and fore are each elements of SoS(IV(G)) and $\alpha$ is a sequence of subsets of S, then the following three properties are equivalent*:

1. $\langle \mathit{aft}, \alpha, \mathit{fore}\rangle$ *is a link of G*
2. *fore* $\leq max^+(G, \mathit{aft}, \alpha)$
3. *aft* $\leq max^-(G, \mathit{fore}, \alpha)$

From Property 3.2, we see that $max^+(G, \mathit{aft}, \alpha)$ is the maximum element $sos_i$ of *SoS(IV(G))* such that $\langle \mathit{aft}, \alpha, sos_i\rangle$ is a link of $G$. Similarly, $max^-(G, \mathit{fore}, \alpha)$ is the maximum element $sos_j$ of *SoS(IV(G))* such that $\langle sos_j, \alpha, \mathit{fore}\rangle$ is a link of $G$. Accordingly, we say that $\langle \mathit{aft}, \alpha, max^+(G, \mathit{aft}, \alpha)\rangle$ is a *forwards-maximal link* of $G$, and that $\langle max^-(G, \mathit{fore}, \alpha), \alpha, \mathit{fore}\rangle$ is a *backwards-maximal link* of $G$.

*Comment*: Although $max^+$ and $max^-$ are symmetrical, our emphasis is on $max^+$ since it is more *intuitive* to work with forwards-maximal links. Note, however, that all of the results and procedures described in this paper – including the normalization process of Section 6 – can be just as easily expressed in terms of $max^-$.

*Example*: Let $G$ be the set graph in Figure 3, let $\mathit{aft} = \{\{v_3, v_{12}\}\}$ and let $\alpha = \langle\{s_0, s_1, s_2, s_3, s_4, s_5, s_6, s_7, s_8, s_9, s_{10}, s_{11}, s_{12}, s_{13}, s_{14}, s_{15}\}\rangle$. Notice that $\alpha$ is of length 1. In this special case, we can calculate $max^+(G, \mathit{aft}, \alpha)$ using the micro inferences of Theorem 3.4. (A detailed algorithm will be described in a future paper.) Starting with initial links of $G$ from Table 3, we can generate a forwards-maximal link of $G$ as follows. Apply Theorem 3.4(b) to two initial links of $G$:

$$\langle\{\{v_3\}\}, \{s_2, s_3, s_6, s_7, s_{10}, s_{11}, s_{14}, s_{15}\}, \{\{\}\}\rangle$$
$$\langle\{\{\}\}, \{s_0, s_1, s_4, s_5, s_8, s_9, s_{12}, s_{13}\}, \{\{v_6\}\}\rangle$$

$$\langle\{\{v_3\}\}, \{s_0, s_1, s_2, s_3, s_4, s_5, s_6, s_7, s_8, s_9, s_{10}, s_{11}, s_{12}, s_{13}, s_{14}, s_{15}\}, \{\{v_6\}\}\rangle$$

Apply Theorem 3.4(b) twice to three initial links of $G$:

$$\langle\{\{v_3\}\}, \{s_2, s_3, s_6, s_7, s_{10}, s_{11}, s_{14}, s_{15}\}, \{\{\}\}\rangle$$

$$\langle\{\{\}\}, \{s_1, s_3, s_5, s_7, s_9, s_{11}, s_{13}, s_{15}\}, \{\{v_{12}\}\}\rangle$$

$$\langle\{\{v_{12}\}\}, \{s_0, s_3, s_4, s_7, s_8, s_{11}, s_{12}, s_{15}\}, \{\{\}\}\rangle$$

$$\langle\{\{v_3, v_{12}\}\}, \{s_0, s_1, s_2, s_3, s_4, s_5, s_6, s_7, s_8, s_9, s_{10}, s_{11}, s_{12}, s_{13}, s_{14}, s_{15}\}, \{\{v_{12}\}\}\rangle$$

Apply Theorem 3.4(c) to the two just-inferred links:

$$\langle\{\{v_3\}\}, \{s_0, s_1, s_2, s_3, s_4, s_5, s_6, s_7, s_8, s_9, s_{10}, s_{11}, s_{12}, s_{13}, s_{14}, s_{15}\}, \{\{v_6\}\}\rangle$$

$$\langle\{\{v_3, v_{12}\}\}, \{s_0, s_1, s_2, s_3, s_4, s_5, s_6, s_7, s_8, s_9, s_{10}, s_{11}, s_{12}, s_{13}, s_{14}, s_{15}\}, \{\{v_{12}\}\}\rangle$$

$$\langle\{\{v_3, v_{12}\}\}, \{s_0, s_1, s_2, s_3, s_4, s_5, s_6, s_7, s_8, s_9, s_{10}, s_{11}, s_{12}, s_{13}, s_{14}, s_{15}\}, \{\{v_6\}, \{v_{12}\}\}\rangle$$

That this last link is a forwards-maximal link can be verified by considering those $sos \in SoS(IV(G))$ such that $\{\{v_6\}, \{v_{12}\}\} < sos$, and determining if $\langle aft, \alpha, sos \rangle$ is a link of $G$. Such an examination reveals that there is indeed no such $sos$, and therefore $max^+(G, aft, \alpha) = \{\{v_6\}, \{v_{12}\}\}$.

In the preceding example, we have sketched a method for calculating $max^+(G, aft, \alpha)$ and $max^-(G, fore, \alpha)$ when the length of $\alpha$ is 1. The next result allows us to calculate $max^+(G, aft, \alpha)$ and $max^-(G, fore, \alpha)$ when the length of $\alpha$ is greater than 1.

THEOREM 3.5. *Let $G = (V, S, A)$ be a set graph, let aft and fore be elements of SoS(IV(G)) and let $\alpha_1$ and $\alpha_2$ each be a non-null sequence of subsets of S. Then*

$$max^+(G, aft, \alpha_1 \bullet \alpha_2) = max^+(G, \sim max^+(G, aft, \alpha_1), \alpha_2)$$

$$max^-(G, fore, \alpha_1 \bullet \alpha_2) = max^-(G, \sim max^-(G, fore, \alpha_2), \alpha_1)$$

PROOF. See Appendix B.

From this result, we see that determining $max^+(G, aft, \alpha_1 \bullet \alpha_2)$ can be reduced to the problem of calculating $max^+(G, aft, \alpha_1)$ and then calculating $max^+(G, aft_2, \alpha_2)$, where $aft_2 = \sim max^+(G, aft, \alpha_1)$.

*Example*: Let $G$ be the set graph in Figure 3 and let

$$aft = \{\{v_3, v_{12}\}\}$$

$$\alpha_1 = \langle\{s_0, s_1, s_2, s_3, s_4, s_5, s_6, s_7, s_8, s_9, s_{10}, s_{11}, s_{12}, s_{13}, s_{14}, s_{15}\}\rangle$$

$$\alpha_2 = \langle\{s_0, s_1, s_2, s_3, s_4, s_5, s_6, s_7, s_8, s_9, s_{10}, s_{11}, s_{12}, s_{13}, s_{14}, s_{15}\}\rangle$$

In order to determine $max^+(G, aft, \alpha_1 \bullet \alpha_2)$, we must first calculate $max^+(G, aft, \alpha_1)$. But from the preceding example, we know that $max^+(G, aft, \alpha_1) = \{\{v_6\}, \{v_{12}\}\}$. Therefore, $max^+(G, aft, \alpha_1 \bullet \alpha_2) = max^+(G, \{\{v_6, v_{12}\}\}, \alpha_2)$. Now since $\alpha_2$, like $\alpha_1$, is of length 1, we can use the micro inferences of Theorem 3.4 to calculate $max^+(G, \{\{v_6, v_{12}\}\}, \alpha_2)$. When we perform those inferences, we find that

$$max^+(G, \{\{v_3, v_{12}\}\}, \alpha_1 \bullet \alpha_2) = max^+(G, \{\{v_6, v_{12}\}\}, \alpha_2) = \{\{v_3\}, \{v_9\}\}$$

### 3.6 Elaborations

We are now ready for the main result at the combinatorics level, a necessary and sufficient condition for a sequence of sets to be an implicant of a set graph. It builds on the machinery developed in the preceding subsections.

*Definition* 3.6. An *elaboration* of a set graph $G = (V_G, S, A_G)$ is a set graph $E = (V_E, S, A_E)$ such that

1. For all $v \in V_E$, $v$ is an ordered pair $\langle aft, fore \rangle$ where $aft, fore \in SoS(IV(G))$
2. For all $v \in V_E$, $v$ is an initial vertex of $E$ if and only if $aft(v) = \{\}$
3. For all $v \in V_E$, $v$ is a terminal vertex of $E$ if and only if $fore(v) = \{\}$
4. For all $v \in V_E$, $\sim aft(v) \leq fore(v)$
5. For all $a \in A_E$, $\langle fore(tail(a)), label(a), aft(head(a)) \rangle$ is a link of length 1 of $G$

The following property follows from Property 2.5 and Conditions 1 – 4 in Definition 3.6.

PROPERTY 3.3. *If E is an elaboration of a set graph, then the unique initial vertex of E is $\langle \{\}, \{\{\}\} \rangle$ and the unique terminal vertex of E is $\langle \{\{\}\}, \{\} \rangle$.*

*Example*: Let $G$ be the set graph in Figure 3 and let $E$ be the set graph in Figure 4. We see that each vertex of $E$ is an ordered pair $\langle aft, fore \rangle$, where $aft, fore \in SoS(IV(G))$. Those ordered pairs are:

$$\langle \{\}, \{\{\}\} \rangle$$
$$\langle \{\{v_6\}, \{v_9\}\}, \{\{v_6, v_9\}\} \rangle$$
$$\langle \{\{v_3\}, \{v_{12}\}\}, \{\{v_3, v_{12}\}\} \rangle$$
$$\langle \{\{v_6\}, \{v_{12}\}\}, \{\{v_6, v_{12}\}\} \rangle$$
$$\langle \{\{v_3\}, \{v_9\}\}, \{\{v_3, v_9\}\} \rangle$$
$$\langle \{\{\}\}, \{\} \rangle$$

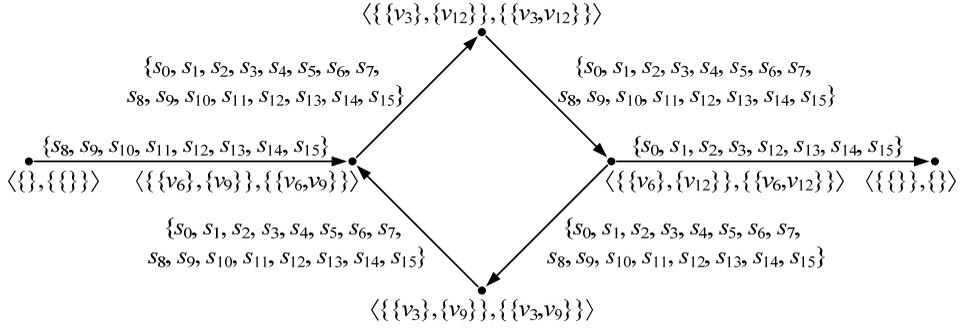

FIG. 4. An Elaboration of the set graph in Figure 3

We also observe that the unique initial vertex of $E$ is $\langle \{\}, \{\{\}\}\rangle$, the unique terminal vertex of $E$ is $\langle \{\{\}\}, \{\}\rangle$ and for each vertex $\langle \textit{aft}, \textit{fore}\rangle$ in $E$, $\sim\!\textit{aft} \leq \textit{fore}$. Finally, we note that for each arc $a$ in $E$, $\langle \textit{fore}(\textit{tail}(a)), \textit{label}(a), \textit{aft}(\textit{head}(a))\rangle$ is a link of length 1 of $G$. Those links of length 1 are:

$$\langle\,\{\{\}\},\ \{s_8, s_9, s_{10}, s_{11}, s_{12}, s_{13}, s_{14}, s_{15}\},\ \{\{v_6\}, \{v_9\}\}\,\rangle$$

$$\langle\,\{\{v_6, v_9\}\},\ \{s_0, s_1, s_2, s_3, s_4, s_5, s_6, s_7, s_8, s_9, s_{10}, s_{11}, s_{12}, s_{13}, s_{14}, s_{15}\},\ \{\{v_3\}, \{v_{12}\}\}\,\rangle$$

$$\langle\,\{\{v_3, v_{12}\}\},\ \{s_0, s_1, s_2, s_3, s_4, s_5, s_6, s_7, s_8, s_9, s_{10}, s_{11}, s_{12}, s_{13}, s_{14}, s_{15}\},\ \{\{v_6\}, \{v_{12}\}\}\,\rangle$$

$$\langle\,\{\{v_6, v_{12}\}\},\ \{s_0, s_1, s_2, s_3, s_4, s_5, s_6, s_7, s_8, s_9, s_{10}, s_{11}, s_{12}, s_{13}, s_{14}, s_{15}\},\ \{\{v_3\}, \{v_9\}\}\,\rangle$$

$$\langle\,\{\{v_3, v_9\}\},\ \{s_0, s_1, s_2, s_3, s_4, s_5, s_6, s_7, s_8, s_9, s_{10}, s_{11}, s_{12}, s_{13}, s_{14}, s_{15}\},\ \{\{v_6\}, \{v_9\}\}\,\rangle$$

$$\langle\,\{\{v_6, v_{12}\}\},\ \{s_0, s_1, s_2, s_3, s_{12}, s_{13}, s_{14}, s_{15}\},\ \{\{\}\}\,\rangle$$

(That these are indeed links of length 1 of $G$ can be confirmed using the micro inferences of Theorem 3.4. Or, alternatively, we can check that each triple satisfies the properties in Definition 3.4.) From these observations, we conclude that $E$ satisfies the five properties listed in Definition 3.6, and that $E$ is therefore an elaboration of $G$.

Of particular interest are the sequences of sets accepted by $E$. In this case, because $E$ contains a (directed) cycle, $E$ accepts an infinite number of sequences. They are all of the form $\langle X, Y^{4n+2}, Z\rangle$, where $n$ is a non-negative integer and

$X = \{s_8, s_9, s_{10}, s_{11}, s_{12}, s_{13}, s_{14}, s_{15}\}$

$Y = \{s_0, s_1, s_2, s_3, s_4, s_5, s_6, s_7, s_8, s_9, s_{10}, s_{11}, s_{12}, s_{13}, s_{14}, s_{15}\}$

$Z = \{s_0, s_1, s_2, s_3, s_{12}, s_{13}, s_{14}, s_{15}\}$

We now turn our attention to characterizing the sequences of sets accepted by an elaboration of a set graph. That characterization is provided by Theorem 3.6, which relies, in part, on Lemmas 3.2 and 3.3. (The wording of Lemma 3.3 was chosen so that the lemma could be used both here and in Section 6.)

LEMMA 3.2. *If E is an elaboration of the set graph G and μ is a path in E, then* ⟨*fore*(*tail*(μ)), *label*(μ), *aft*(*head*(μ))⟩ *is a link of G.*

PROOF. By induction on the length of μ. For paths of length 1 (i.e., arcs), the lemma follows from the definition of an elaboration. Now assume that the lemma is true for all paths of length $n$. Let μ be an arbitrary path in *E* of length $n+1$, let $μ_n$ be the prefix of μ of length $n$ and let *a* be the $n+1$'st (and final) arc of μ. By our hypothesis, ⟨*fore*(*tail*($μ_n$)), *label*($μ_n$), *aft*(*head*($μ_n$))⟩ is a link of *G*, and from the definition of an elaboration ⟨*fore*(*tail*(*a*)), *label*(*a*), *aft*(*head*(*a*))⟩ is a link of *G*. Since *head*($μ_n$) = *tail*(*a*), it follows from the definition of an elaboration that ~*aft*(*head*($μ_n$)) ≤ *fore*(*tail*(*a*)), and from Theorem 3.2 that ⟨*fore*(*tail*($μ_n$)), *label*($μ_n$)•*label*(*a*), *aft*(*head*(*a*))⟩ is a link of *G*, or equivalently that ⟨*fore*(*tail*($μ_n$)), *label*($μ_n$•*a*), *aft*(*head*(*a*))⟩ is a link of *G*. But $μ_n$•*a* = μ. Hence ⟨*fore*(*tail*(μ)), *label*(μ), *aft*(*head*(μ))⟩ is a link of *G*.

LEMMA 3.3. *If G and E are set graphs over the same set of elements and μ is a path in E such that*

1. *All vertices on μ are ordered pairs* ⟨*aft*, *fore*⟩*, where aft, fore* ∈ *SoS*(*IV*(*G*))

2. *For all vertices v on μ,* ~*aft*(*v*) = *fore*(*v*)

3. *For all arcs a on μ, aft*(*head*(*a*)) = *max*$^+$(*G*, *fore*(*tail*(*a*)), *label*(*a*))

*then aft*(*head*(μ)) = *max*$^+$(*G*, *fore*(*tail*(μ)), *label*(μ)).

PROOF. By induction on the length of μ. For paths of length 1 (i.e., arcs), the lemma follows immediately from Property 5 in the definition of a forwards-maximal elaboration. Now assume that the lemma is true for all paths of length $n$. Let μ be an arbitrary path in *E* of length $n+1$, let $μ_n$ be the prefix of μ of length $n$ and let *a* be the $n+1$'st (and final) arc of μ. By Property 5 in the definition of a forwards-maximal elaboration,

$$aft(head(a)) =$$
$$max^+(G, fore(tail(a)), label(a))$$

By construction, *head*($μ_n$) = *tail*(*a*). It follows from Property 4 in the definition of a forwards-maximal elaboration that ~*aft*(*head*($μ_n$)) = *fore*(*tail*(*a*)). Hence,

$$max^+(G, fore(tail(a)), label(a)) =$$
$$max^+(G, \sim aft(head(μ_n)), label(a))$$

By our induction hypothesis, *aft*(*head*($μ_n$)) = *max*$^+$(*G*, *fore*(*tail*($μ_n$)), *label*($μ_n$)). Thus,

$$max^+(G, \sim aft(head(\mu_n)), label(a)) =$$
$$max^+(G, \sim max^+(G, fore(tail(\mu_n)), label(\mu_n)), label(a))$$

Finally, by Theorem 3.5,

$$max^+(G, \sim max^+(G, fore(tail(\mu_n)), label(\mu_n)), label(a)) =$$
$$max^+(G, fore(tail(\mu)), label(\mu))$$

THEOREM 3.6. *Let $G = (V, S, A)$ be a set graph and let $\alpha$ be a sequence of subsets of S. Then $\alpha$ is an implicant of G if and only if a subsequence of $\alpha$ is accepted by an elaboration of G.*

PROOF. Suppose that $\alpha$ is accepted by an elaboration $E$ of $G$. Then there must be a path $\mu$ in $E$ leading from an initial vertex of $E$ to a terminal vertex of $E$ such that $\alpha = label(\mu)$. From the definition of an elaboration, we know that $tail(\mu)$ is $\langle \{\}, \{\{\}\} \rangle$ and that $head(\mu)$ is $\langle \{\{\}\}, \{\} \rangle$. Thus $fore(tail(\mu)) = aft(head(\mu)) = \{\{\}\}$, and by Lemma 3.2, $\langle \{\{\}\}, \alpha, \{\{\}\} \rangle$ is a link of $G$. It follows from Lemma 3.1 that $\alpha$ is an implicant of $G$.

Suppose that $\alpha$ is an implicant of $G$. Let $\alpha'$ be a minimal-length (non-null) subsequence of $\alpha$ such that $\alpha'$ is an implicant of $G$, and let $E$ consist of a single path $\mu$ such that (a) $tail(\mu) = \langle \{\}, \{\{\}\} \rangle$, (b) $label(\mu) = \alpha'$ and (c) for each arc $a$ in $\mu$,

$$head(a) = \langle max^+(G, fore(tail(a)), label(a)), \sim max^+(G, fore(tail(a)), label(a)) \rangle$$

By construction, $E$ satisfies Properties 1, 2, 4 and 5 in Definition 3.6 and Properties 1 – 3 in Lemma 3.3. By Lemma 3.3, $aft(head(\mu)) = max^+(G, fore(tail(\mu)), label(\mu)) = max^+(G, \{\{\}\}, label(\mu))$. But since $label(\mu) = \alpha'$ and $\alpha'$ is an implicant of $G$, it follows from Lemma 3.1 that $aft(head(\mu)) = \{\{\}\}$ and that $head(\mu) = \langle \{\{\}\}, \{\} \rangle$. Now assume that there exists an interior vertex $v$ of $\mu$ such that $v = \langle \{\{\}\}, \{\} \rangle$. By Lemma 3.3, it follows that there is a proper prefix $\mu_p$ of $\mu$ such that $max^+(G, \{\{\}\}, label(\mu_p)) = \{\{\}\}$. Lemma 3.1 then requires that $label(\mu_p)$ be an implicant of $G$, but that contradicts our assumption that $\alpha' = label(\mu)$ is a minimal-length subsequence of $\alpha$ such that $\alpha'$ is an implicant of $G$. We are forced to conclude that no such interior vertex of $\mu$ exists and that Property 3 in Definition 3.6 is satisfied.

Let us consider in detail the meaning of this last result. From Definitions 3.2 and 3.6, we see that Theorem 3.6 can be restated as follows:

Condition 1: *For each sequence of elements ω in the Cartesian product ×α, there exists a subsequence ω′ of ω and a sequence of sets α′ accepted by G such that ω′ is in the Cartesian product ×α′*

is equivalent to

Condition 2: *A subsequence of α is accepted by a set graph E satisfying the five properties:*

(1) *Each vertex of E is an ordered pair ⟨aft, fore⟩, where aft, fore ∈ SoS(IV(G))*
(2) *For each vertex v in E, v is an initial vertex of E if and only if aft(v) = {}*
(3) *For each vertex v in E, v is a terminal vertex of E if and only if fore(v) = {}*
(4) *For each vertex v in E, ~aft(v) ≤ fore(v)*
(5) *For each arc a in E, ⟨fore(tail(a)), label(a), aft(head(a))⟩ is a link of length 1 of G*

Notice that Condition 1 involves Cartesian products and sequences of elements, while Condition 2 involves neither. Condition 2 deals only with sets of sets of vertices of the set graph $G$ and certain structural properties of the set graph $E$. So we have converted the problem of determining whether $\alpha$ is an implicant of $G$ from one that entails exhaustively checking all the sequences in the Cartesian product ×$\alpha$ into one that entails constructing a set graph satisfying certain structural properties.

To make these ideas concrete, consider the set graph $G$ in Figure 3 and the set graph $E$ in Figure 4 which is an elaboration of $G$. Notice that although $G$ accepts only a finite number of sequences – seven, to be exact – $E$ accepts an *infinite* number of sequences. Nevertheless, it follows from Theorem 3.6 that each of these infinitely many sequences is an implicant of $G$ and that each of these sequences therefore satisfies Condition 1 in addition to Condition 2. So we have determined that all of the sequences accepted by $E$ are implicants of $G$ without having to exhaustively verify that each of these sequences satisfies the requirements of Condition 1, which, of course, is an impossible task since there are infinitely many such sequences.

In Section 4, Theorem 3.6 is recast in terms of Boolean graphs, and Sections 5 and 6 provide two different methods for constructing elaborations of such graphs.

## 4. LOGIC LEVEL

The mathematical concepts and results of Section 3 are now reinterpreted in the language of formal logic. Instead of dealing with sets, sequences of sets and set graphs, we will now be dealing with Boolean expressions, sequences of Boolean expressions and Boolean graphs. The theory at the logic level proceeds as follows:

- Section 4.1 introduces the notion of a *generalized Kripke structure* (*S*, *B*, *L*) over a set of *atomic propositions AP*, where *S* is a set of *states*, *B* is a set of *allowed state sequences* (*allowed behaviors*) and *L* is a function that maps each atomic proposition to the set of states in which that proposition is *true*. Through the mapping *L*, each state in *S* defines an *assignment of truth values* to the atomics propositions in *AP*. A *fully populated* Kripke structure is a Kripke structure (*S*, *B*, *L*) such that for each of the $2^{|AP|}$ possible assignments of truth values to the atomic propositions in *AP*, there exists a state in *S* for that assignment of truth values.

- Although Kripke structures are our model of system behavior, the theory at the logic level does not deal directly with such structures. Instead, manipulations on *Boolean expressions*, sequences of Boolean expressions (*Boolean sequences*) and directed graphs in which each arc is labeled with a Boolean expression (*Boolean graphs*) are used to reason about the disallowed behaviors of *infinitely many* Kripke structures. Section 4.2 defines these concepts and shows how the function *L* maps each of these constructs into its counterpart at the combinatorics level. A *sequential constraint* of a Kripke structure (*S*, *B*, *L*) represents a disallowed pattern of behavior and is defined as a Boolean sequence $\alpha$ such that $\times L(\alpha) \cap B$ is empty.

- Section 1.2 defined the notion of an *implicant* of a set of sequences of Boolean expressions *A*. Section 4.3 provides an equivalent definition in the context of a Boolean graph *G* over a set of atomic propositions *AP*: An *implicant* of *G* is a Boolean sequence $\alpha$ such that for all Kripke structures (*S*, *B*, *L*) over *AP*, $L(\alpha)$ is an implicant of $L(G)$.

- In Section 4.4, a *link* of a Boolean graph *G* is defined as a triple $\langle \textit{aft}, \alpha, \textit{fore} \rangle$, where *aft* and *fore* are elements of *SoS*(*IV*(*G*)) and $\alpha$ is a Boolean sequence over *AP*, such that for all Kripke structures (*S*, *B*, *L*) over *AP*, $\langle \textit{aft}, L(\alpha), \textit{fore} \rangle$ is a link of $L(G)$. The links of a Boolean graph *G* are the key to characterizing the implicants of *G* since a Boolean sequence $\alpha$ is an implicant of *G* if and only if $\langle \{\{\}\}, \alpha, \{\{\}\} \rangle$ is a link of *G* (Lemma 4.1). Theorem 4.4, the counterpart to Theorem 3.2, provides a sufficient condition for two links to be *concatenated*: If $\langle \textit{aft}_1, \alpha_1, \textit{fore}_1 \rangle$ and $\langle \textit{aft}_2, \alpha_2, \textit{fore}_2 \rangle$ are links of Boolean graph *G* such that $\sim\textit{fore}_1 \leq \textit{aft}_2$, then $\langle \textit{aft}_1, \alpha_1 \bullet \alpha_2, \textit{fore}_2 \rangle$ is a link of *G*.

- Section 4.5 describes the special properties of those links ⟨*aft*, α, *fore*⟩ such that |α| = 1. But, in contrast to the combinatorics level, two alternate definitions are provided. A *link of length 1* at the logic level is defined in terms of a link of length 1 at the combinatorics level. A *logical link* of the Boolean graph *G* – which is also of the form ⟨*aft*, α, *fore*⟩, where |α| = 1 – is completely equivalent to a link of length 1 at the logic level, but its definition involves only logical and structural properties of *aft*, α, *fore* and *G* – there is no reference to either states or Kripke structures. Logical links permit elaborations (Section 4.7), sequential resolution (Section 5) and normalization (Section 6) to be defined entirely in logical/structural terms. The *initial logical links* of a Boolean graph *G* are derived from the arcs of *G* via Theorem 4.6. Additional logical links are derived from existing logical links through the *micro inferences* described in Theorem 4.7.

- Section 4.6 defines a *forwards-* (*backwards-*) *maximal link* of a Boolean graph *G* as a link ⟨*aft*, α, *fore*⟩ such that *fore* (*aft*) is the maximum element of *SoS*(*IV*(*G*)) – with respect to the partial order ≤ – such that ⟨*aft*, α, *fore*⟩ is a link of *G*. A key result involving such links (Theorem 4.9) allows us to construct for any implicant of *G* an elaboration that accepts a subsequence of that implicant.

- Section 4.7 defines an *elaboration* of a Boolean graph *G* as another Boolean graph *E* in which each vertex is an ordered pair ⟨*aft*, *fore*⟩ satisfying special properties, where *aft* and *fore* are each elements of *SoS*(*IV*(*G*)). The main result at the logic level, and the main result of the paper, is Theorem 4.10 which states that a Boolean sequence α is an implicant of a Boolean graph *G* if and only if a subsequence of α is accepted by an elaboration of *G*.

## 4.1 Kripke Structures

Kripke structures are the model of system behavior used in model checking [Clarke et. al. 2000], and they are also the model of system behavior used in the theory at the logic level that follows. However, we make three modifications to the standard Kripke model, the first two of which are designed to increase the generality of the model while the third, minor, modification simplifies formulation of the theory.

In the standard model, a Kripke structure is a nondeterministic finite state machine whose states are labeled with Boolean variables. More formally, a (*standard*) *Kripke structure* over a set of atomic propositions *AP* is a 3-tuple (*S*, *R*, *L*), where

1. *S* is a finite set of *states*
2. $R \subseteq S \times S$ is a *transition* relation that must be total – that is, for every state $s \in S$, there must exist a $s' \in S$ such that $s \, R \, s'$
3. $L: S \rightarrow 2^{AP}$ is a function that labels each state with the set of atomic propositions *true* in that state

An *allowed state sequence* (*allowed behavior*) of a standard Kripke structure (*S*, *R*, *L*) is a sequence of states $\omega$ such that for all pairs of successive states $\omega(i)$ and $\omega(i+1)$ in $\omega$, $\omega(i) \, R \, \omega(i+1)$. It follows that every subsequence of an allowed behavior of a standard Kripke structure is itself an allowed behavior of that Kripke structure.

*Comment*: Kripke structures can be defined either with or without a set of *initial states* (see [Clarke et. al. 2000]). We choose to omit initial states because it greatly simplifies the theory. There is no loss of generality, however, since we can still introduce a state variable whose assertion causes the system to be initialized. For example, in the counter example described below, asserting *Reset* initializes the counter to a state in which all the *bits* of the counter are 0.

In a *generalized Kripke structure*, we make the following three modifications to the standard model:

- The requirement that the set of states *S* be finite is eliminated since there is nothing at the logic level that requires *S* to be finite, nor is there any requirement for the set of elements *S* of a set graph – the counterpart of *S* at the combinatorics level – to be finite.

- In a standard Kripke structure, the set of allowed system behaviors is defined indirectly using the state-transition relation *R*. In the generalized model, the state-transition relation is replaced by the set of allowed behaviors itself, and this set satisfies the only property we need of allowed behaviors: Every subsequence of an allowed behavior is itself allowed. In other words, the set of allowed behaviors satisfies Axiom 1(a).

- The third modification, though minor, helps to simplify the theory that follows. Instead of defining *L* as a function that labels each state with the set of atomic propositions true in that state, we define *L* as a function that labels each atomic proposition with the set of states in which that proposition is true.

Taken together, these modifications give us the following generalized model of system behavior.

*Definition* 4.1. A (*generalized*) *Kripke structure* over a set of atomic propositions *AP* is a 3-tuple (*S*, *B*, *L*), where

1. *S* is a set of *states*
2. $B \subseteq S^*$ is a set of *allowed state sequences* (*allowed behaviors*) such that if $\omega$ is in *B*, then every subsequence of $\omega$ is in *B*
3. $L: AP \rightarrow 2^S$ is a function that labels each atomic proposition with the set of states in which that proposition is *true*

Consider now the function *L* in this definition. It tells us the set of states in which each atomic proposition is *true*. There is, however, an equivalent way of conveying the same information, and that is by defining for each state an *assignment of truth values* to the atomic propositions in *AP*. This notion, in turn, allows to us to introduce the concept of a *fully populated* Kripke structure. As we see in later sections, a fully populated Kripke structure has special properties that make it representative of an entire class of Kripke structures.

*Definition* 4.2. Let $K = (S, B, L)$ be a Kripke structure over a set of atomic propositions *AP*. For each state $s \in S$, $\pi_s: AP \rightarrow \{true, false\}$ is an *assignment of truth values* to the atomic propositions in *AP* such that $ap \in AP$ is assigned the value *true* if $s \in L(ap)$ and the value *false* otherwise. *K* is said to be *fully populated* if and only if for each of the $2^{|AP|}$ assignments of truth values to the atomic propositions in *AP*, there exists a state in *S* that defines that assignment of truth values.

The following property, which applies to both standard and generalized Kripke structures, is the key to the theory that follows. It is equivalent to Axiom 1(b).

PROPERTY 4.1. If $\omega$ is a disallowed state sequence of the Kripke structure *K*, then every state sequence of *K* containing $\omega$ as a subsequence is a disallowed state sequence of *K*.

## 4.2 Boolean Expressions, Sequences and Graphs

Although Kripke structures are our model of system behavior, the theory at the logic level does not deal directly with such structures. Instead, manipulations on *Boolean expressions*, sequences of Boolean expressions (*Boolean sequences*) and directed graphs in which each arc is labeled with a Boolean expression (*Boolean graphs*) are used to reason about the disallowed behaviors of *infinitely many* Kripke structures. To understand how these manipulations allow us to reason simultaneously about the disallowed behavior infinitely many Kripke structures, we must first understand the connections between these three constructs at the logic level and their counterparts at the combinatorics level.

For a particular Kripke structure $(S, B, L)$, the function $L$ is the bridge between the logic level and the combinatorics level. Not only does $L$ map an atomic proposition into a set of states, but extensions of $L$ map: a Boolean expression into a set of states, a Boolean sequence into a sequence of sets of states and a Boolean graph into a set graph in which each arc is labeled with a set of states.

While we need not concern ourselves with the exact syntax of Boolean expressions, certain manipulations in the theory – in particular, the three forms of *micro inference* and *sequential resolution* – do assume that Boolean conjunction, denoted by $\wedge$, and Boolean disjunction, denoted by $\vee$, are among the Boolean operations used to construct Boolean expressions. Additionally, sequential resolution assumes that Boolean negation, denoted by $\neg$, is among the Boolean operations used to construct Boolean expressions.

With these points in mind, we define Boolean expressions, Boolean sequences and Boolean graphs.

*Definition* 4.3. A *Boolean expression* over a set of atomic propositions $AP$ is either an atomic proposition in $AP$ or is an expression constructed from Boolean expressions over $AP$ using a Boolean operation. *BooleanExpressions*($AP$) denotes the set of Boolean expressions over $AP$. If $(S, B, L)$ is a Kripke structure over $AP$ and $BE$ a Boolean expression over $AP$, then $L(BE)$ is the set of $s \in S$ such that the assignment of truth values to the atomic propositions in $AP$ defined by $s$ causes $BE$ to evaluate to *true*. For the Boolean operations $\wedge$ (*AND*), $\vee$ (*OR*) and $\neg$ (*NOT*),

$$\begin{aligned} L(BE_1 \wedge BE_2) &= L(BE_1) \cap L(BE_2) \\ L(BE_1 \vee BE_2) &= L(BE_1) \cup L(BE_2) \\ L(\neg BE_1) &= S - L(BE_1) \end{aligned}$$

*Definition* 4.4. A *Boolean sequence* over a set of atomic propositions $AP$ is a sequence of Boolean expressions over $AP$. If $K = (S, B, L)$ is a Kripke structure over $AP$ and $\alpha$ a Boolean sequence over $AP$, then $L(\alpha)$ is the sequence of subsets of $S$ obtained from $\alpha$ by replacing each Boolean expression $BE$ in $\alpha$ with $L(BE)$. If $T$ is a set of Boolean sequences over $AP$, then $L(T) = \{L(\alpha) \mid \alpha \in T\}$. A (*sequential*) *constraint* of $K$ is a Boolean sequence $\alpha$ over $AP$ such that $\times L(\alpha) \cap B$ is empty.

*Definition* 4.5. A *Boolean graph* over a set of atomic propositions *AP* is an ordered pair $G = (V, A)$, where

1. *V* is a finite set of *vertices*
2. $A \subseteq V \times BooleanExpressions(AP) \times V$ is a finite set of labeled *arcs*

If $K = (S, B, L)$ is a Kripke structure over *AP*, then $L(G)$ is the set graph $(V, S, A_S)$ where $A_S$ is obtained from *A* by replacing each arc $\langle v_1, BE, v_2 \rangle$ in *A* with $\langle v_1, L(BE), v_2 \rangle$. A *constraint graph* of *K* is a Boolean graph over *AP* that accepts only sequential constraints of *K*.

To see how manipulations on these three types of structures allow us to reason about the disallowed behavior of infinitely many Kripke structures, consider an arbitrary Kripke structure *K* over a set of atomic propositions *AP* such that *G* is a constraint graph of *K*. In subsequent sections, we will see that reasoning about the disallowed behavior of *K* entails constructing an *elaboration* of *G*. But the process of constructing an elaboration of *G* is applicable not only to *K* but also to *all Kripke structures over AP for which G is a constraint graph*, of which there are infinitely many. In other words, in reasoning about the disallowed behavior of *K*, we are also reasoning about the disallowed behavior of *K*'s brethren who share with *K* the property that *G* is a constraint graph.

*Comment*: A constraint graph *G* used to describe the behavior of the Kripke structure $K = (S, B, L)$ is not required to characterize all of the disallowed behaviors of *K*. More precisely, it is not necessary that for each disallowed state sequence $\omega$ of *K* – each state sequence not in *B* – there exist a subsequence $\omega'$ of $\omega$ and a Boolean sequence $\alpha$ accepted by *G* such that $\omega' \in \times L(\alpha)$. This characteristic of constraint graphs gives a programmer or designer the flexibility to specify just those aspects of a system's excluded behavior that are needed to solve the problem at hand.

*Example*: Consider a 2-bit counter with binary state variables *Q0*, *Q1*, *Reset* and *Carry*, where: *Q0* is the least-significant bit and *Q1* the most significant bit of the counter, *Reset* is an input that, when asserted, causes both *Q0* and *Q1* to be reset to 0 and *Carry* is the carry output from the counter. Let *AP* be the set of atomic propositions {*Q0* = 1, *Q1* = 1, *Reset* = 1, *Carry* = 1}, which we abbreviate as simply {*Q0*, *Q1*, *Reset*, *Carry*}. Let *S* be the set of states $\{s_0, s_1, s_2, s_3, s_4, s_5, s_6, s_7, s_8, s_9, s_{10}, s_{11}, s_{12}, s_{13}, s_{14}, s_{15}\}$ and let $L: AP \rightarrow 2^S$ be defined such that

$L(Q0) \quad = \{s_2, s_3, s_6, s_7, s_{10}, s_{11}, s_{14}, s_{15}\}$

$L(Q1)$ = $\{s_1, s_2, s_5, s_6, s_9, s_{10}, s_{13}, s_{14}\}$

$L(Reset)$ = $\{s_8, s_9, s_{10}, s_{11}, s_{12}, s_{13}, s_{14}, s_{15}\}$

$L(Carry)$ = $\{s_4, s_5, s_6, s_7, s_8, s_9, s_{10}, s_{11}\}$

From $L$ we derive the assignment of truth values to the atomic propositions in *AP* listed in Table 4. An examination of this table reveals that for each of the $2^4 = 16$ possible assignments of truth values to the atomic propositions in *AP*, there exists a state in *S* that defines that assignment of truth values. Any Kripke structure over *AP* that has *S* as its state set and *L* as its mapping from *AP* to $2^S$ is therefore fully populated.

TABLE 4. Assignments of Truth Values to the Atomic Propositions in *AP*

| State | Atomic Propositions | | | |
| --- | --- | --- | --- | --- |
| | Q0 | Q1 | Reset | Carry |
| $s_0$ | false | false | false | false |
| $s_1$ | false | true | false | false |
| $s_2$ | true | true | false | false |
| $s_3$ | true | false | false | false |
| $s_4$ | false | false | false | true |
| $s_5$ | false | true | false | true |
| $s_6$ | true | true | false | true |
| $s_7$ | true | false | false | true |
| $s_8$ | false | false | true | true |
| $s_9$ | false | true | true | true |
| $s_{10}$ | true | true | true | true |
| $s_{11}$ | true | false | true | true |
| $s_{12}$ | false | false | true | false |
| $s_{13}$ | false | true | true | false |
| $s_{14}$ | true | true | true | false |
| $s_{15}$ | true | false | true | false |

To complete the definition of a (generalized) Kripke structure (*S*, *B*, *L*) over *AP*, we need only specify a set of allowed behaviors *B* for the counter. But we forego specifying *B* directly, and instead specify a set of disallowed behaviors for the counter using the constraint graph *G* in Figure 5, and we reason in later sections about the behavior of the counter based on the sequential constraints accepted by *G*. So although we may not be specifying *B* directly (or even completely), we are declaring that *none* of the state sequences on which *G* holds tightly is in *B* and we are declaring that *no* supersequences of these disallowed state sequences are in *B*. (Boolean graph *G* holds tightly on a state sequence $\omega$ if and only if the set of Boolean sequences accepted by *G* holds tightly on $\omega$.)

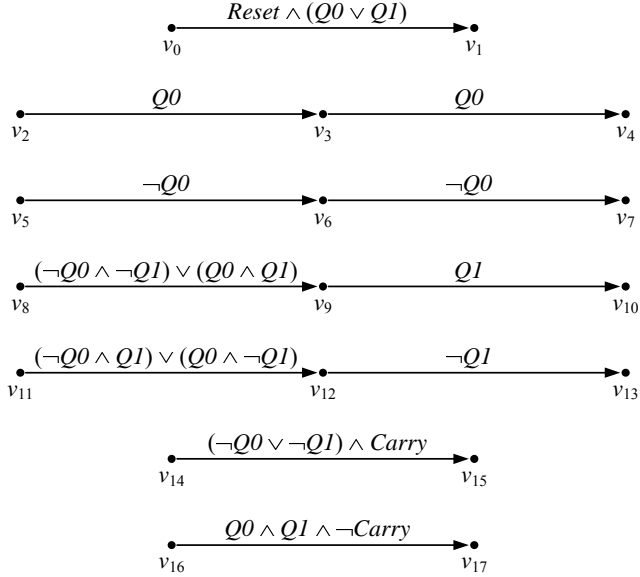

FIG. 5. Constraint Graph for a 2-Bit Counter

To understand what *G* says about the counter's behavior, we consider the meaning of the seven sequential constraints accepted by *G*. The constraint

$$\langle Reset \wedge (Q0 \vee Q1) \rangle$$

says that neither $Q0 = 1$ nor $Q1 = 1$ in the same state in which $Reset = 1$. In other words, if $Reset = 1$, then both $Q0 = 0$ and $Q1 = 0$. The two constraints

$$\langle Q0, \; Q0 \rangle$$

$$\langle \neg Q0, \; \neg Q0 \rangle$$

say that $Q0$ cannot have the same value in successive states. In other words, $Q0$ *toggles* in successive states. The two constraints

$$\langle ((\neg Q0 \wedge \neg Q1) \vee (Q0 \wedge Q1)), \; Q1 \rangle$$

$$\langle ((\neg Q0 \wedge Q1) \vee (Q0 \wedge \neg Q1)), \; \neg Q1 \rangle$$

say, in effect, that the value of *Q1* in a state is the exclusive OR (XOR) of the values of *Q0* and *Q1* in the preceding state. Lastly, the two constraints

$$\langle (\neg Q0 \vee \neg Q1) \wedge Carry \rangle$$

$$\langle Q0 \wedge Q1 \wedge \neg Carry \rangle$$

say that $Carry = (Q0 \wedge Q1)$, where $Carry$, $Q0$ and $Q1$ are all evaluated in the same state. (The theory can also describe this counter using functional notation, but *functions*, *formal variables* and *temporal offsets* are advanced topics and are deferred to a future paper.)

To see the parallels between the logic level and combinatorics level as we reason in later sections about the behavior of this counter, we first must determine $L(G)$, the set-graph counterpart to $G$. That is accomplished using an extension to the function $L: AP \to 2^S$ defined above. When this function, as extended in Definition 4.3, is applied to the Boolean expressions labeling the arcs of $G$, we find that $L(G)$ is, in fact, the set graph in Figure 3.

### 4.3 Implicants

In Section 1.1, we posed the question:

*How do we know whether a logical/temporal dependency*
*follows as a logical consequence from a set of logical/temporal dependencies?*

We saw that this question is equivalent to the following question expressed in terms of sequential constraints:

*How do we know whether a sequence of Boolean expressions*
*is a sequential constraint as a result of a set of sequential constraint?*

In Section 1.2, we answered this question in terms of a generalization of Boolean implicant:

*A sequence of Boolean expressions $\alpha$ is a sequential constraint*
*as a consequence of a set of sequential constraints $A$*
*if and only if*
*$\alpha$ is an implicant of $A$*

This notion of a *sequential implicant* is now recast in the context of Kripke structures.

*Definition* 4.6. An *implicant* of a set $T$ of Boolean sequences over a set of atomic propositions $AP$ is a Boolean sequence $\alpha$ over $AP$ such that for all Kripke structures $(S, B, L)$ over $AP$, $L(\alpha)$ is an implicant of $L(T)$. An *implicant* of a Boolean graph $G$ over $AP$ is a Boolean sequence $\alpha$ over $AP$ such that for all Kripke structures $(S, B, L)$ over $AP$, $L(\alpha)$ is an implicant of $L(G)$.

From the definition of *implicant* at the combinatorics level (Definition 3.2), we see that an implicant of a set $T$ of Boolean sequences over a set of atomic propositions $AP$ is a Boolean sequence $\alpha$ over $AP$ such that for all Kripke structures $(S, B, L)$ over $AP$, for all state sequences $\omega$ in the Cartesian product $\times L(\alpha)$, there exists a subsequence $\omega'$ of $\omega$ and a Boolean sequence $\alpha'$ in $T$ such that $\omega'$ is in the Cartesian product $\times L(\alpha')$.

This definition may seem unsatisfying since it is expressed in terms of an infinite number of Kripke structures. There is, however, an equivalent definition expressed in terms of a single, *fully populated* Kripke structure.

THEOREM 4.1. *Let G be a Boolean graph over a set of atomic propositions AP, let $\alpha$ be a Boolean sequence over AP and let $(S, B, L)$ be a fully populated Kripke structure over AP. Then $\alpha$ is an implicant of G if and only if $L(\alpha)$ is an implicant of $L(G)$.*

PROOF. See Appendix C.

The next theorem captures the relationship between Kripke structures, implicants and sequential constraints.

THEOREM 4.2. *If G is a constraint graph of the Kripke structure K, then the implicants of G are sequential constraints of K.*

PROOF. Suppose that $G$ is a constraint graph of the Kripke structure $K$. Let $\alpha$ be an arbitrary implicant of $G$. It follows that for all state sequences $\omega \in \times L(\alpha)$, there exists a subsequence $\omega'$ of $\omega$ and a Boolean sequence $\alpha'$ accepted by $G$ such that $\omega' \in \times L(\alpha')$. Since $\alpha'$ is accepted by $G$ and $G$ is a constraint graph of $K$, $\alpha'$ must be a constraint of $K$, which means that all state sequences in $\times L(\alpha')$ are disallowed state sequences of $K$. Hence $\omega'$ is a disallowed state sequence of $K$. From Property 4.1, it follows that $\omega$ is also a disallowed state sequence of $K$, which means that all of the states sequences in $\times L(\alpha)$ are disallowed state sequences of $K$. $\alpha$ is therefore a constraint of $K$.

## 4.4 Links

At the combinatorics level, links are the key to characterizing the implicants of a set graph. At the logic level, the logic counterparts to combinatorial links are the key to characterizing the implicants of a Boolean graph. We begin with the counterparts to (*general*) *links* (Definition 3.3) in this section, and then in the next section discuss the counterparts to *links of length 1* (Definition 3.4).

*Definition* 4.7. A *link* of a Boolean graph *G* over a set of atomic propositions *AP* is a triple ⟨*aft*, $\alpha$, *fore*⟩, where *aft* and *fore* are elements of *SoS*(*IV*(*G*)) and $\alpha$ is a Boolean sequence over *AP*, such that for all Kripke structures (*S*, *B*, *L*) over *AP*, ⟨*aft*, *L*($\alpha$), *fore*⟩ is a link of *L*(*G*).

Links at the logic level – like implicants at the logic level – are thus defined in terms of *all* Kripke structures over a set of atomic propositions. But as with implicants, there is an equivalent definition involving just a single, fully populated Kripke structure.

THEOREM 4.3. *Let G be a Boolean graph over a set of atomic propositions AP, let aft and fore be elements of SoS*(*IV*(*G*)), *let $\alpha$ be a Boolean sequence over AP and let* (*S*, *B*, *L*) *be a fully populated Kripke structure over AP. Then* ⟨*aft*, $\alpha$, *fore*⟩ *is a link of G if and only if* ⟨*aft*, *L*($\alpha$), *fore*⟩ *is a link of L*(*G*).

PROOF. See Appendix D.

The following results for logic links parallel those for combinatorial links. The first result is the counterpart to Lemma 3.1.

LEMMA 4.1. *Let G be a Boolean graph over a set of atomic propositions AP and let $\alpha$ be a Boolean sequence over AP. Then $\alpha$ is an implicant of G if and only if* ⟨{{}}, $\alpha$, {{}}⟩ *is a link of G.*

PROOF. Suppose that ⟨{{}}, $\alpha$, {{}}⟩ is a link of *G*. Then for all for all Kripke structures (*S*, *B*, *L*) over *AP*, ⟨{{}}, *L*($\alpha$), {{}}⟩ is a link of *L*(*G*). By Lemma 3.1, *L*($\alpha$) is an implicant of *L*(*G*), and by the definition of an implicant at the logic level (Definition 4.6), $\alpha$ is an implicant of *G*. A reverse argument shows that if $\alpha$ is an implicant of *G*, then ⟨{{}}, $\alpha$, {{}}⟩ is a link of *G*.

The next property, the counterpart to Property 3.1, states, in effect, that *weakening* any, or all, of the components of a link yields another link.

PROPERTY 4.2. *Let G be a Boolean graph over the set of atomic propositions AP, let $\langle aft_1, \alpha_1, fore_1 \rangle$ be a link of G, let $aft_2$ and $fore_2$ be elements of SoS(IV(G)) and let $\alpha_2$ be a Boolean sequence over AP such that $|\alpha_2| = |\alpha_1|$. If each of the following three properties holds*

1. $aft_2 \leq aft_1$
2. *For all Kripke structures (S, B, L) over AP:* $L(\alpha_2(i)) \subseteq L(\alpha_1(i))$ *for $0 \leq i < |\alpha_1|$*
3. $fore_2 \leq fore_1$

*then $\langle aft_2, \alpha_2, fore_2 \rangle$ is a link of G.*

Theorem 4.4, the counterpart to Theorem 3.2, is the main result for links.

THEOREM 4.4. *If $\langle aft_1, \alpha_1, fore_1 \rangle$ and $\langle aft_2, \alpha_2, fore_2 \rangle$ are links of a Boolean graph G over a set of atomic propositions AP such that $\sim fore_1 \leq aft_2$, then $\langle aft_1, \alpha_1 \bullet \alpha_2, fore_2 \rangle$ is a link of G.*

PROOF. Suppose that $\langle aft_1, \alpha_1, fore_1 \rangle$ and $\langle aft_2, \alpha_2, fore_2 \rangle$ are links of G and that $\sim fore_1 \leq aft_2$. From Definition 4.7, it follows that for all Kripke structures (S, B, L) over AP, $\langle aft_1, L(\alpha_1), fore_1 \rangle$ and $\langle aft_2, L(\alpha_2), fore_2 \rangle$ are links of L(G). From Theorem 3.2, it then follows that $\langle aft_1, L(\alpha_1) \bullet L(\alpha_2), fore_2 \rangle$ is a link of L(G). But $L(\alpha_1) \bullet L(\alpha_2) = L(\alpha_1 \bullet \alpha_2)$. Hence, $\langle aft_1, L(\alpha_1 \bullet \alpha_2), fore_2 \rangle$ is a link of L(G). Thus for all Kripke structures (S, B, L) over AP, $\langle aft_1, L(\alpha_1 \bullet \alpha_2), fore_2 \rangle$ is a link of L(G). Theorem 4.4 follows.

## 4.5 Links of Length 1

We now describe the special properties of those links $\langle aft, \alpha, fore \rangle$ such that $|\alpha| = 1$. But, in contrast to the combinatorics level, two alternate definitions are provided. A *link of length 1* at the logic level is defined in terms of a link of length 1 at the combinatorics level. A *logical link* of the Boolean graph G – which is also of the form $\langle aft, \alpha, fore \rangle$, where $|\alpha| = 1$ – is completely equivalent to a link of length 1 at the logic level, but its definition involves only logical and structural properties of *aft*, $\alpha$, *fore* and G – there is no reference, either directly or indirectly, to either states or Kripke structures. Logical links permit elaborations (Section 4.7), sequential resolution (Section 5) and normalization (Section 6) to be defined entirely in logical/structural terms.

*Definition* 4.8. A *link of length 1* of a Boolean graph G over a set of atomic propositions AP is a triple $\langle aft, BE, fore \rangle$, where *aft* and *fore* are elements of *SoS(IV(G))* and *BE* is a Boolean expression over *AP*, such that for all Kripke structures (*S*, *B*, *L*) over *AP*, ⟨*aft*, *L*(*BE*), *fore*⟩ is a link of length 1 of *L*(*G*).

*Definition* 4.9. A *logical link* of a Boolean graph $G = (V, A)$ over a set of atomic propositions *AP* is a triple ⟨*aft*, *BE*, *fore*⟩, where *aft* and *fore* are elements of *SoS*(*IV*(*G*)) and *BE* is a Boolean expression over *AP*, such that for all $set_a \in aft$, for all $set_f \in fore$,

$$BE \rightarrow \bigvee_{\substack{a \in A \text{ and} \\ tail(a) \text{ is an initial vertex of } G \text{ or is in } set_a \text{ and} \\ head(a) \text{ is a terminal vertex of } G \text{ or is in } set_f}} label(a)$$

A logical link is thus a triple ⟨*aft*, *BE*, *fore*⟩ such that for all $set_a \in aft$, for all $set_f \in fore$, the Boolean expression *BE* implies the disjunction (*OR*) of those Boolean expressions labeling arcs *a* in *G* such that: (1) *tail*(*a*) is an initial vertex of *G* or is in $set_a$ and (2) *head*(*a*) is a terminal vertex of *G* or is in $set_f$.

THEOREM 4.5. *Let G be a Boolean graph over a set of atomic propositions AP, let aft and fore be elements of SoS(IV(G)) and let α be a Boolean sequence over AP. Then ⟨aft, BE, fore⟩ is a link of length 1 of G if and only if ⟨aft, BE, fore⟩ is a logical link of G.*

PROOF. Suppose that ⟨*aft*, *BE*, *fore*⟩ is a link of length 1 of *G*. Let $G = (V, A)$ and let (*S*, *B*, *L*) be a fully populated Kripke structure over *AP*. By Theorem 4.3 and Definition 4.8, ⟨*aft*, *L*(*BE*), *fore*⟩ is a link of length 1 of *L*(*G*). It follows that for all $set_a \in aft$, for all $set_f \in fore$, for all $s \in L(BE)$, there exists $a \in A$ such that (1) *tail*(*a*) is an initial vertex of *G* or is in $set_a$, (2) *head*(*a*) is a terminal vertex of *G* or is in $set_f$ and (3) $s \in L(label(a))$. But since *L*(*BE*) is the set of states in which *BE* evaluates to *true* and *L*(*label*(*a*)) is the set of states in which *label*(*a*) evaluates to *true* (Definition 4.3), it must be that for all $set_a \in aft$, for all $set_f \in fore$, the set of states in which *BE* evaluates to *true* is a subset of the set of states in which

$$\bigvee_{\substack{a \in A \text{ and} \\ tail(a) \text{ is an initial vertex of } G \text{ or is in } set_a \text{ and} \\ head(a) \text{ is a terminal vertex of } G \text{ or is in } set_f}} label(a)$$

evaluates to true. Hence for all $set_a \in aft$, for all $set_f \in fore$,

$$BE \;\to\; \bigvee_{\substack{a \in A \text{ and} \\ \textit{tail}(a) \text{ is an initial vertex of } G \text{ or is in } \textit{set}_a \text{ and} \\ \textit{head}(a) \text{ is a terminal vertex of } G \text{ or is in } \textit{set}_f}} \textit{label}(a)$$

In other words, $\langle \textit{aft}, BE, \textit{fore} \rangle$ is a logical link of $G$.

A reverse argument shows that if $\langle \textit{aft}, BE, \textit{fore} \rangle$ is a logical link of $G$, then $\langle \textit{aft}, BE, \textit{fore} \rangle$ is a link of length 1 of $G$.

The question now arises, as it did at the combinatorics level: Where do links of length 1 – and their identical twins, logical links – come from? The answer, as before, is twofold: (1) the *initial links of length 1* and *initial logical links* of a Boolean graph $G$ are derived from the arcs of $G$ via Theorem 4.6; (2) additional links of length 1 and logical links of $G$ are derived from existing links of length 1 through *micro inferences* as described in Theorem 4.7.

THEOREM 4.6. *Let $G$ be a Boolean graph over a set of atomic propositions and let $\langle v_i, BE, v_j \rangle$ be an arc of $G$.*

(a) *If $v_i$ is an initial vertex of $G$ and $v_j$ is a terminal vertex of $G$, then $\langle \{\{\}\}, BE, \{\{\}\} \rangle$ is both a link of length 1 and logical link of $G$*

(b) *If $v_i$ is an initial vertex of $G$ and $v_j$ is an interior vertex of $G$, then $\langle \{\{\}\}, BE, \{\{v_j\}\} \rangle$ is both a link of length 1 and logical link of $G$*

(c) *If $v_i$ is an interior vertex of $G$ and $v_j$ is a terminal vertex of $G$, then $\langle \{\{v_i\}\}, BE, \{\{\}\} \rangle$ is both a link of length 1 and logical link of $G$*

(d) *If $v_i$ and $v_j$ are interior vertices of $G$, then $\langle \{\{v_i\}\}, BE, \{\{v_j\}\} \rangle$ is both a link of length 1 and logical link of $G$*

PROOF. (a) Let $G$ be a Boolean graph over the set of atomic propositions $AP$, let $\langle v_i, BE, v_j \rangle$ be an arc of $G$ such that $v_i$ is an initial vertex of $G$ and $v_j$ is a terminal vertex of $G$ and let $(S, B, L)$ be an arbitrary Kripke structure over $AP$. From Theorem 3.3, if follows that $\langle \{\{\}\}, L(BE), \{\{\}\} \rangle$ is a link of length 1 of the set graph $L(G)$, and from Definition 4.8, it follows that $\langle \{\{\}\}, BE, \{\{\}\} \rangle$ is a link of length 1 of the Boolean graph $G$. Similar arguments apply to (b), (c) and (d).

The application of this theorem to the Boolean graph of Figure 5 is illustrated in Table 5. Column (a) lists the arcs of the Boolean graph, while Column (b) lists for each arc the corresponding link of length 1 / logical link.

TABLE 5. Links of length 1 / Logical links derived from the arcs in Figure 5

| (a) Arc | (b) Link of Length 1 / Logical link |
|---|---|
| $\langle v_0, (Reset \wedge (Q0 \vee Q1)), v_1 \rangle$ | $\langle \{\{\}\}, (Reset \wedge (Q0 \vee Q1)), \{\{\}\} \rangle$ |
| $\langle v_2, Q0, v_3 \rangle$ | $\langle \{\{\}\}, Q0, \{\{v_3\}\} \rangle$ |
| $\langle v_3, Q0, v_4 \rangle$ | $\langle \{\{v_3\}\}, Q0, \{\{\}\} \rangle$ |
| $\langle v_5, \neg Q0, v_6 \rangle$ | $\langle \{\{\}\}, \neg Q0, \{\{v_6\}\} \rangle$ |
| $\langle v_6, \neg Q0, v_7 \rangle$ | $\langle \{\{v_6\}\}, \neg Q0, \{\{\}\} \rangle$ |
| $\langle v_8, ((\neg Q0 \wedge \neg Q1) \vee (Q0 \wedge Q1)), v_9 \rangle$ | $\langle \{\{\}\}, ((\neg Q0 \wedge \neg Q1) \vee (Q0 \wedge Q1)), \{\{v_9\}\} \rangle$ |
| $\langle v_9, Q1, v_{10} \rangle$ | $\langle \{\{v_9\}\}, Q1, \{\{\}\} \rangle$ |
| $\langle v_{11}, ((\neg Q0 \wedge Q1) \vee (Q0 \wedge \neg Q1)), v_{12} \rangle$ | $\langle \{\{\}\}, ((\neg Q0 \wedge Q1) \vee (Q0 \wedge \neg Q1)), \{\{v_{12}\}\} \rangle$ |
| $\langle v_{12}, \neg Q1, v_{13} \rangle$ | $\langle \{\{v_{12}\}\}, \neg Q1, \{\{\}\} \rangle$ |
| $\langle v_{14}, ((\neg Q0 \vee \neg Q1) \wedge Carry), v_{15} \rangle$ | $\langle \{\{\}\}, ((\neg Q0 \vee \neg Q1) \wedge Carry), \{\{\}\} \rangle$ |
| $\langle v_{16}, (Q0 \wedge Q1 \wedge \neg Carry), v_{17} \rangle$ | $\langle \{\{\}\}, (Q0 \wedge Q1 \wedge \neg Carry), \{\{\}\} \rangle$ |

THEOREM 4.7. *If G is a Boolean graph over a set of atomic propositions and $\langle aft_1, BE_1, fore_1 \rangle$ and $\langle aft_2, BE_2, fore_2 \rangle$ are logical links of G, then each of the following is both a link of length 1 and logical link of G*:

(a) $\langle\ aft_1 \vee aft_2,\ BE_1 \wedge BE_2,\ fore_1 \wedge fore_2\ \rangle$

(b) $\langle\ aft_1 \wedge aft_2,\ BE_1 \vee BE_2,\ fore_1 \wedge fore_2\ \rangle$

(c) $\langle\ aft_1 \wedge aft_2,\ BE_1 \wedge BE_2,\ fore_1 \vee fore_2\ \rangle$

PROOF. (a) Let *G* be a Boolean graph over the set of atomic propositions *AP*, let $\langle aft_1, BE_1, fore_1 \rangle$ and $\langle aft_2, BE_2, fore_2 \rangle$ be links of length 1 of *G* and let (*S*, *B*, *L*) be an arbitrary Kripke structure over *AP*. From Definition 4.8, it follows that $\langle aft_1, L(BE_1), fore_1 \rangle$ and $\langle aft_2, L(BE_2), fore_2 \rangle$ are links of length 1 of the set graph *L*(*G*), and from Theorem 3.4, it follows that $\langle aft_1 \vee aft_2, L(BE_1) \cap L(BE_2), fore_1 \wedge fore_2 \rangle$ is a link of length 1 of *L*(*G*). But from Definition 4.3, we know that $L(BE_1 \wedge BE_2) = L(BE_1) \cap L(BE_2)$. Hence, $\langle aft_1 \vee aft_2, L(BE_1 \wedge BE_2), fore_1 \wedge fore_2 \rangle$ is a link of length 1 of *L*(*G*), and by Definition 4.8, $\langle aft_1 \lor aft_2, BE_1 \land BE_2, fore_1 \land fore_2 \rangle$ is a link of length 1 of $G$. (b) and (c) are proved in a similar fashion.

This theorem is illustrated by applying each of the three forms of micro inference to links of length 1 / logical links from Table 5. The following is an example of a micro inference according to Theorem 4.7(a):

$$\langle \{\{v_6\}\}, \neg Q0, \{\{\}\} \rangle$$
$$\langle \{\{v_9\}\}, Q1, \{\{\}\} \rangle$$
$$\Downarrow$$
$$\langle \{\{v_6\}, \{v_9\}\}, (\neg Q0 \land Q1), \{\{\}\} \rangle$$

The following is an example of a micro inference according to Theorem 4.7(b):

$$\langle \{\{\}\}, Q0, \{\{v_3\}\} \rangle$$
$$\langle \{\{v_6\}\}, \neg Q0, \{\{\}\} \rangle$$
$$\Downarrow$$
$$\langle \{\{v_6\}\}, true, \{\{v_3\}\} \rangle$$

The following is an example of a micro inference according to Theorem 4.7(c):

$$\langle \{\{\}\}, Q0, \{\{v_3\}\} \rangle$$
$$\langle \{\{\}\}, ((\neg Q0 \land Q1) \lor (Q0 \land \neg Q1)), \{\{v_{12}\}\} \rangle$$
$$\Downarrow$$
$$\langle \{\{\}\}, (Q0 \land \neg Q1), \{\{v_3\}, \{v_{12}\}\} \rangle$$

### 4.6 Maximal Links

The *max* function – the counterpart to the *max* function defined at the combinatorics level (Definition 3.5) – is used in the normalization process described in Section 6.

*Definition* 4.10. Let $G$ be a Boolean graph over a set of atomic propositions *AP*, let *aft* and *fore* be elements of $SoS(IV(G))$ and let $\alpha$ be a Boolean sequence over *AP*. Then

$$max^+(G, aft, \alpha) = \bigwedge_{\text{For all Kripke structures } (S, B, L) \text{ over } AP} max^+(L(G), aft, L(\alpha))$$

$$max^-(G, fore, \alpha) = \bigwedge_{\text{For all Kripke structures } (S, B, L) \text{ over } AP} max^-(L(G), fore, L(\alpha))$$

$max^+(G, aft, \alpha)$ is thus the greatest lower bound for all Kripke structures $(S, B, L)$ over $AP$ of $max^+(L(G), aft, L(\alpha))$. Similarly, $max^-(G, fore, \alpha)$ is the greatest lower bound for all Kripke structures $(S, B, L)$ over $AP$ of $max^-(L(G), fore, L(\alpha))$.

So we see that the *max* function, like the notions of *implicant* and *link* above, is defined in terms of *all* Kripke structures over a set of atomic propositions. But as with implicants and links, there is an equivalent definition involving just a single, fully populated Kripke structure.

THEOREM 4.8. *Let G be a Boolean graph over a set of atomic propositions AP, let aft and fore be elements of SoS(IV(G)), let $\alpha$ be a Boolean sequence over AP and let $(S, B, L)$ be a fully populated Kripke structure over AP. Then*

$$max^+(G, aft, \alpha) = max^+(L(G), aft, L(\alpha))$$
$$max^-(G, fore, \alpha) = max^-(L(G), fore, L(\alpha))$$

PROOF. See Appendix E.

The next result is the counterpart to Property 3.2.

PROPERTY 4.3. *Let G be a Boolean graph over a set of atomic propositions AP, let aft and fore be elements of SoS(IV(G)) and let $\alpha$ be a Boolean sequence over AP. Then the following three properties are equivalent*:

1. $\langle aft, \alpha, fore \rangle$ *is a link of G*
2. $fore \leq max^+(G, aft, \alpha)$
3. $aft \leq max^-(G, fore, \alpha)$

From Property 4.3, we see that $max^+(G, aft, \alpha)$ is the maximum element $sos_i$ of $SoS(IV(G))$ such that $\langle aft, \alpha, sos_i \rangle$ is a link of $G$. Similarly, $max^-(G, fore, \alpha)$ is the maximum element $sos_j$ of $SoS(IV(G))$ such that $\langle sos_j, \alpha, fore \rangle$ is a link of $G$. Accordingly, we say that $\langle aft, \alpha, max^+(G, aft, \alpha) \rangle$ is a *forwards-maximal link* of $G$, and that $\langle max^-(G, fore, \alpha), \alpha, fore \rangle$ is a *backwards-maximal link* of $G$.

*Comment*: Although $max^+$ and $max^-$ are symmetrical, our emphasis is on $max^+$ since it is more *intuitive* to work with forwards-maximal links. Note, however, that all of the results and procedures described in this paper – including the normalization process of Section 6 – can be just as easily expressed in terms of $max^-$.

*Example*: Let $G$ be the Boolean graph in Figure 5, let $\mathit{aft} = \{\{v_3, v_{12}\}\}$ and let $\mathit{BE} =$ *true*. We can calculate $\mathit{max}^+(G, \mathit{aft}, \mathit{BE})$ using the micro inferences of Theorem 4.7. (A detailed algorithm will be described in a future paper.) Starting with initial links of $G$ from Table 5, we can generate a forwards-maximal link of $G$ as follows. Apply Theorem 4.7(b) to two initial links of $G$:

$$\langle \{\{v_3\}\}, Q0, \{\{\}\} \rangle$$
$$\langle \{\{\}\}, \neg Q0, \{\{v_6\}\} \rangle$$
$$\Downarrow$$
$$\langle \{\{v_3\}\}, \mathit{true}, \{\{v_6\}\} \rangle$$

Apply Theorem 4.7(b) twice to three initial links of $G$:

$$\langle \{\{v_3\}\}, Q0, \{\{\}\} \rangle$$
$$\langle \{\{\}\}, ((\neg Q0 \wedge Q1) \vee (Q0 \wedge \neg Q1)), \{\{v_{12}\}\} \rangle$$
$$\langle \{\{v_{12}\}\}, \neg Q1, \{\{\}\} \rangle$$
$$\Downarrow$$
$$\langle \{\{v_3, v_{12}\}\}, \mathit{true}, \{\{v_{12}\}\} \rangle$$

Apply Theorem 4.7(c) to the two just-inferred links:

$$\langle \{\{v_3\}\}, \quad \mathit{true}, \{\{v_6\}\} \rangle$$
$$\langle \{\{v_3, v_{12}\}\}, \mathit{true}, \{\{v_{12}\}\} \rangle$$
$$\Downarrow$$
$$\langle \{\{v_3, v_{12}\}\}, \quad \mathit{true}, \{\{v_6\}, \{v_{12}\}\} \rangle$$

That this last link is a forwards-maximal link can be verified by considering those $\mathit{sos} \in \mathit{SoS}(\mathit{IV}(G))$ such that $\{\{v_6\}, \{v_{12}\}\} < \mathit{sos}$, and determining, via Theorem 4.3, if $\langle \mathit{aft}, \alpha, \mathit{sos} \rangle$ is a link of $G$. Such an examination reveals that there is indeed no such $\mathit{sos}$, and therefore $\mathit{max}^+(G, \mathit{aft}, \alpha) = \{\{v_6\}, \{v_{12}\}\}$.

In the preceding example, we have sketched a method for calculating $\mathit{max}^+(G, \mathit{aft}, \alpha)$ and $\mathit{max}^-(G, \mathit{fore}, \alpha)$ when the length of $\alpha$ is 1. The next result allows us to calculate $\mathit{max}^+(G, \mathit{aft}, \alpha)$ and $\mathit{max}^-(G, \mathit{fore}, \alpha)$ when the length of $\alpha$ is greater than 1.

THEOREM 4.9. *Let G be a Boolean graph over a set of atomic propositions AP, let aft and fore be elements of SoS(IV(G)) and let $\alpha_1$ and $\alpha_2$ each be a Boolean sequence over AP. Then*

$$max^+(G, aft, \alpha_1 \bullet \alpha_2) = max^+(G, \sim max^+(G, aft, \alpha_1), \alpha_2)$$

$$max^-(G, fore, \alpha_1 \bullet \alpha_2) = max^-(G, \sim max^-(G, fore, \alpha_2), \alpha_1)$$

PROOF. Let (*S*, *B*, *L*) be an arbitrary fully populated Kripke structure over *AP*. From Theorem 4.8, we know that

$$max^+(G, aft, \alpha_1 \bullet \alpha_2) = max^+(L(G), aft, L(\alpha_1 \bullet \alpha_2))$$

But from Definition 4.4, it follows that $L(\alpha_1 \bullet \alpha_2) = L(\alpha_1) \bullet L(\alpha_2)$. Thus

$$max^+(L(G), aft, L(\alpha_1 \bullet \alpha_2)) = max^+(L(G), aft, L(\alpha_1) \bullet L(\alpha_2))$$

By Theorem 3.5,

$$max^+(L(G), aft, L(\alpha_1) \bullet L(\alpha_2)) = max^+(L(G), \sim max^+(L(G), aft, L(\alpha_1)), L(\alpha_2))$$

And by Theorem 4.8,

$$max^+(L(G), \sim max^+(L(G), aft, L(\alpha_1)), L(\alpha_2)) = max^+(G, \sim max^+(G, aft, \alpha_1), \alpha_2)$$

Hence $max^+(G, aft, \alpha_1 \bullet \alpha_2) = max^+(G, \sim max^+(G, aft, \alpha_1), \alpha_2)$. A similar proof applies to $max^-$.

From this result, we see that determining $max^+(G, aft, \alpha_1 \bullet \alpha_2)$ can be reduced to the problem of calculating $max^+(G, aft, \alpha_1)$ and then calculating $max^+(G, aft_2, \alpha_2)$, where $aft_2 = \sim max^+(G, aft, \alpha_1)$.

## 4.7 Elaborations

The main result at the logic level, and the main result of the paper, is a necessary and sufficient condition for a sequence of Boolean expressions to be an implicant of a Boolean graph. It builds on the machinery developed in Section 3 and the preceding subsections.

*Definition* 4.11. An *elaboration* of a Boolean graph *G* over a set of atomic propositions *AP* is a Boolean graph $E = (V_E, A_E)$ over *AP* such that

1. For all $v \in V_E$, *v* is an ordered pair $\langle aft, fore \rangle$ where $aft, fore \in SoS(IV(G))$
2. For all $v \in V_E$, *v* is an initial vertex of *E* if and only if $aft(v) = \{\}$
3. For all $v \in V_E$, *v* is a terminal vertex of *E* if and only if $fore(v) = \{\}$

4. For all $v \in V_E$, $\sim aft(v) \leq fore(v)$

5. For all $a \in A_E$, $\langle fore(tail(a)), label(a), aft(head(a))\rangle$ is a logical link of $G$

The following property follows from Property 2.5 and Conditions 1 – 4 in Definition 4.11.

PROPERTY 4.4. *If E is an elaboration of a Boolean graph, then the unique initial vertex of E is $\langle\{\}, \{\{\}\}\rangle$ and the unique terminal vertex of E is $\langle\{\{\}\}, \{\}\rangle$.*

The next property follows from the definitions of an elaboration at the combinatorics level (Definition 3.6) and at the logic level (Definition 4.11). It simply states that if $E$ is an elaboration of a Boolean graph $G$ over a set of atomic propositions $AP$ and if $(S, B, L)$ is a Kripke structure over $AP$, then replacing each Boolean expression $BE$ appearing in $G$ and in $E$ with the set of states $L(BE)$ yields two structures, the set graphs $L(G)$ and $L(E)$, with the property that $L(E)$ is an elaboration of $L(G)$.

PROPERTY 4.5. *If E is an elaboration of a Boolean graph G over a set of atomic propositions AP and if $(S, B, L)$ is a Kripke structure over AP, then $L(E)$ is an elaboration of the set graph $L(G)$.*

The following result – the logic counterpart to Theorem 3.6 – is the main result at the logic level, and the main result of the paper.

THEOREM 4.10. *A Boolean sequence $\alpha$ over a set of atomic propositions AP is an implicant of a Boolean graph G over AP if and only if a subsequence of $\alpha$ is accepted by an elaboration of G.*

PROOF. Suppose that a subsequence $\alpha'$ of $\alpha$ is accepted by an elaboration $E$ of $G$. Let $(S, B, L)$ be an arbitrary Kripke structure over $AP$. From the definitions of $L(\alpha')$ (Definition 4.4) and $L(E)$ (Definition 4.5), it follows that the sequence of sets $L(\alpha')$ is accepted by $L(E)$, and from Property 4.5, we know that $L(E)$ is an elaboration of the set graph $L(G)$. From Theorem 3.6, it follows that $L(\alpha')$ is an implicant of $L(G)$, but that means that $\alpha'$, and also $\alpha$, are implicants of $G$ (Definition 4.6).

Suppose that $\alpha$ is an implicant of $G$. Let $(S, B, L)$ be an arbitrary fully populated Kripke structure over $AP$. By Definition 4.6, the sequence of sets of states $L(\alpha)$ is an implicant of the set graph $L(G)$. It follows from Theorem 3.6 that a subsequence of $L(\alpha)$ is accepted by an elaboration of $L(G)$. Let $\alpha'$ be the subsequence of $\alpha$ that corresponds to such a subsequence of $L(\alpha)$, and let $E_L = (V, S, A_L)$ be a minimal elaboration of $L(G)$ that accepts $L(\alpha')$. This means that all of the vertices in $V$ and arcs in $A_L$ are on a path $\mu$ in the set graph $E_L$ such that: (1) $tail(\mu)$ is an initial vertex of $E_L$, (2) $label(\mu) = L(\alpha')$ and (3) $head(\mu)$ is a terminal vertex of $E_L$. Finally, let $A$ be obtained from $A_L$ by replacing the set of states labeling each arc in $A$ with the Boolean expression in $\alpha$ corresponding to that set of states. Now consider the structure $E = (V, A)$. By construction, $E$ is a Boolean graph over $AP$ that accepts a subsequence of $\alpha$. Moreover, because $E_L$ is an elaboration of $L(G)$, it follows immediately that $E$ satisfies Properties 1 – 4 in Definition 4.11. Property 5 follows with assistance from Theorem 4.3. $E$ is therefore an elaboration of $G$ that accepts a subsequence of $\alpha$.

COROLLARY 4.1. *An elaboration of Boolean graph G accepts only implicants of G.*

PROOF. Suppose that a Boolean sequence $\alpha$ is accepted by an elaboration of $G$. Since $\alpha$ is a subsequence of itself, it follows from Theorem 4.10 that $\alpha$ is an implicant of $G$.

COROLLARY 4.2. *An elaboration of a constraint graph of Kripke structure K is itself a constraint graph of K.*

PROOF. Suppose that $G$ is a constraint graph of $K$ and that $E$ is an elaboration of $G$. By Corollary 4.1, we know that $E$ accepts only implicants of $G$. It follows from Theorem 4.2, that $E$ accepts only sequential constraints of $G$. $E$ is therefore a constraint graph of $K$.

*Example*: Let $G$ be the constraint graph in Figure 5, and let $E$ be the elaboration of $G$ in Figure 6. To see that $E$ is indeed an elaboration of $G$, we observe first that each vertex of $E$ is an ordered pair $\langle aft, fore \rangle$, where $aft, fore \in SoS(IV(G))$. Those ordered pairs are:

$$\langle \{\}, \{\{\}\} \rangle$$
$$\langle \{\{v_6\}, \{v_9\}\}, \{\{v_6, v_9\}\} \rangle$$
$$\langle \{\{v_3\}, \{v_{12}\}\}, \{\{v_3, v_{12}\}\} \rangle$$
$$\langle \{\{v_6\}, \{v_{12}\}\}, \{\{v_6, v_{12}\}\} \rangle$$
$$\langle \{\{v_3\}, \{v_9\}\}, \{\{v_3, v_9\}\} \rangle$$
$$\langle \{\{\}\}, \{\} \rangle$$

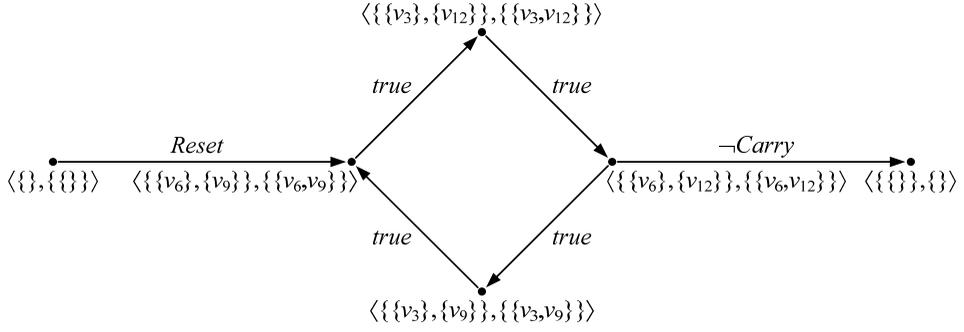

FIG. 6. An Elaboration of the Boolean graph in Figure 5

We also observe that the unique initial vertex of $E$ is $\langle\{\}, \{\{\}\}\rangle$, the unique terminal vertex of $E$ is $\langle\{\{\}\}, \{\}\rangle$ and for each vertex $\langle \textit{aft, fore}\rangle$ in $E$, $\sim\!\textit{aft} \leq \textit{fore}$. Finally, we note that for each arc $a$ in $E$, $\langle \textit{fore}(\textit{tail}(a)), \textit{label}(a), \textit{aft}(\textit{head}(a))\rangle$ is a logical link of $G$. Those logical links are:

$$\langle\ \{\{\}\}, \textit{Reset}, \{\{v_6\}, \{v_9\}\}\ \rangle$$
$$\langle\ \{\{v_6, v_9\}\}, \textit{true}, \{\{v_3\}, \{v_{12}\}\}\ \rangle$$
$$\langle\ \{\{v_3, v_{12}\}\}, \textit{true}, \{\{v_6\}, \{v_{12}\}\}\ \rangle$$
$$\langle\ \{\{v_6, v_{12}\}\}, \textit{true}, \{\{v_3\}, \{v_9\}\}\ \rangle$$
$$\langle\ \{\{v_3, v_9\}\}, \textit{true}, \{\{v_6\}, \{v_9\}\}\ \rangle$$
$$\langle\ \{\{v_6, v_{12}\}\}, \neg\textit{Carry}, \{\{\}\}\ \rangle$$

(That these are indeed logical links of $G$ can be confirmed using the micro inferences of Theorem 4.7.) From these observations, we conclude that $E$ satisfies all the properties listed in Definition 4.11, and that $E$ is therefore an elaboration of $G$. And from Corollary 4.2, we conclude that $E$, like $G$, is a constraint graph of the Kripke structure described in Section 4.2.

Now consider the Boolean sequences accepted by $E$. Because $E$ contains a (directed) cycle, $E$ accepts an infinite number of sequences – each of the form $\langle \textit{Reset}, \textit{true}^{4n+2}, \neg\textit{Carry}\rangle$, where $n$ is a non-negative integer. But since $E$ is a constraint graph, each of these Boolean sequences must be a sequential constraint of the Kripke structure in Section 4.2. This set of sequential constraints tells us that

*A Carry occurs 3 states following Reset and every $4^{th}$ state thereafter*

Let us now consider in detail the meaning of Theorem 4.10. From Definitions 3.2, 4.6 and 4.11, we see that Theorem 4.10 can be restated as follows:

Condition 1: *For all Kripke structures (S, B, L) over AP and for each state sequence ω in the Cartesian product ×L(α), there exists a subsequence ω′ of ω and a Boolean sequence α′ accepted by G such that ω′ is in the Cartesian product ×L(α′)*

is equivalent to

Condition 2: *A subsequence of α is accepted by a Boolean graph E satisfying the five properties:*

(1) *Each vertex of E is an ordered pair ⟨aft, fore⟩, where aft, fore ∈ SoS(IV(G))*
(2) *For each vertex v of E, v is an initial vertex of E if and only if aft(v) = {}*
(3) *For each vertex v of E, v is a terminal vertex of E if and only if fore(v) = {}*
(4) *For each vertex v of E, ~aft(v) ≤ fore(v)*
(5) *For each arc a of E, ⟨fore(tail(a)), label(a), aft(head(a))⟩ is a logical link of G*

Notice that Condition 1 involves states, sequences of states, Kripke structures and Cartesian products, while Condition 2 involves none of these. Condition 2 deals only with logical/structural properties of the Boolean graphs $G$ and $E$. So we have converted the problem of determining whether $\alpha$ is an implicant of $G$ from one that entails exhaustively checking all the sequences in the Cartesian product ×L(α) into one that entails constructing a Boolean graph satisfying certain logical/structural properties.

To make these ideas concrete, consider the Boolean graph $G$ in Figure 5 and the Boolean graph $E$ in Figure 6 which is an elaboration of $G$. Notice that although $G$ accepts only a finite number of sequences – seven, to be exact – $E$ accepts an *infinite* number of sequences. Nevertheless, it follows from Theorem 4.10 that each of these infinitely many sequences is an implicant of $G$ and that each of these sequences therefore satisfies Condition 1 in addition to Condition 2. So we have determined that all of the sequences accepted by $E$ are implicants of $G$ without having to exhaustively verify that each of these sequences satisfies the requirements of Condition 1, which, of course, is an impossible task since there are infinitely many such sequences.

The next two sections describe two methods – sequential resolution and normalization – for constructing elaborations of a Boolean graph.

## 5. SEQUENTIAL RESOLUTION

Boolean resolution is a powerful inference rule in Boolean logic, and comes in two forms. The *disjunctive* form [Blake 1937; Quine 1952] – which is sometimes called *consensus* [Tison 1967] – is applied to a sum of products of literals, while the *conjunctive* form [Robinson 1965] is applied to a product of sums of literals.

In the disjunction form, if $C_1$ and $C_2$ are conjunctions of literals such that exactly one Boolean variable $x$ appears negated in one conjunction and not negated in the other, then the conjunction obtained from $C_1$ and $C_2$ by deleting $x$ and $\neg x$ and omitting repetitions of any other literals is called the *resolvent* of $C_1$ and $C_2$. For example, the resolvant of the conjunctions $a \wedge x$ and $b \wedge \neg x$ is the conjunction $a \wedge b$. Depending on the objective of the resolution, the resolvant may be either added to the sum of products, or it may replace the conjunctions $C_1$ and $C_2$ in that sum.

Sequential resolution is a generalization of the disjunctive form of Boolean resolution. It is applied to a succession of elaborations of a Boolean graph $G$ starting with an *initial elaboration* that is isomorphic to $G$. Each instance of sequential resolution is performed on two equal-length paths in an elaboration, and yields a new path that is the same length as the two resolved paths. This *inferred path* is added to the existing elaboration to create a new elaboration which accepts an expanded set of sequences of Boolean expressions. These added sequences represent logical/temporal dependencies that are *inferred* from the dependencies associated with the previous elaboration.

## 5.1 The Initial Elaboration

In order for sequential resolution to be applied, there must first be an elaboration. The function *elaboration*($G$) provides the *initial elaboration* for a Boolean graph $G$. It is defined with the aid of the function $e$ which maps each vertex and each arc of a Boolean graph into its counterpart in this initial elaboration.

*Definition* 5.1. Let $G = (V, A)$ be a Boolean graph over a set of atomic propositions. For $v \in V$ and $\langle v_i, BE, v_j \rangle \in A$,

$$e(G, v) = \begin{cases} \langle \{\}, \{\{\}\} \rangle & \text{if } v \text{ is an initial vertex of } G \\ \langle \{\{\}\}, \{\} \rangle & \text{if } v \text{ is a terminal vertex of } G \\ \langle \{\{v\}\}, \{\{v\}\} \rangle & \text{if } v \text{ is an interior vertex of } G \end{cases}$$

$$e(G, \langle v_i, BE, v_j \rangle) = \langle e(G, v_i), BE, e(G, v_j) \rangle$$

*Definition* 5.2. Let $G = (V, A)$ be a Boolean graph over a set of atomic propositions. Then *elaboration*$(G) = (V_E, A_E)$, where

$$V_E = \{e(G, v) \mid v \in V\}$$
$$A_E = \{e(G, a) \mid a \in A\}$$

THEOREM 5.1. *If G is a Boolean graph over a set of atomic propositions, then elaboration(G) is an elaboration of G.*

PROOF. By construction, *elaboration*$(G)$ satisfies Properties 1 – 4 of Definition 4.10. Property 5 follows from Theorem 3.3.

## 5.2 Sequential Resolution

Sequential resolution is illustrated in Figure 7. Figure 7(a) shows two equal-length paths – the two upper paths – in an existing elaboration that are resolved to produce (infer) a *resolvent path* – the lower path – which is added to the existing elaboration to create a new elaboration. This resolvent path consists of a (possibly null) sequence of *predecessor arcs*, followed by a single *resolvent arc*, followed by a (possibly null) sequence of *successor arcs*. Figure 7(b), 7(c) and 7(d) show, respectively, how predecessor arcs, the single resolvant arc and successor arcs are created.

In Figures 7(b) and 7(d), we see that the Boolean expression labeling either a predecessor arc or successor arc is the conjunction ($\wedge$) of the Boolean expressions labeling the corresponding arcs in the two resolved paths. While in Figure 7(c), we see that the Boolean expression labeling the resolvant arc is the disjunction ($\vee$) of the Boolean expressions labeling the corresponding arcs in the two resolved paths. We also observe that the vertices in the resolvant path are created by two different methods. Both the head and the tail of each predecessor arc is of the form

$$\langle \mathit{aft}(v_1) \vee \mathit{aft}(v_2),\ \mathit{fore}(v_1) \wedge \mathit{fore}(v_2) \rangle$$

where $v_1$ and $v_2$ are the corresponding vertices in the two resolved paths, while both the head and the tail of each successor arc is of the form

$$\langle \mathit{aft}(v_1) \wedge \mathit{aft}(v_2),\ \mathit{fore}(v_1) \vee \mathit{fore}(v_2) \rangle$$

where, as before, $v_1$ and $v_2$ are the corresponding vertices in the two resolved paths.

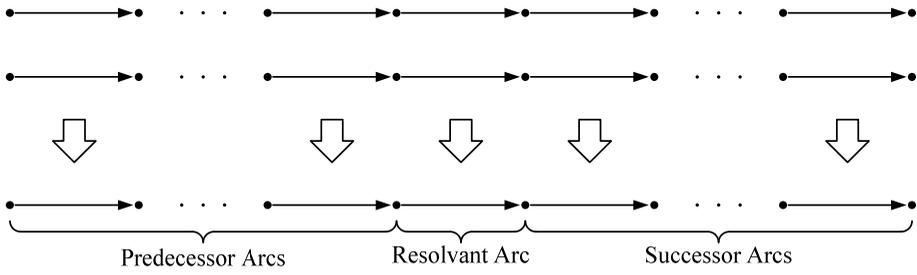

(a) Resolving Two Equal-Length Paths in an Elaboration

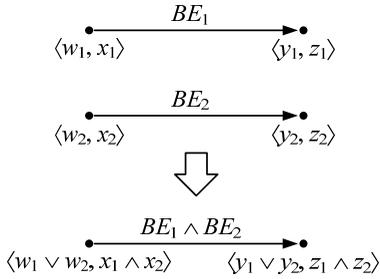

(b) Creating a Predecessor Arc

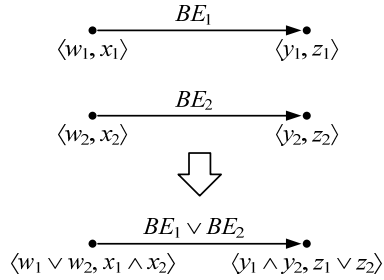

(c) Creating the Resolvant Arc

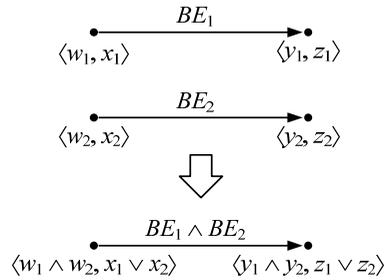

(d) Creating a Successor Arc

FIG. 7. Sequential Resolution

This construction is formalized as follows.

*Definition* 5.3. For a finite set of elements $V$ and for $aft_1, fore_1, aft_2, fore_2 \in SoS(V)$,

$$pre(\langle aft_1, fore_1 \rangle, \langle aft_2, fore_2 \rangle) = \langle (aft_1 \vee aft_2), (fore_1 \wedge fore_2) \rangle$$

$$post(\langle aft_1, fore_1 \rangle, \langle aft_2, fore_2 \rangle) = \langle (aft_1 \wedge aft_2), (fore_1 \vee fore_2) \rangle$$

*Definition* 5.4. Let $E$ be an elaboration of a Boolean graph, and let $a_1$ and $a_2$ be arcs in $E$. Then

$$predecessor(a_1, a_2) = \langle v_t, (label(a_1) \wedge label(a_2)), v_h \rangle$$

where $v_t = pre(tail(a_1), tail(a_2))$ and $v_h = pre(head(a_1), head(a_2))$,

$$resolvant(a_1, a_2) = \langle v_t, (label(a_1) \vee label(a_2)), v_h \rangle$$

where $v_t = pre(tail(a_1), tail(a_2))$ and $v_h = post(head(a_1), head(a_2))$,

$$successor(a_1, a_2) = \langle v_t, (label(a_1) \wedge label(a_2)), v_h \rangle$$

where $v_t = post(tail(a_1), tail(a_2))$ and $v_h = post(head(a_1), head(a_2))$.

*Definition* 5.5. Let $E$ be an elaboration of a Boolean graph, let $\mu_1$ and $\mu_2$ be equal-length paths in $E$ and let $k$ be an integer such that $0 \leq k < |\mu_1|$. Then $resolve(E, \mu_1, \mu_2, k)$ is the Boolean graph obtained by adding to $E$ the following arcs and associated vertices. For each $0 \leq i < k$, add the arc

$$predecessor(\mu_1(i), \mu_2(i))$$

Add the arc

$$resolvant(\mu_1(k), \mu_2(k))$$

For each $k < j < |\mu_1|$, add the arc

$$successor(\mu_1(j), \mu_2(j))$$

THEOREM 5.2. *If $E$ is an elaboration of Boolean graph $G$, $\mu_1$ and $\mu_2$ are equal-length paths in $E$ such that $pre(tail(\mu_1), tail(\mu_2))$ and $post(head(\mu_1), head(\mu_2))$ are vertices of $E$ and $k$ is an integer such that $0 \leq k < |\mu_1|$, then $resolve(E, \mu_1, \mu_2, k)$ is an elaboration of $G$.*

PROOF. By construction, each newly created arc in $resolve(E, \mu_1, \mu_2, k)$ is labeled with a Boolean expression over $AP$. $resolve(E, \mu_1, \mu_2, k)$ is therefore a Boolean graph over $AP$. We now show that all five properties required for $resolve(E, \mu_1, \mu_2, k)$ to be an elaboration of $G$ are satisfied by each newly created vertex and each newly created arc in $resolve(E, \mu_1, \mu_2, k)$.

1. *Each vertex is an ordered pair $\langle \textit{aft}, \textit{fore} \rangle$ where $\textit{aft}, \textit{fore} \in \textit{SoS}(IV(G))$* – Each newly created vertex is either of the form $\langle (\textit{aft}_1 \vee \textit{aft}_2), (\textit{fore}_1 \wedge \textit{fore}_2) \rangle$ or of the form $\langle (\textit{aft}_1 \wedge \textit{aft}_2), (\textit{fore}_1 \vee \textit{fore}_2) \rangle$ where $\textit{aft}_1, \textit{fore}_1, \textit{aft}_2, \textit{fore}_2 \in \textit{SoS}(IV(G))$. In both cases, the required property is satisfied.

2. *Each vertex $v$ is an initial vertex of $E$ if and only if $\textit{aft}(v) = \{\}$* – By construction, the newly created arcs in $\textit{resolve}(E, \mu_1, \mu_2, k)$ form a path with $\textit{pre}(\textit{tail}(\mu_1), \textit{tail}(\mu_2))$ as its tail. By assumption, this vertex is a pre-existing vertex in $E$, and therefore no new initial vertices are created by the resolution operation. Furthermore, since $\mu_1$ and $\mu_2$ are paths in $E$ and $E$ is an elaboration of $G$, for each arc $a$ in $\mu_1$ and $\mu_2$, $\textit{aft}(\textit{head}(a)) \neq \{\}$. It follows from Property 2.5(d) that for each newly created arc $a$, $\textit{aft}(\textit{head}(a)) \neq \{\}$. Hence no newly created arc is incident on the initial vertex of $E$, and the initial vertex of $E$ remains an initial vertex in $\textit{resolve}(E, \mu_1, \mu_2, k)$.

3. *Each vertex $v$ is a terminal vertex of $E$ if and only if $\textit{fore}(v) = \{\}$* – Argument is similar to that for Property 2.

4. *For each vertex $v$, $\sim\!\textit{aft}(v) \leq \textit{fore}(v)$* – Each newly created vertex is either of the form $\langle (\textit{aft}_1 \vee \textit{aft}_2), (\textit{fore}_1 \wedge \textit{fore}_2) \rangle$ or of the form $\langle (\textit{aft}_1 \wedge \textit{aft}_2), (\textit{fore}_1 \vee \textit{fore}_2) \rangle$ where $\langle \textit{aft}_1, \textit{fore}_1 \rangle$ and $\langle \textit{aft}_2, \textit{fore}_2 \rangle$ are pre-existing vertices in $E$. Since $E$ is an elaboration, it must be that $\sim\!\textit{aft}_1(v) \leq \textit{fore}_1(v)$ and $\sim\!\textit{aft}_2(v) \leq \textit{fore}_2(v)$. It follows from Property 2.1(b) that $\sim\!(\textit{aft}_1 \wedge \textit{aft}_2) \leq (\textit{fore}_1 \vee \textit{fore}_2)$ and from Property 2.1(c) that $\sim\!(\textit{aft}_1 \vee \textit{aft}_2) \leq (\textit{fore}_1 \wedge \textit{fore}_2)$.

5. *For each arc $a$, $\langle \textit{fore}(\textit{tail}(a)), \textit{label}(a), \textit{aft}(\textit{head}(a)) \rangle$ is a logical link of $G$* – By construction, each newly created predecessor arc $a$ is of the form $\langle v_t, (\textit{label}(a_1) \wedge \textit{label}(a_2)), v_h \rangle$, where $v_t = \textit{pre}(\textit{tail}(a_1), \textit{tail}(a_2))$, $v_h = \textit{pre}(\textit{head}(a_1), \textit{head}(a_2))$ and $a_1$ and $a_2$ are pre-existing arcs in $E$. That means that $v_t = \langle (\textit{aft}(\textit{tail}(a_1)) \vee \textit{aft}(\textit{tail}(a_2))), (\textit{fore}(\textit{tail}(a_1)) \wedge \textit{fore}(\textit{tail}(a_2))) \rangle$ and $v_h = \langle (\textit{aft}(\textit{head}(a_1)) \vee \textit{aft}(\textit{head}(a_2))), (\textit{fore}(\textit{head}(a_1)) \wedge \textit{fore}(\textit{head}(a_2))) \rangle$. Since $a_1$ and $a_2$ are arcs in $E$ and $E$ is an elaboration of $G$, we know that $\langle \textit{fore}(\textit{tail}(a_1)), \textit{label}(a_1), \textit{aft}(\textit{head}(a_1)) \rangle$ and $\langle \textit{fore}(\textit{tail}(a_2)), \textit{label}(a_2), \textit{aft}(\textit{head}(a_2)) \rangle$ are both logical links of $G$. From Theorem 4.7(c), it follows that $\langle (\textit{fore}(\textit{tail}(a_1)) \wedge \textit{fore}(\textit{tail}(a_2))), (\textit{label}(a_1) \wedge \textit{label}(a_2)), (\textit{aft}(\textit{head}(a_1)) \vee \textit{aft}(\textit{head}(a_2))) \rangle$ is a logical link of $G$. But that means that $\langle \textit{fore}(\textit{tail}(a)), \textit{label}(a), \textit{aft}(\textit{head}(a)) \rangle$ is a logical link of $G$. A similar argument, relying on Theorem 4.7(b), shows that the property is satisfied for the newly created resolvent arc. And an argument, relying on Theorem 4.7(a), shows that the property is satisfied for each newly created successor arc.

## 5.3 Combining Sequential Resolution and Boolean Resolution

The definition of sequential resolution in Section 5.2 may seem at odds with the notion of Boolean resolution. Specifically, there is nothing in the definition of sequential resolution corresponding to the elimination of a Boolean variable that appears negated in one term of a Boolean sum of products and not negated in another term.

But consider the special case where the Boolean expression labeling Arc $a$ in an elaboration of Boolean graph $G$ is of the form

$$(C_1 \wedge P) \vee (C_2 \wedge \neg P)$$

where $P$ is a Boolean variable and $C_1$ and $C_2$ are conjunctions of literals such that no Boolean variable appears negated in $C_1$ and not negated in $C_2$, or vice versa. Recall that in the definition of an elaboration (Definition 4.10), the only requirement on the Boolean expression labeling Arc $a$ is that $\langle fore(tail(a)), label(a), aft(head(a)) \rangle$ be a logical link of $G$. It follows from Property 4.2 that $(C_1 \wedge P) \vee (C_2 \wedge \neg P)$ can be replaced by

$$C_1 \wedge C_2$$

since this replacement serves only to *weaken* the link $\langle fore(tail(a)), label(a), aft(head(a)) \rangle$.

This result provides the foundation for a variant of sequential resolution that is used when the Boolean expressions labeling the arcs of an elaboration are all products of literals. This variation is identical to sequential resolution except for the Boolean expression labeling the resolvant arc. In contrast to Figure 7(c), there are now requirements on the Boolean expressions labeling the two arcs used to create the resolvant arc. One must be of the form $(C_1 \wedge P)$ and the other of the form $(C_2 \wedge \neg P)$, where $P$, $C_1$ and $C_2$ are as described above. The construction of the label for the resolvant arc from these two expressions is illustrated in Figure 8 as a two-step process (although in practice, these two steps are combined into one.) In the first step, sequential resolution is applied to $(C_1 \wedge P)$ and $(C_2 \wedge \neg P)$ to obtain $(C_1 \wedge P) \vee (C_2 \wedge \neg P)$, while in the second step, Boolean resolution is applied to $(C_1 \wedge P) \vee (C_2 \wedge \neg P)$ to obtain the label for the resolvant arc, $C_1 \wedge C_2$. The resulting operation on equal-length paths in an elaboration is a generalization of Boolean resolution in which conjunctions in *space* are replaced with conjunctions in both *space* and *time*. Sections 5.4 and 5.5 provide examples of this new form of resolution.

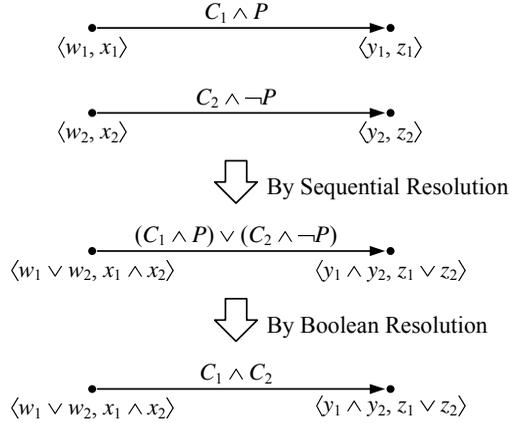

FIG. 8. Combining Sequential Resolution and Boolean Resolution

## 5.4 A Simple Example

In Section 1.2, we showed through an ad hoc argument that the sequence of Boolean expressions

$$\alpha = \langle (P \wedge R), true, \neg T \rangle$$

is an implicant of the set of Boolean sequences

$$A = \{ \langle P, \neg Q \rangle, \langle R, \neg S \rangle, \langle (Q \wedge S), \neg T \rangle \}$$

We now show how to achieve the same result using the variant of sequential resolution described in Section 5.3. First, we convert the set of sequences $A$ into the Boolean graph shown in Figure 9, and then construct from this graph the initial elaboration shown in Figure 10. (For graphical convenience, an elaboration is often depicted with multiple initial and terminal vertices even though there is just one initial vertex, $\langle \{\}, \{\{\}\} \rangle$, and one terminal vertex, $\langle \{\{\}\}, \{\} \rangle$.) Two sequential resolutions are then performed. The first resolution, shown in Figure 11(a), is performed on the initial elaboration and causes a single arc from vertex $\langle \{\{v_1\}\}, \{\{v_1\}\} \rangle$ to vertex $\langle \{\{v_7\}\}, \{\{v_7\}\} \rangle$ and labeled with the expression $S$ to be added to the initial elaboration thereby yielding the elaboration of Figure 11(b). The second resolution, shown in Figure 12(a), is performed on the elaboration of Figure 11(b) and causes a path containing a predecessor arc and a resolvant arc to be added to this elaboration. The predecessor arc leads from vertex $\langle \{\}, \{\{\}\} \rangle$ to vertex $\langle \{\{v_1\}, \{v_4\}\}, \{\{v_1, v_4\}\} \rangle$ and is labeled with $P \wedge R$, while the resolvant arc leads from vertex $\langle\{\{v_1\},\{v_4\}\},\{\{v_1,v_4\}\}\rangle$ to vertex $\langle\{\{v_7\}\},\{\{v_7\}\}\rangle$ and is labeled with *true*.

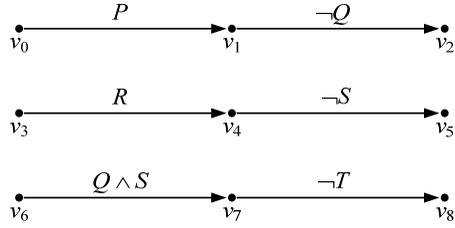

FIG. 9. Boolean Graph *G*

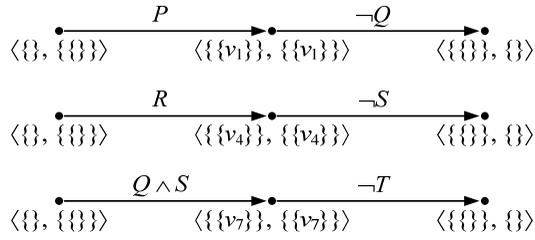

FIG. 10. Initial Elaboration of *G*

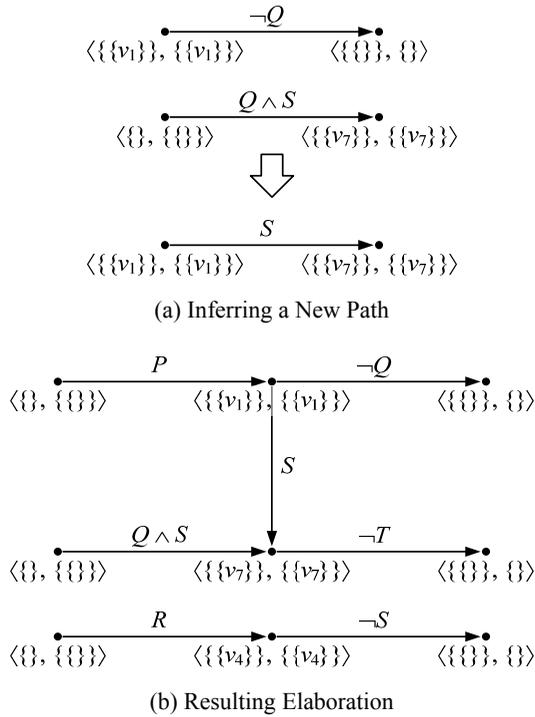

(a) Inferring a New Path

(b) Resulting Elaboration

FIG. 11. First Resolution

Now notice that the resulting elaboration in Figure 12(b) contains the path shown in Figure 13 which begins at the initial vertex of the elaboration, ends at the terminal vertex and is labeled with ⟨(P ∧ R), *true*, ¬T⟩. The elaboration therefore accepts ⟨(P ∧ R), *true*, ¬T⟩, and by Theorems 4.9 and 5.2 it follows that this sequence is an implicant of the Boolean graph in Figure 9. But since this graph accepts the set of sequences {⟨P, ¬Q⟩, ⟨R, ¬S⟩, ⟨(Q ∧ S), ¬T⟩}, it must be the case that

$$\langle (P \wedge R), \textit{true}, \neg T \rangle$$

*is an implicant of*

$$\{ \langle P, \neg Q \rangle, \langle R, \neg S \rangle, \langle (Q \wedge S), \neg T \rangle \}$$

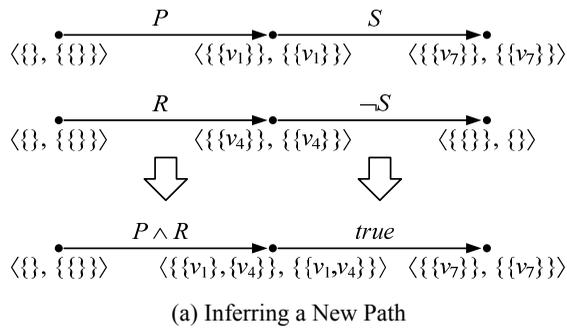

(a) Inferring a New Path

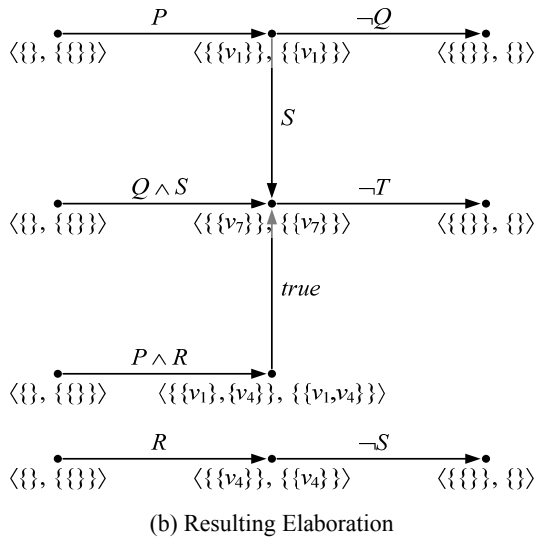

(b) Resulting Elaboration

FIG. 12. Second Resolution

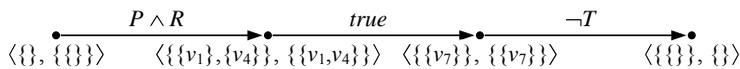

FIG. 13. Path from the Initial Vertex to the Terminal Vertex in the Final Elaboration

## 5.5 An Example of Induction

Mathematical *induction* is a method of mathematical proof used to establish that a given statement is true for all natural numbers $n$. It consists of two steps:

1. *Basis step*: Showing that the statement holds for $n = 0$

2. *Inductive step*: Showing that if the statement holds for $n = m$, where $m$ is any natural number, then the same statement also holds for $n = m + 1$

We now apply the induction principle to the following problem. We are given the logical/temporal dependency

*If P and Q hold in a state, then Q holds in the next state*

and wish to prove that the dependency

*For all natural numbers n,*
*if P and Q hold in a state and P holds in the next n states,*
*then Q holds in the state following this sequence of n states*

follows as a logical consequence. The proof by induction is as follows:

1. *Basis step*: For $n = 0$, the two statements are identical, and so the second statement follows trivially from the first.

2. *Inductive step*: Assume that the second statement is true for $n = m$. Suppose that $P$ and $Q$ hold in State 0 and that $P$ holds in States 1 to $m$. By our assumption, $Q$ must hold in State $m + 1$. Suppose, furthermore, that $P$ also holds in State $m + 1$. It then follows from the first statement that $Q$ holds in State $m + 2$. We have thus shown that if that $P$ and $Q$ hold in State 0 and that $P$ holds in States 1 to $m + 1$, then $Q$ holds in State $m + 2$. So the second statement is proved for the case where $n = m + 1$.

Let us now consider an alternative approach to proving that the second dependency follows from the first, one based on sequential resolution. We begin by converting the first statement into the sequential constraint $\langle (P \wedge Q), \neg Q \rangle$. Next, we construct a Boolean graph that accepts just that sequence (Figure 14), and then construct an initial elaboration from this graph (Figure 15). We then perform the sequential resolution shown in Figure 16(a) which produces a path consisting of a single arc, an arc that both begins and ends at the vertex $\langle \{\{v_1\}\}, \{\{v_1\}\} \rangle$. Figure 16(b) shows the graph that results when this arc is added to the initial elaboration in Figure 15. Notice that the infinite set of sequential constraints accepted by this elaboration correspond exactly to the second dependency. So we have achieved the same result as the induction argument above through a single sequential resolution.

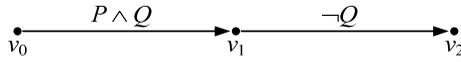

FIG. 14. Boolean Graph *G*

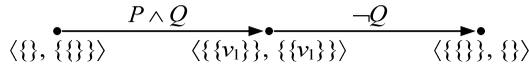

FIG. 15. Initial Elaboration of *G*

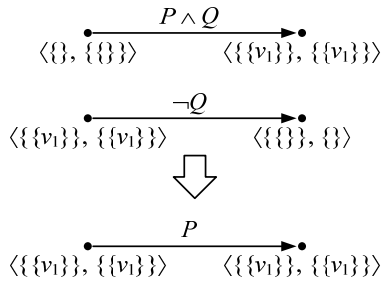

(a) Inferring a New Path

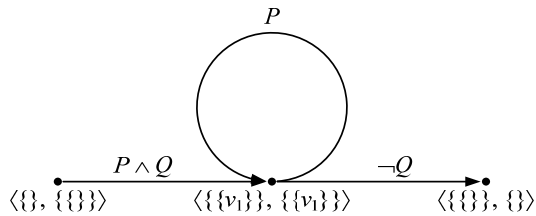

(b) Resulting Elaboration

FIG. 16. Induction Example

This example illustrates the principle that sequential resolution can deal with situations that require proving that a logical/temporal dependency spanning an unbounded number of states follows as a logical consequence from dependencies spanning a strictly bounded number of states. And sequential resolution does so without resorting to a classical, two-step induction argument.

## 6. NORMALIZATION

Normalization, the second method for constructing elaborations, starts with two Boolean graphs: (1) a graph representing a regular set of *known* sequential constraints and (2) a graph representing a set of *conjectured* sequential constraints. The first graph typically represents a system (model), while the second represents logical/temporal dependencies that one conjectures about the behavior of the system. Normalization determines which of those conjectured dependencies are satisfied by the system (model). The process involves transforming the conjectured graph, using the $max^+$ function defined above, into an elaboration of the system graph. The resulting *verified* graph satisfies two properties:

1. The verified graph is an elaboration of the system graph

2. For each sequence of Boolean expressions $\alpha$ that is (a) an implicant of the system graph and (b) accepted by the conjectured graph, there exists a subsequence of $\alpha$ that is accepted by the verified graph

The process of normalization is thus able to extract from a regular set of Boolean sequences those sequences that are sequential constraints as a consequence of a set of known sequential constraints. This capability means that someone who is unsure about a system's exact behavior can make an overly broad conjecture about that behavior – a conjecture known to be *false* – in order to find a version of the conjecture that is *true*.

### 6.1 Forwards-Maximal Elaborations

In the proof of Theorem 3.6, the $max^+$ function was used to construct a special type of elaboration of the set graph $G$ from an implicant of $G$. Normalization applies the same principle to construct a special type of elaboration of the Boolean graph $G$ from the set of implicants of $G$ accepted by a Boolean graph $E$. That special type of elaboration is called a *forwards-maximal elaboration*.

*Definition* 6.1. A *forwards-maximal elaboration* of a Boolean graph $G$ over a set of atomic propositions is an elaboration $(V, A)$ of $G$ such that

1. For all $v \in V$, $\sim aft(v) = fore(v)$

2. For all $a \in A$, $aft(head(a)) = max^+(G, fore(tail(a)), label(a))$

LEMMA 6.1. *Let $G$ and $E$ be Boolean graphs over the same set of atomic propositions and let $\mu$ be a path in $E$ such that*

1. *All vertices on $\mu$ are ordered pairs $\langle aft, fore \rangle$, where $aft, fore \in SoS(IV(G))$*

2. *For all vertices v on μ, ~aft(v) = fore(v)*

3. *For all arcs a on μ, aft(head(a)) = max$^+$(G, fore(tail(a)), label(a))*

Then *aft(head(μ)) = max$^+$(G, fore(tail(μ)), label(μ))*.

PROOF. Let ($S$, $B$, $L$) be a fully populated Kripke structure over the same set of atomic propositions as $G$ and $E$, and let $\mu_L$ be the image of $\mu$ under the mapping $L$. $\mu_L$ is thus a path in $L(G)$ such that

1. $tail(\mu_L) = tail(\mu)$
2. $label(\mu_L) = L(label(\mu))$
3. $head(\mu_L) = head(\mu)$
4. All vertices on $\mu_L$ are ordered pairs $\langle aft, fore \rangle$, where $aft, fore \in SoS(IV(G))$
5. For all vertices $v$ on $\mu_L$, $\sim aft(v) = fore(v)$
6. For all arcs $a$ on $\mu_L$, $aft(head(a)) = max^+(L(G), fore(tail(a)), label(a))$  (by Theorem 4.8)

Since $head(\mu_L) = head(\mu)$, $aft(head(\mu)) = aft(head(\mu_L))$. By Lemma 3.3,

$$aft(head(\mu_L)) = max^+(L(G), fore(tail(\mu_L)), label(\mu_L))$$

Since $tail(\mu_L) = tail(\mu)$ and $label(\mu_L) = L(label(\mu))$,

$$max^+(L(G), fore(tail(\mu_L)), label(\mu_L)) = max^+(L(G), fore(tail(\mu)), L(label(\mu)))$$

Finally, by Theorem 4.8,

$$max^+(L(G), fore(tail(\mu)), L(label(\mu))) = max^+(G, fore(tail(\mu)), label(\mu))$$

COROLLARY 6.1. *If E is a forwards-maximal elaboration of the Boolean graph G and μ is a path in E, then aft(head(μ)) = max$^+$(G, fore(tail(μ)), label(μ))*.

## 6.2 Normalization

The normalization process begins with two Boolean graphs, $G$ and $E$, with $E$ satisfying the requirement that none of the vertices of $E$ be of the form $\langle aft, fore \rangle$, where $aft, fore \in SoS(IV(G))$. In the first step of the process, all initial vertices of $E$ are merged into the *initial vertex* $\langle \{\}, \{\{\}\} \rangle$. That step is followed by the main phase of normalization, the *updating* of arcs $a$ in $E$ such that

$$tail(a) \in (SoS(IV(G)) \times SoS(IV(G))) \text{ and } head(a) \notin (SoS(IV(G)) \times SoS(IV(G)))$$

Each such update involves *splitting* the head of arc $a$ from its current location and *merging* it with the vertex

$$\langle \mathit{max}^+(G, \mathit{fore}(\mathit{tail}(a)), \mathit{label}(a)),\ \sim\!\mathit{max}^+(G, \mathit{fore}(\mathit{tail}(a)), \mathit{label}(a))\rangle$$

In addition, if the new location of *head*(*a*) is not $\langle\{\{\}\},\{\}\rangle$, then those arcs emerging from the former location of *head*(*a*) are copied and the tails of the copied arcs are merged with the new location of *head*(*a*). The requirement that $\mathit{head}(a) \neq \langle\{\{\}\},\{\}\rangle$ guarantees that no arcs are created emerging from the *terminal vertex* $\langle\{\{\}\},\{\}\rangle$.

If, during the course of updating arcs, an arc *a* is encountered such that each of the following properties holds

$$\mathit{tail}(a) \in (\mathit{SoS}(\mathit{IV}(G)) \times \mathit{SoS}(\mathit{IV}(G)))$$

$$\mathit{head}(a) \text{ is a terminal vertex of } E$$

$$\mathit{max}^+(G, \mathit{fore}(\mathit{tail}(a)), \mathit{label}(a)) \neq \{\{\}\}$$

then that arc is deleted since it cannot ever be on a path leading to the terminal vertex $\langle\{\{\}\},\{\}\rangle$.

When there are no further arcs to update, the *cleanup* phase begins. In the first part of this phase, all arcs *a* in *E* such that such that $\mathit{head}(a) = \langle\{\},\{\{\}\}\rangle$ are deleted, and in the second part, all arcs and vertices not on a path in *E* from the initial vertex $\langle\{\},\{\{\}\}\rangle$ to the terminal vertex $\langle\{\{\}\},\{\}\rangle$ are deleted.

These ideas are formalized in Definition 6.3. The following two abbreviations help simplify that definition.

*Definition* 6.2. Let *G* be a Boolean graph over a set of atomic propositions *AP*, $\mathit{aft} \in \mathit{SoS}(\mathit{IV}(G))$ and *BE* a Boolean expression over *AP*. Then

$$\mathit{vertices}(G) = \mathit{SoS}(\mathit{IV}(G)) \times \mathit{SoS}(\mathit{IV}(G))$$

$$\mathit{vertex}^+(G, \mathit{aft}, \mathit{BE}) = \langle \mathit{max}^+(G, \mathit{aft}, \mathit{BE}), \sim\!\mathit{max}^+(G, \mathit{aft}, \mathit{BE})\rangle$$

*Definition* 6.3. Let *G* and $E = (V_E, A_E)$ be Boolean graphs over the same set of atomic propositions such that $V_E \cap \mathit{vertices}(G)$ is empty. *normalize*(*G*, *E*) is the Boolean graph *E* after it has been transformed by the following algorithm.

1. Add $\langle\{\},\{\{\}\}\rangle$ to $V_E$

2. For each arc $\langle v_t, \mathit{BE}, v_h\rangle \in A_E$ such that $v_t$ is an initial vertex of *E*, replace that arc in $A_E$ with $\langle\langle\{\},\{\{\}\}\rangle, \mathit{BE}, v_h\rangle$

3. While there exists an arc $\langle v_t, \mathit{BE}, v_h\rangle \in A_E$ such that $v_t \in \mathit{vertices}(G)$ and $v_h \notin \mathit{vertices}(G)$,

(a) If $v_h$ is a terminal vertex of $E$ and $vertex^+(G, fore(v_t), BE) \neq \langle\{\{\}\},\{\}\rangle$,

   i. Delete $\langle v_t, BE, v_h \rangle$

(b) Else

   i. Add $vertex^+(G, fore(v_t), BE)$ to $V_E$

   ii. Replace $\langle v_t, BE, v_h \rangle$ in $A_E$ with $\langle v_t, BE, vertex^+(G, fore(v_t), BE) \rangle$

   iii. If $vertex^+(G, fore(v_t), BE) \neq \langle\{\{\}\},\{\}\rangle$, then for each arc $\langle u_t, BE', u_h \rangle \in A_E$ such that $u_t = v_h$, add $\langle vertex^+(G, fore(v_t), BE), BE', u_h \rangle$ to $A_E$ if it has not been previously added to $A_E$

4. Delete all arcs $a \in A_E$ such that $head(a) = \langle\{\},\{\{\}\}\rangle$

5. Delete all arcs in $A_E$ and vertices in $V_E$ that are not on a path in $E$ from the vertex $\langle\{\},\{\{\}\}\rangle$ to the vertex $\langle\{\{\}\},\{\}\rangle$

LEMMA 6.2. Let $G$ and $E = (V_E, A_E)$ be Boolean graphs over the same set of atomic propositions such that $V_E \cap vertices(G)$ is empty. Then $normalize(G, E)$ produces a result in a finite number of steps.

PROOF. To prove the lemma, we need to show that there can only be a finite number of iterations of the *while* loop in Step 3 of Definition 6.3. To accomplish that, we first observe that each such iteration requires an arc $\langle v_t, BE, v_h \rangle \in A_E$ such that $v_t \in vertices(G)$ and $v_t \neq \langle\{\{\}\},\{\}\rangle$ and $v_h \notin vertices(G)$. Since (the pre-normalized) $E$ is a finite, there can only be a finite number of such arcs at the beginning of the algorithm. Additional arcs satisfying these conditions are created only via Step 3(b)(iii), but the requirement that any new arc must not have been previously added to $A_E$ means that only a finite number of such arcs can be added to $A_E$. Therefore, there can only be a finite number of iterations of the *while* loop in Step 3.

LEMMA 6.3. Let $G$ and $E = (V_E, A_E)$ be Boolean graphs over the same set of atomic propositions such that $V_E \cap vertices(G)$ is empty and let $\mu$ be a path in $E$ such that

1. $tail(\mu)$ is an initial vertex of $E$

2. For all proper prefixes $\mu_P$ of $\mu$, $max^+(G, \{\{\}\}, label(\mu_P)) \neq \{\{\}\}$

3. $head(\mu)$ is not a terminal vertex of $E$ or $max^+(G, \{\{\}\}, label(\mu)) = \{\{\}\}$

Then in Steps 1, 2 and 3 of Definition 6.3, $\mu$ is transformed into a new path $\mu_T$ in $E$ such that

4. $label(\mu_T) = label(\mu)$

5. For all vertices $v$ on $\mu_T$, $v \in vertices(G)$

6. $tail(\mu_T) = \langle \{\}, \{\{\}\} \rangle$

7. For all interior vertices $v$ on $\mu_T$, $v \neq \langle \{\{\}\}, \{\} \rangle$

PROOF. See Appendix F.

LEMMA 6.4. Let $G$ and $E = (V_E, A_E)$ be Boolean graphs over the same set of atomic propositions such that $V_E \cap vertices(G)$ is empty. Then $normalize(G, E)$ produces a unique result.

PROOF. To prove the lemma, we must show that $normalize(G, E)$ produces the same result regardless of the order in which arcs are *processed* in Step 3 of Definition 6.3. To that end, let $M$ be the set of paths in (the pre-normalized) $E$ satisfying Properties 1 – 3 in Lemma 6.3. By Lemma 6.3, the paths in $M$ are transformed in Steps 1, 2 and 3 of Definition 6.3 into a set of paths $N$ satisfying Properties 4 – 7 in Lemma 6.3. Let $V$ be the set of vertices appearing on a path in $N$ and let $A$ be the set of arcs appearing on a path in $N$. Notice that both $V$ and $A$ are independent of the order in which arcs are *processed* in Step 3 of Definition 6.3. Moreover, it follows from Definition 6.3 that at the end of Step 3 the only vertices $v$ of $E$ such that $v \in vertices(G)$ are those in $V$ and the only arcs $a$ of $E$ such that $tail(a) \in vertices(G)$ and $head(a) \in vertices(G)$ are those in $A$. Since these are the only types of vertices and arcs remaining after the deletions of Steps 4 and 5, we conclude that $normalize(G, E)$ produces the same result regardless of the order in which the heads of arcs are updated in Step 3 of Definition 6.3.

THEOREM 6.1. Let $G$ and $E = (V_E, A_E)$ be Boolean graphs over the same set of atomic propositions such that $V_E \cap vertices(G)$ is empty. Then $normalize(G, E)$ is well defined.

PROOF. A consequence of Lemmas 6.2 and 6.4.

THEOREM 6.2. Let $G$ and $E = (V_E, A_E)$ be Boolean graphs over the same set of atomic propositions such that $V_E \cap vertices(G)$ is empty. Then $normalize(G, E)$ is a forwards-maximal elaboration of $G$.

PROOF. By construction, each newly created arc in $normalize(G, E)$ is labeled with a Boolean expression over $AP$. $normalize(G, E)$ is therefore a Boolean graph over $AP$. We now show that all five properties required for *normalize*(G, E) to be an elaboration of G are satisfied.

1. *Each vertex is an ordered pair ⟨aft, fore⟩ where aft, fore ∈ SoS(IV(G))* – The deletions in Step 5 of Definition 6.3 ensure that only *updated* vertices – those vertices that are of the form ⟨*aft, fore*⟩, where *aft, fore* ∈ SoS(IV(G)) – remain at the completion of the algorithm.

2. *Each vertex v is an initial vertex of E if and only if aft(v) = {}* – Step 4 in Definition 6.3 ensures that ⟨{},{{}}⟩, if it exists, is an initial vertex of E. Step 5 ensures that there are no other initial vertices of E.

3. *Each vertex v is a terminal vertex of E if and only if fore(v) = {}* – For each arc *a* created in the normalization process, *tail*(*a*) ≠ ⟨{{}},{}⟩. So ⟨{{}},{}⟩, if it exists, is a terminal vertex. Step 5 in Definition 6.3 ensures that there are no terminal vertices other than ⟨{{}},{}⟩.

4. *For each vertex v, ~aft(v) = fore(v)* – Each updated vertex is either the initial vertex ⟨{},{{}}⟩ or is of the form ⟨*max*$^+$(G, *fore*(v), BE), ~*max*$^+$(G, *fore*(v), BE)⟩.

5. *For each arc a, aft(head(a)) = max$^+$(G, fore(tail(a)), label(a))* – Each arc created in the normalization process is of the form ⟨$v_t$, BE, *vertex*$^+$(G, *fore*($v_t$), BE)⟩. The required property follows.

THEOREM 6.3. Let G and E = ($V_E$, $A_E$) be Boolean graphs over the same set of atomic propositions such that $V_E$ ∩ *vertices*(G) is empty. Then for each sequence of Boolean expressions $\alpha$ that is accepted by E and is an implicant of G, there exists a subsequence of $\alpha$ that is accepted by *normalize*(G, E).

PROOF. Suppose that the sequence of Boolean expressions $\alpha$ is accepted by E and is an implicant of G. Because $\alpha$ is accepted by E, there exists a path $\mu$ in E such that *tail*($\mu$) is an initial vertex of E, *label*($\mu$) = $\alpha$ and *head*($\mu$) is a terminal vertex of E. Because $\alpha$ is an implicant of G, it follows from Lemma 4.1 and Property 4.3 that *max*$^+$(G, {{}}, $\alpha$) = {{}}. Therefore *max*$^+$(G, {{}}, *label*($\mu$)) = {{}}. Let $\mu_P$ be the minimum-length prefix of $\mu$ such that *max*$^+$(G, {{}}, *label*($\mu_P$)) = {{}}. Thus for all proper prefixes $\mu_{PP}$ of $\mu_P$, *max*$^+$(G, {{}}, *label*($\mu_{PP}$)) ≠ {{}}. It follows from Lemma 6.3 that $\mu_P$ is transformed into a new path $\mu_{PT}$ in E such that

1. *label*($\mu_{PT}$) = *label*($\mu_P$)

2. For all vertices $v$ on $\mu_{PT}$, $v \in vertices(G)$

3. $tail(\mu_{PT}) = \langle\{\},\{\{\}\}\rangle$

4. For all interior vertices $v$ of $\mu_{PT}$, $v \neq \langle\{\{\}\},\{\}\rangle$

Furthermore, since $max^+(G, \{\{\}\}, label(\mu_P)) = \{\{\}\}$, it must be that $max^+(G, \{\{\}\}, label(\mu_{PT})) = \{\{\}\}$. It follows from Lemma 6.1 that $head(\mu_{PT}) = \langle\{\{\}\},\{\}\rangle$. Now consider Step 4 of Definition 6.3. In this step, all arcs $a$ in $\mu_{PT}$ such that $head(a) = \langle\{\},\{\{\}\}\rangle$ are deleted. But that still leaves a suffix $\mu_{PTS}$ of $\mu_{PT}$ – containing, at a minimum, the last arc of $\mu_{PT}$ – such that

1. $tail(\mu_{PTS}) = \langle\{\},\{\{\}\}\rangle$

2. For all interior vertices $v$ of $\mu_{PTS}$, $v \neq \langle\{\{\}\},\{\}\rangle$ and $v \neq \langle\{\{\}\},\{\}\rangle$

3. $head(\mu_{PTS}) = \langle\{\{\}\},\{\}\rangle$

Finally, consider Step 5 of Definition 6.3. Since all the vertices and arcs of $\mu_{PTS}$ are on a path in $E$ from $\langle\{\},\{\{\}\}\rangle$ to $\langle\{\{\}\},\{\}\rangle$, $\mu_{PTS}$ is left untouched. So we have shown that there exists a subsequence of $\alpha$ – namely, $label(\mu_{PTS})$ – that is accepted by $normalize(G, E)$.

## 6.2 An Example

In Section 4, we illustrated the concept of a constraint graph with the 2-bit-counter example in Figure 5, and illustrated the concept of an elaboration with the Boolean graph in Figure 6. But there was no explanation in Section 4 of how the elaboration in Figure 6 was obtained from the constraint graph in Figure 5. Figures 17(a) – 17(h) show how that was accomplished by normalizing the Boolean graph $E = (V_E, A_E)$ in Figure 17(a) using the Boolean graph $G$ in Figure 5.

(a) Figure 17(a) depicts the initial version of Boolean graph $E$. When interpreted as a constraint graph, it says that *Carry* = 1 in all future states following *Reset*. But this statement is clearly false; *Carry* = 1 only in certain states following *Reset*. The Boolean graph obtained by normalizing $E$ using $G$ tells us exactly what those states are.

(b) Figure 17(b) shows the Boolean graph $E$ after all initial vertices of $E$ (there is only one in this example) are merged into the single vertex $\langle\{\},\{\{\}\}\rangle$.

(c) Figure 17(c) shows the outcome from *updating* the head of arc $\langle\langle\{\},\{\{\}\}\rangle, Reset, u_1\rangle$ in Step 3(b). The outcome is three new arcs. The arc $\langle\langle\{\},\{\{\}\}\rangle, Reset,$

⟨{{v₆},{v₉}},{{v₆,v₉}}⟩⟩ replaces the arc ⟨⟨{},{{}}⟩, *Reset*, $u_1$⟩ in Step 3(b)(ii). The two arcs ⟨⟨{{v₆},{v₉}},{{v₆,v₉}}⟩, *true*, $u_1$⟩ and ⟨⟨{{v₆},{v₉}},{{v₆,v₉}}⟩, ¬*Carry*, $u_2$⟩ are created in Step 3(b)(iii). The second of these two arcs, however, is ultimately deleted in a future Step 3(a), and that fact is indicated with an **X** through the arc.

(d) Figure 17(d) shows the result of updating the head of arc ⟨⟨{{v₆},{v₉}},{{v₆,v₉}}⟩, *true*, $u_1$⟩ in Step 3(b). The outcome is similar to that of Figure 17(c).

(e) Figure 17(e) shows the result of updating the head of arc ⟨⟨{{v₃},{v₁₂}},{{v₃,v₁₂}}⟩, *true*, $u_1$⟩ in Step 3(b). The outcome is similar to that of Figures 17(c) and 17(d).

(f) Figure 17(f) shows the result of updating the heads of two arcs, ⟨⟨{{v₆},{v₁₂}},{{v₆,v₁₂}}⟩, *true*, $u_1$⟩ and ⟨⟨{{v₆},{v₁₂}},{{v₆,v₁₂}}⟩, ¬*Carry*, $u_2$⟩, in two iterations of Step 3(b). The outcome is similar to that of Figures 17(c) – 17(e), except that the arc ⟨⟨{{v₆},{v₁₂}},{{v₆,v₁₂}}⟩, ¬*Carry*, ⟨{{}},{}⟩⟩ is not deleted in a future Step 3(a) since it does not satisfy the condition in that step.

(g) Figure 17(g) shows the result of *updating* the head of arc ⟨⟨{{v₃},{v₉}},{{v₃,v₉}}⟩, *true*, $u_1$⟩ in Step 3(b), but unlike all previous updates, there are no arcs created in Step 3(b)(iii). That's because the two arcs ⟨⟨{{v₆},{v₉}},{{v₆,v₉}}⟩, *true*, $u_1$⟩ and ⟨⟨{{v₆},{v₉}},{{v₆,v₉}}⟩, ¬*Carry*, $u_2$⟩ were previously added to $A_E$ in Figure 17(c).

(h) Finally, Figure 17(h) shows the result of deleting in Step 4 all arcs incident on the vertex ⟨{},{{}}⟩ (there are none) and deleting in Step 5 all vertices $v \in V_E$ and arcs $a \in A_E$ that are not on a path in $E$ from the vertex ⟨{},{{}}⟩ to the vertex ⟨{{}},{}⟩. These final deletions eliminate the subgraph indicated in Figure 17(g). Notice that the resulting graph is identical to the one in Figure 6. When interpreted as a constraint graph, it tells us that *a Carry occurs 3 states following Reset and every 4$^{th}$ state thereafter*.

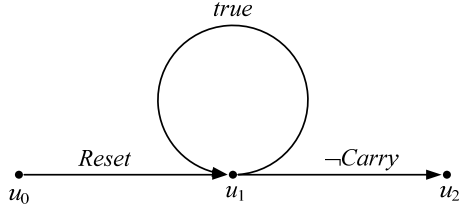

(a) Initial Boolean Graph *E*

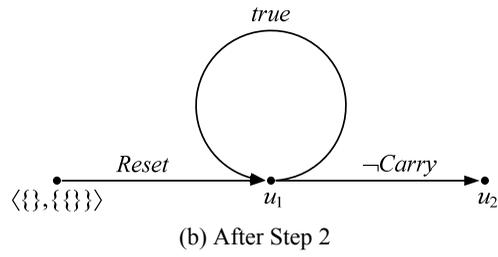

(b) After Step 2

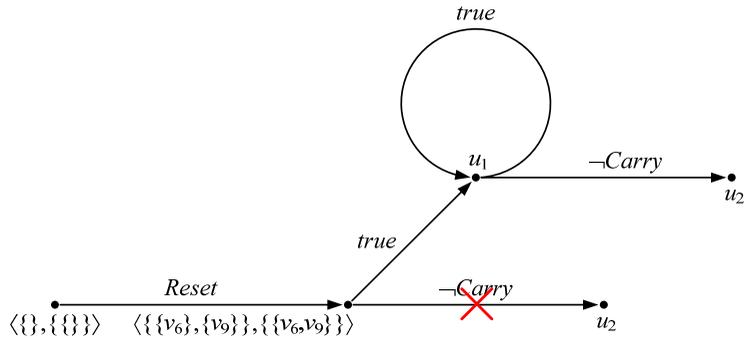

(c) After First Arc Update (✗ Indicates Arc to be Eventually Deleted in Step 3(a))

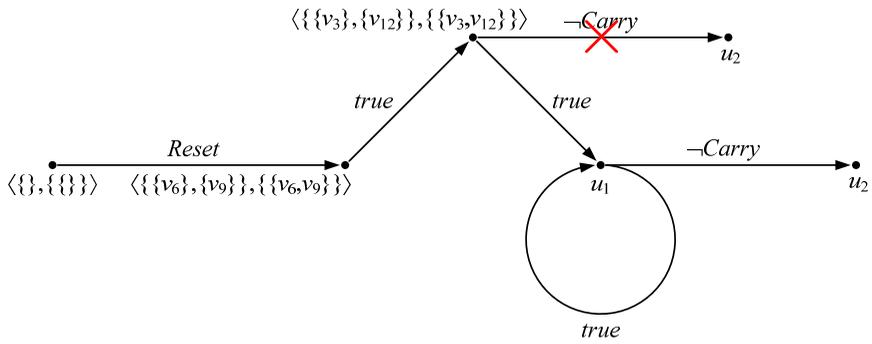

(d) After Second Arc Update (✗ Indicates Arc to be Eventually Deleted in Step 3(a))

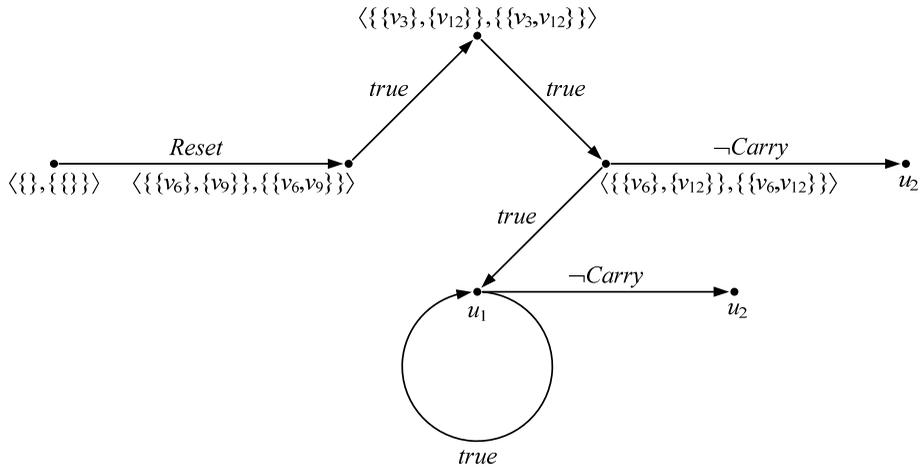

(e) After Third Arc Update

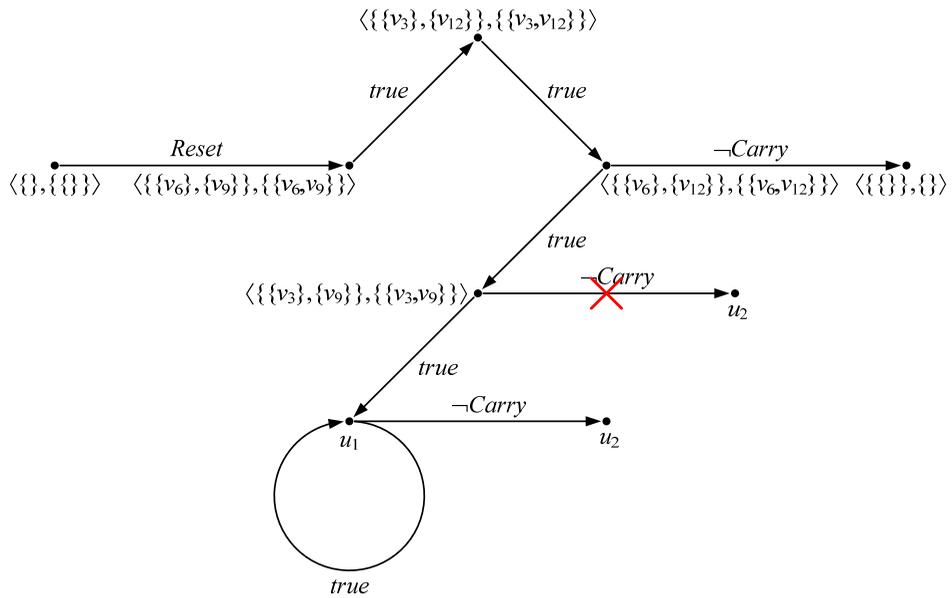

(f) After Fourth and Fifth Arc Updates (✗ Indicates Arc to be Eventually Deleted in Step 3(a))

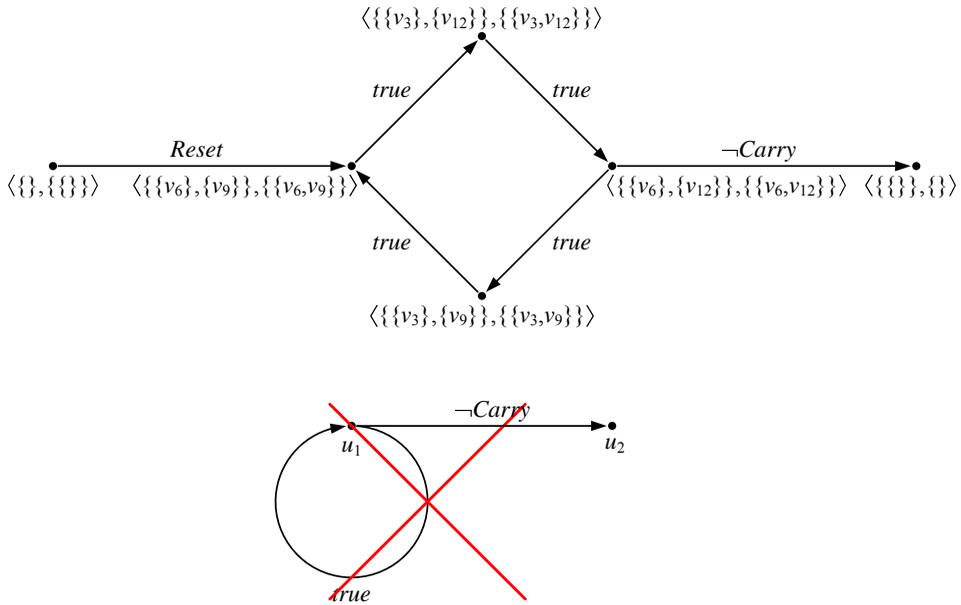

(g) After Sixth Arc Update (✗ Indicates Subgraph to be Deleted in Step 5)

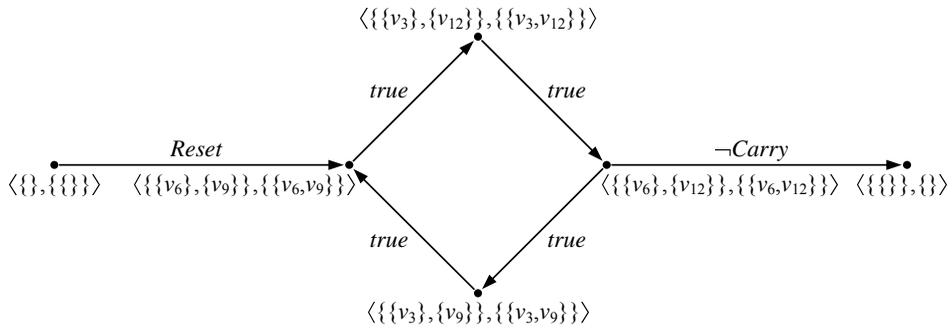

(h) After Step 5 – The Final Normalized Graph

FIG. 17. Normalization Example

## 7. CONCLUSIONS

Reasoning about sequential behavior is a fundamental and inescapable part of digital design, but for too long, this reasoning has been guided by informal, and highly error-prone, mental models. The mathematical theory and calculus described in the preceding sections hopefully contribute towards an eventual design methodology that is both mathematically rigorous and accessible to the average designer/programmer.

## 7.1 Distinguishing Characteristics of the Theory

The theory is distinguished from other approaches to formal verification by the following characteristics:

- The theory is primarily *mathematical*, with the formal/symbolic aspects of the theory playing a relatively minor role.

- The theory has only one type of construct for describing both systems and logical/temporal dependencies: a *regular set of sequential constraints* represented by either a regular expression or finite state automaton.

- Proofs are obtained through *deductive reasoning* entirely within the realm of logical/temporal dependencies. No attempt is made to model a system's state-transition function, and no attempt is made to explore, traverse or enumerate a system's state space.

- There are two proof methods: *Sequential resolution*, a generalization of Boolean resolution, allows new logical/temporal dependencies to be inferred from existing dependencies. *Normalization* starts with a model (system) and a set of logical/temporal dependencies and determines which of those dependencies are satisfied by the model.

- *Finite state automata* play a central role in the theory, but, in contrast to the usual practice, each FSA describes a set of *disallowed* system state sequences – but not necessarily *all* disallowed state sequences. This last point is significant. Because the theory relies on deductive reasoning, ignoring disallowed behaviors affects what is provable but does not affect the soundness of proofs obtained via either of the two proof methods.

- When a new component or instruction is added to a system, the sequential constraints associated with that component or instruction are added to the set of sequential constraints defining the system. The set of sequential constraints defining a system thus grows *linearly*, not *exponentially*, with the size of a system. A combinatorial explosion is still possible, but if it occurs, it is only through repeated applications of sequential resolution or in the normalization process.

- The assumption that a system state is *total* – that is, the current state completely determines the set of possible next states – is replaced by a more fundamental assumption (axiom): *every subsequence of an allowed state sequence is allowed*. The increased generality afforded by this axiom means that the theory can describe and reason about the *partial states* associated with the *visible* (*black box*) *behavior* of a system.

- Through the normalization process, someone who is unsure about a system's exact behavior can make an overly broad conjecture about that behavior – a conjecture known to be *false* – in order to find a version of the conjecture that is *true*.

## 7.2 Topics Not Covered

Although a lot of ground has been covered in this paper, a number of topics have been deferred to future articles.

- Boolean expressions with *uninterpreted functions*
- *Temporal offsets* appearing as arguments of uninterpreted functions which permits the concise representation of non-recursive dependencies
- *Formal variables* appearing as arguments of uninterpreted functions which permits the representation of recursive dependencies
- *Prime* (*sequential*) *implicants*, the sequential counterpart to prime implicants in Boolean logic
- A *completeness* theorem for sequential resolution that mirrors the completeness theorem for Boolean resolution
- *Self-normalization*, whereby a Boolean graph is normalized with itself to produce a graph in *canonical*/*normal form*
- Algorithms for computing $max^+$ and $max^-$
- An algorithm for deriving the *input/output* (*black-box*) *behavior* of a system
- *Heuristics* that reduce the chances for a combinatorial explosion in sequential resolution and in normalization
- A *constraint-based simulator* that behaves like a conventional cycle-accurate simulator except that it provides visibility into *cause and effect* by allowing a user to determine why Signal *S* has Value *V* at Time *T*.

## 7.3 Future Research

The following are suggestions for future research.

- While we may have solved the *state-space-explosion problem*, we have not completely solved the *combinatorial-explosion problem*. Using sequential resolution to generate the implicants of a Boolean graph, in particular, is prone to such an explosion. But that should not come as a surprise since using Boolean resolution – a special form of sequential resolution – to generate the implicants of a Boolean sum of products is also prone to such an explosion. Fortunately, there are a host of techniques for dealing with the Boolean problem, and many of these should be applicable to the sequential case. In fact, several heuristics – including the pruning of extraneous arcs – have already been incorporated into the sequential resolution algorithm. More work needs to be done in this area.

- *Hierarchy* has historically played an important role in managing complexity. The theory described here needs to be extended to encompass both different *granularities of time* and different *levels of abstraction*.

- Since the output of sequential resolution and normalization is ultimately intended for human consumption, there needs to more work done in making the output of these algorithms more *readable*. Temporal logics, like PSL [Accellera 2004], and languages for describing regular expressions can play an important role here.

- Currently, only sequential resolution can deal with uninterpreted functions. Normalization also needs to be made compatible with uninterpreted functions.

- Proving properties about *allowed* (*permitted*) behavior has been mentioned in passing, but this is an important area that deserves considerably more attention.

- The theory described here deals only with *regular* sets of disallowed sequences, but what interesting results are there for *context-free*, *context-sensitive* and *recursively-enumerable* sets of disallowed sequences? And what role do uninterpreted functions play in the expressiveness of the theory?

- A basic assumption of our theory – and many other theories in computer science – is that it is meaningful and productive to represent system behavior in terms of *total orderings of states*, but Petri [1962, 1986], Holt [1968, 1971] and others have stressed the fundamental nature of *concurrency*. In their models of system behavior, total orderings of states are replaced by *partial orderings* on either *condition holdings* or *event occurrences*. How do we extend the theory of sequential constraints to deal with such partial orderings?

- The theory described here is essentially an extension of propositional logic to handle sequential behavior, and although this logic has been further extended with uninterpreted functions, it will be necessary to incorporate techniques from theorem proving [Owre et. al. 1992; Owre et. al. 1998] in order to achieve the power and expressiveness of theorem proving together with the automated deduction supported by the present approach.

Appendix A: Proof of Theorem 3.2

THEOREM 3.2. *If $\langle aft_1, \alpha_1, fore_1\rangle$ and $\langle aft_2, \alpha_2, fore_2\rangle$ are links of the set graph G such that ~$fore_1 \leq aft_2$, then $\langle aft_1, \alpha_1 \bullet \alpha_2, fore_2\rangle$ is a link of G.*

PROOF. Suppose that $\langle aft_1, \alpha_1, fore_1\rangle$ and $\langle aft_2, \alpha_2, fore_2\rangle$ are links of the set graph $G$ such that ~$fore_1 \leq aft_2$. Let $A$ be the set of ordered pairs $\langle aset_1, \omega_1\rangle$, where $aset_1 \in aft_1$ and $\omega_1 \in \times\alpha_1$, such that there does NOT exist a path $\mu$ in $G$ such that at least one of the following two properties holds:

1. (a) $\times label(\mu)$ contains a subsequence of $\omega_1$ and (b) $tail(\mu)$ is an initial vertex of $G$ and (c) $head(\mu)$ is a terminal vertex of $G$

2. (a) $\times label(\mu)$ contains a prefix of $\omega_1$ and (b) $tail(\mu) \in aset_1$ and (c) $head(\mu)$ is a terminal vertex of $G$

Let $B$ denote the interior vertices of $G$. Let $C$ be the set of ordered pairs $\langle \omega_2, fset_2\rangle$, where $\omega_2 \in \times\alpha_2$ and $fset_2 \in fore_2$, such that there does NOT exist a path $\mu$ in $G$ such that at least one of the following two properties holds:

3. (a) $\times label(\mu)$ contains a subsequence of $\omega_2$ and (b) $tail(\mu)$ is an initial vertex of $G$ and (c) $head(\mu)$ is a terminal vertex of $G$

4. (a) $\times label(\mu)$ contains a suffix of $\omega_2$ and (b) $tail(\mu)$ is an initial vertex of $G$ and (c) $head(\mu) \in fset_2$

Let the relation $R_{AB} \subseteq A \times B$ be defined such that $\langle aset_1, \omega_1\rangle \ R_{AB}\ v$ if and only if there exists a path $\mu$ in $G$ such that at least one of the following two properties holds:

5. (a) $\times label(\mu)$ contains a suffix of $\omega_1$ and (b) $tail(\mu)$ is an initial vertex of $G$ and (c) $head(\mu) = v$

6. (a) $\times label(\mu)$ contains $\omega_1$ and (b) $tail(\mu) \in aset_1$ and (c) $head(\mu) = v$

Let the relation $R_{BC} \subseteq B \times C$ be defined such that $v\ R_{BC}\ \langle \omega_2, fset_2\rangle$ if and only if there exists a path $\mu$ in $G$ such that at least one of the following two properties holds:

7. (a) $\times label(\mu)$ contains a prefix of $\omega_2$ and (b) $tail(\mu) = v$ and (c) $head(\mu)$ is a terminal vertex of $G$

8. (a) $\times label(\mu)$ contains $\omega_2$ and (b) $tail(\mu) = v$ and (c) $head(\mu) \in fset_2$

Now consider an arbitrary ordered pair $\langle aset_1, \omega_1\rangle$ in $A$ and an arbitrary $fset_1$ in $fore_1$. Because $\langle aft_1, \alpha_1, fore_1\rangle$ is a link of $G$, we know that there exists a path $\mu$ in $G$ such that at least one of the following four properties holds:

9. (a) ×*label*($\mu$) contains a subsequence of $\omega_1$ and (b) *tail*($\mu$) is an initial vertex of $G$ and (c) *head*($\mu$) is a terminal vertex of $G$

10. (a) ×*label*($\mu$) contains a prefix of $\omega_1$ and (b) *tail*($\mu$) $\in$ *aset*$_1$ and (c) *head*($\mu$) is a terminal vertex of $G$

11. (a) ×*label*($\mu$) contains a suffix of $\omega_1$ and (b) *tail*($\mu$) is an initial vertex of $G$ and (c) *head*($\mu$) $\in$ *fset*$_1$

12. (a) ×*label*($\mu$) contains $\omega_1$ and (b) *tail*($\mu$) $\in$ *aset*$_1$ and (c) *head*($\mu$) $\in$ *fset*$_1$

However, because of the way in which $A$ is defined, neither Property 9 nor Property 10 can hold. Therefore, either Property 11 or Property 12 must hold. But that means that there exists $v \in$ *fset*$_1$ – namely, *head*($\mu$) – such that for all $\langle$*aset*$_1$, $\omega_1\rangle \in A$: $\langle$*aset*$_1$, $\omega_1\rangle$ $R_{AB}$ $v$. Hence, for all *fset*$_1 \in$ *fore*$_1$: $R_{AB}^{-1}$(*fset*$_1$) = $A$. Thus, *fore*$_1 \subseteq \{P \subseteq B \mid R_{AB}^{-1}(P) = A\}$. Using a similar argument, we have *aft*$_2 \subseteq \{Q \subseteq B \mid R_{BC}(Q) = C\}$. From Property 2.4, it follows that

$$fore_1 \leq min_\subseteq(\{P \subseteq B \mid R_{AB}^{-1}(P) = A\})$$
$$aft_2 \leq min_\subseteq(\{Q \subseteq B \mid R_{BC}(Q) = C\})$$

and from Property 2.1(a) and the fact that ~*fore*$_1 \leq$ *aft*$_2$, it follows that

$$\sim min_\subseteq(\{P \subseteq B \mid R_{AB}^{-1}(P) = A\}) \leq min_\subseteq(\{Q \subseteq B \mid R_{BC}(Q) = C\})$$

Applying the Fundamental Theorem (Theorem 3.1), we see that for all $\langle$*aset*$_1$, $\omega_1\rangle \in A$ and for all $\langle\omega_2$, *fset*$_2\rangle \in C$, there exists $v \in B$ such that $\langle$*aset*$_1$, $\omega_1\rangle$ $R_{AB}$ $v$ and $v$ $R_{BC}$ $\langle\omega_2$, *fset*$_2\rangle$. It follows that if none of Properties 1 – 4 holds, there must exist a path $\mu$ in $G$ such that at least one of the following four properties holds:

13. (a) ×*label*($\mu$) contains a subsequence of $\omega_1\bullet\omega_2$ and (b) *tail*($\mu$) is an initial vertex of $G$ and (c) *head*($\mu$) is a terminal vertex of $G$

14. (a) ×*label*($\mu$) contains a prefix of $\omega_1\bullet\omega_2$ and (b) *tail*($\mu$) $\in$ *aset* and (c) *head*($\mu$) is a terminal vertex of $G$

15. (a) ×*label*($\mu$) contains a suffix of $\omega_1\bullet\omega_2$ and (b) *tail*($\mu$) is an initial vertex of $G$ and (c) *head*($\mu$) $\in$ *fset*

16. (a) ×*label*($\mu$) contains $\omega_1\bullet\omega_2$ and (b) *tail*($\mu$) $\in$ *aset* and (c) *head*($\mu$) $\in$ *fset*

So either: one of Properties 1 – 4 holds or one of Properties 13 – 16 holds. It follows from Definition 3.3 that $\langle aft_1, \alpha_1\bullet\alpha_2, fore_2\rangle$ is a link of $G$. QED

Appendix B: Proof of Theorem 3.5

THEOREM 3.5. *Let $G = (V, S, A)$ be a set graph, let aft and fore be elements of SoS(IV(G)) and let $\alpha_1$ and $\alpha_2$ each be a non-null sequence of subsets of S. Then*

$$max^+(G, aft, \alpha_1 \bullet \alpha_2) = max^+(G, {\sim}max^+(G, aft, \alpha_1), \alpha_2)$$

$$max^-(G, fore, \alpha_1 \bullet \alpha_2) = max^-(G, {\sim}max^-(G, fore, \alpha_2), \alpha_1)$$

PROOF. Let $B$ denote the interior vertices of $G$. Suppose that $P \subseteq B$ and that $\langle {\sim}max^+(G, aft, \alpha_1), \alpha_2, \{P\} \rangle$ is a link of $G$. From Property 3.2, we know that $\langle aft, \alpha_1, max^+(G, aft, \alpha_1) \rangle$ is a link of $G$. It follows from Theorem 3.2 that $\langle aft, \alpha_1 \bullet \alpha_2, \{P\} \rangle$ is a link of $G$. Therefore

$$\{P \subseteq B \mid \langle {\sim}max^+(G, aft, \alpha_1), \alpha_2, \{P\} \rangle \text{ is a link of } G\} \subseteq$$
$$\{P \subseteq B \mid \langle aft, \alpha_1 \bullet \alpha_2, \{P\} \rangle \text{ is a link of } G\}$$

From Property 2.4, it follows that

$$min_\subseteq(\{P \subseteq B \mid \langle {\sim}max^+(G, aft, \alpha_1), \alpha_2, \{P\} \rangle \text{ is a link of } G\}) \leq$$
$$min_\subseteq(\{P \subseteq B \mid \langle aft, \alpha_1 \bullet \alpha_2, \{P\} \rangle \text{ is a link of } G\}) \qquad (1)$$

We now use the Fundamental Theorem (Theorem 3.1) to show that

$$min_\subseteq(\{P \subseteq B \mid \langle aft, \alpha_1 \bullet \alpha_2, \{P\} \rangle \text{ is a link of } G\}) \leq$$
$$min_\subseteq(\{P \subseteq B \mid \langle {\sim}max^+(G, aft, \alpha_1), \alpha_2, \{P\} \rangle \text{ is a link of } G\})$$

Suppose that $P \subseteq B$ and that

$$\langle aft, \alpha_1 \bullet \alpha_2, \{P\} \rangle \text{ is a link of } G \qquad (B)$$

Let $A$ be the set of ordered pairs $\langle aset, \omega_1 \rangle$, where $aset \in aft$ and $\omega_1 \in \times \alpha_1$, such that there does NOT exist a path $\mu$ in $G$ such that at least one of the following two properties holds:

1. (a) $\times label(\mu)$ contains a subsequence of $\omega_1$ and (b) $tail(\mu)$ is an initial vertex of $G$ and (c) $head(\mu)$ is a terminal vertex of $G$

2. (a) $\times label(\mu)$ contains a prefix of $\omega_1$ and (b) $tail(\mu) \in aset$ and (c) $head(\mu)$ is a terminal vertex of $G$

Let $C$ be the set of ordered pairs $\langle \omega_2, P \rangle$, where $\omega_2 \in \times \alpha_2$, such that there does NOT exist a path $\mu$ in $G$ such that at least one of the following two properties holds:

3. (a) $\times label(\mu)$ contains a subsequence of $\omega_2$ and (b) $tail(\mu)$ is an initial vertex of $G$ and (c) $head(\mu)$ is a terminal vertex of $G$

4. (a) ×*label*($\mu$) contains a suffix of $\omega_2$ and (b) *tail*($\mu$) is an initial vertex of $G$ and (c) *head*($\mu$) ∈ $P$

Let the relation $R_{AB} \subseteq A \times B$ be defined such that $\langle aset, \omega_1 \rangle \, R_{AB} \, v$ if and only if there exists a path $\mu$ in $G$ such that at least one of the following two properties holds:

5. (a) ×*label*($\mu$) contains a suffix of $\omega_1$ and (b) *tail*($\mu$) is an initial vertex of $G$ and (c) *head*($\mu$) = $v$

6. (a) ×*label*($\mu$) contains $\omega_1$ and (b) *tail*($\mu$) ∈ *aset* and (c) *head*($\mu$) = $v$

Let the relation $R_{BC} \subseteq B \times C$ be defined such that $v \, R_{BC} \, \langle \omega_2, P \rangle$ if and only if there exists a path $\mu$ in $G$ such that at least one of the following two properties holds:

7. (a) ×*label*($\mu$) contains a prefix of $\omega_2$ and (b) *tail*($\mu$) = $v$ and (c) *head*($\mu$) is a terminal vertex of $G$

8. (a) ×*label*($\mu$) contains $\omega_2$ and (b) *tail*($\mu$) = $v$ and (c) *head*($\mu$) ∈ $P$

Now consider an arbitrary ordered pair $\langle aset, \omega_1 \rangle$ in $A$ and an arbitrary ordered pair $\langle \omega_2, P \rangle$ in $C$. From our assumption that $\langle aft, \alpha_1 \bullet \alpha_2, \{P\} \rangle$ is a link of $G$ and from the fact that $aset \in aft$, $\omega_1 \in \times \alpha_1$, $\omega_2 \in \times \alpha_2$, we know that there exists a path $\mu$ in $G$ such that at least one of the following four properties must hold:

9. (a) ×*label*($\mu$) contains a subsequence of $\omega_1 \bullet \omega_2$ and (b) *tail*($\mu$) is an initial vertex of $G$ and (c) *head*($\mu$) is a terminal vertex of $G$

10. (a) ×*label*($\mu$) contains a prefix of $\omega_1 \bullet \omega_2$ and (b) *tail*($\mu$) ∈ *aset* and (c) *head*($\mu$) is a terminal vertex of $G$

11. (a) ×*label*($\mu$) contains a suffix of $\omega_1 \bullet \omega_2$ and (b) *tail*($\mu$) is an initial vertex of $G$ and (c) *head*($\mu$) ∈ $P$

12. (a) ×*label*($\mu$) contains $\omega_1 \bullet \omega_2$ and (b) *tail*($\mu$) ∈ *aset* and (c) *head*($\mu$) ∈ $P$

However, because of the way in which $A$ and $C$ are defined, the path $\mu$ in Properties 9 – 12 must *overlap* both $\omega_1$ and $\omega_2$. That is, $\mu$ must be partitionable into two subpaths $\mu_1$ and $\mu_2$ such that *head*($\mu_1$) = *tail*($\mu_2$) and at least one of the following four properties holds:

13. (a) ×*label*($\mu_1$) contains a suffix of $\omega_1$ and (b) ×*label*($\mu_2$) contains a prefix of $\omega_2$ and (c) *tail*($\mu_1$) is an initial vertex of $G$ and (d) *head*($\mu_2$) is a terminal vertex of $G$

14. (a) ×*label*($\mu_1$) contains $\omega_1$ and (b) ×*label*($\mu_2$) contains a prefix of $\omega_2$ and (c) *tail*($\mu_1$) ∈ *aset* and (d) *head*($\mu_2$) is a terminal vertex of $G$

15. (a) ×*label*($\mu_1$) contains a suffix of $\omega_1$ and (b) ×*label*($\mu_2$) contains $\omega_2$ and (c) *tail*($\mu_1$) is an initial vertex of $G$ and (d) *head*($\mu_2$) ∈ $P$

16. (a) ×*label*($\mu_1$) contains $\omega_1$ and (b) ×*label*($\mu_2$) contains $\omega_2$ and (c) *tail*($\mu_1$) ∈ *aset* and (d) *head*($\mu_2$) ∈ $P$

But this fact means that for all ⟨*aset*, $\omega_1$⟩ ∈ $A$, for all ⟨$\omega_2$, $P$⟩ ∈ $C$, there exists a vertex $v$ in $B$ – namely, *head*($\mu_1$) and *tail*($\mu_2$) – such that ⟨*aset*, $\omega_1$⟩ $R_{AB}$ $v$ and $v$ $R_{BC}$ ⟨$\omega_2$, $P$⟩. Applying the Fundamental Theorem, we see that:

$$\sim min_\subseteq(\{Q \subseteq B \mid R_{AB}^{-1}(Q) = A\}) \leq min_\subseteq(\{Q \subseteq B \mid R_{BC}(Q) = C\}) \quad (C)$$

Now consider all $Q \subseteq B$ such that $R_{AB}^{-1}(Q) = A$, and consider all ordered pairs ⟨*aset*, $\omega_1$⟩ such that *aset* ∈ *aft* and $\omega_1$ ∈ ×$\alpha_1$. Either ⟨*aset*, $\omega_1$⟩ ∈ $A$ or ⟨*aset*, $\omega_1$⟩ ∉ $A$. If ⟨*aset*, $\omega_1$⟩ ∈ $A$, then because $R_{AB}^{-1}(Q) = A$ there exists a path $\mu$ in $G$ such that at least one of the following two properties holds:

17. (a) ×*label*($\mu$) contains a suffix of $\omega_1$ and (b) *tail*($\mu$) is an initial vertex of $G$ and (c) *head*($\mu$) ∈ $Q$

18. (a) ×*label*($\mu$) contains $\omega_1$ and (b) *tail*($\mu$) ∈ *aset* and (c) *head*($\mu$) ∈ $Q$

If ⟨*aset*, $\omega_1$⟩ ∉ $A$, then there must exist a path $\mu$ in $G$ such that at least one of Property 1 or Property 2 holds. Thus for all $\omega_1$ ∈ ×$\alpha_1$, for all *aset* ∈ *aft*, there exists a path $\mu$ in $G$ such that at least one of the following four properties holds:

19. (a) ×*label*($\mu$) contains a subsequence of $\omega_1$ and (b) *tail*($\mu$) is an initial vertex of $G$ and (c) *head*($\mu$) is a terminal vertex of $G$

20. (a) ×*label*($\mu$) contains a prefix of $\omega_1$ and (b) *tail*($\mu$) ∈ *aset* and (c) *head*($\mu$) is a terminal vertex of $G$

21. (a) ×*label*($\mu$) contains a suffix of $\omega_1$ and (b) *tail*($\mu$) is an initial vertex of $G$ and (c) *head*($\mu$) ∈ $Q$

22. (a) ×*label*($\mu$) contains $\omega_1$ and (b) *tail*($\mu$) ∈ *aset* and (c) *head*($\mu$) ∈ $Q$

In other words, ⟨*aft*, $\alpha_1$, {$Q$}⟩ is a link of $G$. Thus

$$\{Q \subseteq B \mid R_{AB}^{-1}(Q) = A\} \subseteq max^+(G, aft, \alpha_1)$$

and

$$min_\subseteq(\{Q \subseteq B \mid R_{AB}^{-1}(Q) = A\}) \leq max^+(G, aft, \alpha_1) \quad (D)$$

A similar argument shows that

$$\{Q \subseteq B \mid R_{BC}(Q) = C\} \subseteq max^-(G, \{P\}, \alpha_2)$$

and

$$min_\subseteq(\{Q \subseteq B \mid R_{BC}(Q) = C\}) \leq max^-(G, \{P\}, \alpha_2) \qquad (E)$$

From Property D and Property 2.1(a), it follows that

$$\sim max^+(G, \mathit{aft}, \alpha_1) \leq \sim min_\subseteq(\{Q \subseteq B \mid R_{AB}^{-1}(Q) = A\}) \qquad (F)$$

From Properties C, E and F, it follows that

$$\sim max^+(G, \mathit{aft}, \alpha_1) \leq max^-(G, \{P\}, \alpha_2) \qquad (G)$$

However, from Property 3.2 we know that $\langle max^-(G, \{P\}, \alpha_2), \alpha_2, \{P\}\rangle$ is a link of $G$, and therefore from Property G and Property 3.1 it follows that

$$\langle \sim max^+(G, \mathit{aft}, \alpha_1), \alpha_2, \{P\}\rangle \text{ is a link of } G \qquad (H)$$

We have thus shown that Property B implies Property H. Therefore

$$\{P \subseteq B \mid \langle \mathit{aft}, \alpha_1 \bullet \alpha_2, \{P\}\rangle \text{ is a link of } G\} \subseteq$$
$$\{P \subseteq B \mid \langle \sim max^+(G, \mathit{aft}, \alpha_1), \alpha_2, \{P\}\rangle \text{ is a link of } G\}$$

and

$$min_\subseteq(\{P \subseteq B \mid \langle \mathit{aft}, \alpha_1 \bullet \alpha_2, \{P\}\rangle \text{ is a link of } G\}) \leq$$
$$min_\subseteq(\{P \subseteq B \mid \langle \sim max^+(G, \mathit{aft}, \alpha_1), \alpha_2, \{P\}\rangle \text{ is a link of } G\}) \qquad (I)$$

From Properties A and I it then follows that

$$min_\subseteq(\{P \subseteq B \mid \langle \mathit{aft}, \alpha_1 \bullet \alpha_2, \{P\}\rangle \text{ is a link of } G\}) =$$
$$min_\subseteq(\{P \subseteq B \mid \langle \sim max^+(G, \mathit{aft}, \alpha_1), \alpha_2, \{P\}\rangle \text{ is a link of } G\})$$

But this means that

$$max^+(G, \mathit{aft}, \alpha_1 \bullet \alpha_2) = max^+(G, \sim max^+(G, \mathit{aft}, \alpha_1), \alpha_2)$$

A similar argument shows that

$$max^-(G, \mathit{fore}, \alpha_1 \bullet \alpha_2) = max^-(G, \sim max^-(G, \mathit{fore}, \alpha_2), \alpha_1) \qquad \text{QED}$$

Appendix C: Proof of Theorem 4.1

THEOREM 4.1. *Let G be a Boolean graph over a set of atomic propositions AP, let $\alpha$ be a Boolean sequence over AP and let (S, B, L) be a fully populated Kripke structure over AP. Then $\alpha$ is an implicant of G if and only if $L(\alpha)$ is an implicant of $L(G)$.*

PROOF. Suppose that $\alpha$ is an implicant of *G*. By Definition 4.6, $L(\alpha)$ is an implicant of $L(G)$.

Suppose that $L(\alpha)$ is an implicant of $L(G)$. To show that $\alpha$ is an implicant of *G*, we need to show that for an arbitrary Kripke structure $(S', B', L')$ over *AP*, $L'(\alpha)$ is an implicant of $L'(G)$. To that end, assume that $\omega' \in \times L'(\alpha)$ and consider an arbitrary state $\omega'(i)$ in $\omega'$. It follows from Definition 4.3 that

$$\text{The assignment of truth values to the atomic propositions in } AP$$
$$\text{defined by } \omega'(i) \text{ causes } \alpha(i) \text{ to evaluate to } true \qquad (1)$$

Now observe that because $(S, B, L)$ is fully populated, each state $s' \in S'$ can be mapped to a state $s \in S$ that has the same assignment of truth values to the atomic propositions in *AP* as $s'$. Let $\phi: S' \rightarrow S$ be such a mapping, and let $\phi$ be extended to sequences of states in the obvious way. Now consider $\omega = \phi(\omega')$. By construction,

$$\omega(i) \text{ and } \omega'(i) \text{ define the same assignment of truth values}$$
$$\text{to the atomic propositions in } AP \qquad (2)$$

From (1) and (2), we see that

$$\text{The assignment of truth values to the atomic propositions in } AP$$
$$\text{defined by } \omega(i) \text{ causes } \alpha(i) \text{ to evaluate to } true \qquad (3)$$

From (3) and Definition 4.3, it follows that $\omega(i) \in L(\alpha(i))$ and that $\omega \in \times L(\alpha)$. But because $L(\alpha)$ is an implicant of $L(G)$,

$$\text{There exists a subsequence } \psi \text{ of } \omega \text{ and}$$
$$\text{a sequence of sets } \sigma \text{ accepted by } L(G) \text{ such that } \psi \in \times \sigma \qquad (4)$$

Since $\psi$ is a subsequence of $\omega$ and $\omega = \phi(\omega')$,

$$\text{There exists a subsequence } \psi' \text{ of } \omega' \text{ such that } \psi = \phi(\psi') \qquad (5)$$

From (2) and (5), it follows that

$$\psi(i) \text{ and } \psi'(i) \text{ define the same assignment of truth values}$$
$$\text{to the atomic propositions in } AP \qquad (6)$$

Moreover, because $\sigma$ is accepted by $L(G)$,

$$\text{There exists a Boolean sequence } \beta \text{ accepted by } G \text{ such that } \sigma = L(\beta) \qquad (7)$$

And from (4) and (7), we see that

$$\psi \in \times L(\beta) \qquad (8)$$

From (6) and (8) and Definitions 4.3 and 4.4, it follows that

$$\psi' \in \times L'(\beta) \qquad (9)$$

Now consider $L'(\beta)$. Because $\beta$ accepted by $G$, it must be that

$$L'(\beta) \text{ is accepted by } L'(G) \qquad (10)$$

From (9) and (10) it follows that for all $\omega' \in \times L'(\alpha)$, there exists a subsequence $\psi'$ of $\omega'$ and a sequence of sets $L'(\beta)$ that is accepted by $L'(G)$ such that $\psi' \in \times L'(\beta)$. $L'(\alpha)$ is therefore an implicant of $L'(G)$.

Appendix D: Proof of Theorem 4.3

THEOREM 4.3. *Let G be a Boolean graph over a set of atomic propositions AP, let aft and fore be elements of SoS(IV(G)), let $\alpha$ be a Boolean sequence over AP and let (S, B, L) be a fully populated Kripke structure over AP. Then $\langle aft, \alpha, fore \rangle$ is a link of G if and only if $\langle aft, L(\alpha), fore \rangle$ is a link of L(G).*

PROOF. Suppose that $\langle aft, \alpha, fore \rangle$ is a link of $G$. By Definition 4.7, $\langle aft, L(\alpha), fore \rangle$ is a link of $L(G)$.

Suppose that $\langle aft, L(\alpha), fore \rangle$ is a link of $L(G)$. To show that $\langle aft, \alpha, fore \rangle$ is a link of $G$, we need to show that for an arbitrary Kripke structure $(S', B', L')$ over $AP$, $\langle aft, L'(\alpha), fore \rangle$ is a link of $L'(G)$. That means showing that for each $set_a \in aft$, for each $\omega' \in \times L'(\alpha)$, for each $set_f \in fore$, there exists a path $\mu'$ in $L'(G)$ such that at least one of the four properties listed in Definition 3.3 holds. To see that this is the case, we first observe that for each state $\omega'(i)$ in $\omega'$, $\omega'(i) \in L'(\alpha(i))$. It follows from Definition 4.3 that

$$\text{The assignment of truth values to the atomic propositions in } AP$$
$$\text{defined by } \omega'(i) \text{ causes } \alpha(i) \text{ to evaluate to } true \qquad (1)$$

Now observe that because $(S, B, L)$ is fully populated, each state $s' \in S'$ can be mapped to a state $s \in S$ that has the same assignment of truth values to the atomic propositions in $AP$ as $s'$. Let $\phi: S' \to S$ be such a mapping, and let $\phi$ be extended to sequences of states in the obvious way. Now consider $\omega = \phi(\omega')$. By construction,

$$\omega(i) \text{ and } \omega'(i) \text{ define the same assignment of truth values}$$
$$\text{to the atomic propositions in } AP \qquad (2)$$

From (1) and (2), it follows that

$$\text{The assignment of truth values to the atomic propositions in } AP$$
$$\text{defined by } \omega(i) \text{ causes } \alpha(i) \text{ to evaluate to } true \qquad (3)$$

That means that $\omega(i) \in L(\alpha(i))$ and that $\omega \in \times L(\alpha)$. But because $\langle aft, L(\alpha), fore \rangle$ is a link of $L(G)$,

There exists a path $\mu$ in $L(G)$ and $\psi \in \times label(\mu)$ such that $\psi$ is a subsequence
of $\omega$ that satisfies at least one of the four properties listed in Definition 3.3 (4)

Since $\psi$ is a subsequence of $\omega$ and $\omega = \phi(\omega')$,

$$\text{There must exist a subsequence } \psi' \text{ of } \omega' \text{ such that } \psi = \phi(\psi')$$
$$\text{and } \psi' \text{ is in the same position of } \omega' \text{ that } \psi \text{ is in } \omega \tag{5}$$

Moreover from (2), it follows that

$$\psi(i) \text{ and } \psi'(i) \text{ define the same assignment of truth values}$$
$$\text{to the atomic propositions in } AP \tag{6}$$

Now since $\mu$ is a path in $L(G)$, it must be the image under $L$ of a path $\nu$ in $G$. Let $\beta = label(\nu)$. $\beta$ is thus the sequence of Boolean expressions labeling $\nu$. From Definitions 4.3 and 4.4, it follows that

$$\psi(i) \text{ causes } \beta(i) \text{ to evaluate to true} \tag{7}$$

From (6) and (7), it follows that

$$\psi'(i) \text{ causes } \beta(i) \text{ to evaluate to true} \tag{8}$$

From (8) and Definitions 4.3 and 4.4, we see that

$$\psi'(i) \in L'(\beta(i)) \tag{9}$$

Now let $\mu'$ be the path in $L'(G)$ that is the image under $L'$ of $\nu$. That means that

$$L'(\beta(i)) = label(\mu'(i)) \tag{10}$$

From (9) and (10), we have

$$\psi' \in \times label(\mu') \tag{11}$$

Finally, from (4), (5) and (11), it follows that

There exists a path $\mu'$ in $L'(G)$ and $\psi' \in \times label(\mu')$ such that $\psi'$ is a subsequence

of $\omega'$ that satisfies at least one of the four properties listed in Definition 3.3

So we have shown that for an arbitrary Kripke structure $(S', B', L')$ over $AP$, $\langle aft, L'(\alpha), fore \rangle$ is a link of $L'(G)$. $\langle aft, \alpha, fore \rangle$ is therefore a link of $G$.

## Appendix E: Proof of Theorem 4.8

THEOREM 4.8. *Let G be a Boolean graph over a set of atomic propositions AP, let aft and fore be elements of SoS(IV(G)), let $\alpha$ be a Boolean sequence over AP and let (S, B, L) be a fully populated Kripke structure over AP. Then*

$$max^+(G, aft, \alpha) = max^+(L(G), aft, L(\alpha))$$

$$max^-(G, fore, \alpha) = max^-(L(G), fore, L(\alpha))$$

PROOF. By Definition 4.9,

$$max^+(G, aft, \alpha) = \bigwedge_{\text{For all Kripke structures } (S, B, L) \text{ over } AP} max^+(L(G), aft, L(\alpha))$$

By Definition 3.5,

$$max^+(L(G), aft, L(\alpha)) = min_\subseteq(\{U \subseteq IV(G) \mid \langle aft, L(\alpha), \{U\}\rangle \text{ is a link of } L(G)\})$$

From these two equalities and the definition of $\wedge$ (Definition 2.5), it follows that $max^+(G, aft, \alpha)$ is the set of minimal $U \subseteq IV(G)$, with respect to set inclusion, such that

$$\langle aft, L(\alpha), \{U\}\rangle \text{ is a link of } L(G) \text{ for all Kripke structures } (S, B, L) \text{ over } AP$$

But from the definition of a link at the logic level (Definition 4.7), we see that this last property is equivalent to

$$\langle aft, \alpha, \{U\}\rangle \text{ is a link of } G$$

Thus

$$max^+(G, aft, \alpha) = min_\subseteq(\{U \subseteq IV(G) \mid \langle aft, \alpha, \{U\}\rangle \text{ is a link of } G\})$$

From Theorem 4.3, we know that $\langle aft, \alpha, \{U\}\rangle$ is a link of $G$ if and only if $\langle aft, L(\alpha), \{U\}\rangle$ is a link of $L(G)$. Therefore

$$max^+(G, aft, \alpha) = min_\subseteq(\{U \subseteq IV(G) \mid \langle aft, L(\alpha), \{U\}\rangle \text{ is a link of } L(G)\})$$

But by Definition 3.5,

$$max^+(L(G), aft, L(\alpha)) = min_\subseteq(\{U \subseteq IV(G) \mid \langle aft, L(\alpha), \{U\}\rangle \text{ is a link of } L(G)\})$$

Hence $max^+(G, aft, \alpha) = max^+(L(G), aft, L(\alpha))$. A similar proof applies to $max^-$.

Appendix F:  Proof of Lemma 6.3

LEMMA 6.3. Let $G$ and $E = (V_E, A_E)$ be Boolean graphs over the same set of atomic propositions such that $V_E \cap vertices(G)$ is empty and let $\mu$ be a path in $E$ such that

1. $tail(\mu)$ is an initial vertex of $E$

2. For all proper prefixes $\mu_P$ of $\mu$, $max^+(G, \{\{\}\}, label(\mu_P)) \neq \{\{\}\}$

3. $head(\mu)$ is not a terminal vertex of $E$ or $max^+(G, \{\{\}\}, label(\mu)) = \{\{\}\}$

Then in Steps 1, 2 and 3 of Definition 6.3, $\mu$ is transformed into a new path $\mu_T$ in $E$ such that

4. $label(\mu_T) = label(\mu)$

5. For all vertices $v$ on $\mu_T$, $v \in vertices(G)$

6. $tail(\mu_T) = \langle\{\},\{\{\}\}\rangle$

7. For all interior vertices $v$ on $\mu_T$, $v \neq \langle\{\{\}\},\{\}\rangle$

PROOF. By induction on the length of $\mu$. Let $\mu$ be a path in $E$ of length 1 (i.e., an arc) satisfying Properties 1 – 3 in the lemma. Since $tail(\mu)$ is an initial vertex of $E$, the tail of $\mu$ is replaced with $\langle\{\},\{\{\}\}\rangle$ in Step 2 of Definition 6.3. Then since $tail(\mu) \in vertices(G)$ and $head(\mu) \notin vertices(G)$ and either $head(\mu)$ is not a terminal vertex of $E$ or $max^+(G, \{\{\}\}, label(\mu)) = \{\{\}\}$, it follows that the head of $\mu$ is eventually *updated* in Step 3(b) (Lemma 6.2). The resulting arc/path satisfies Properties 4 – 7 in the lemma.

Now assume that the lemma is true for all paths of length $n$. Let $\mu$ be a path in $E$ of length $n+1$ satisfying Properties 1 – 3 in the lemma, let $\mu_n$ be the prefix of $\mu$ of length $n$ and let $a$ be the $n+1$'st (and final) arc of $\mu$. By our hypothesis, $\mu_n$ is transformed into a path $\mu_{nT}$ in $E$ satisfying Properties 4 – 7. Upon completion of that transformation, we know that since $\mu$ satisfies Property 2, $max^+(G, \{\{\}\}, label(\mu_{nT})) \neq \{\{\}\}$. It follows from Lemma 6.1 that $head(\mu_{nT}) \neq \langle\{\{\}\},\{\}\rangle$, and therefore it must have been the case that when the last arc $\langle v_t, BE, v_h \rangle$ in $\mu_{nT}$ was updated in Step 3 of Definition 6.3, $vertex^+(G, fore(v_t), BE) \neq \langle\{\{\}\},\{\}\rangle$. It follows that the arc

$$\langle head(\mu_{nT}), label(a), head(a) \rangle$$

must have been added to $A_E$ in Step 3(b)(iii) if it had not been previously added, and, as a result, this arc is eventually *processed* in Step 3. In that processing, since $head(a) = head(\mu)$ and $\mu$ satisfies Property 3, either $head(a)$ is not a terminal vertex of $E$ or $max^+(G,$

{{}}, $label(\mu)$) = {{}}. If $head(a)$ is not a terminal vertex of $E$, then the condition in Step 3(a) evaluates to false and the head of $\langle head(\mu_{nT}), label(a), head(a)\rangle$ is updated in Step 3(b). If $max^+(G, \{\{\}\}, label(\mu)) = \{\{\}\}$, then by Lemma 6.1,

$$vertex^+(G, fore(head(\mu_{nT})), label(a)) = \langle\{\{\}\},\{\}\rangle$$

And again the condition in Step 3(a) evaluates to false and the head of $\langle head(\mu_{nT}), label(a), head(a)\rangle$ is updated in Step 3(b). Now consider the transformed path $\mu_T$ and Properties 4 – 7 in the lemma.

4. By construction, $\mu_T$ is the concatenation of $\mu_{nT}$ and the arc $\langle head(\mu_{nT}), label(a), vertex^+(G, fore(head(\mu_{nT})), label(a))\rangle$. Since $\mu_n$ is a path of length $n$, it follows by hypothesis (Property 4) that $label(\mu_{nT}) = label(\mu_n)$. Thus $label(\mu_T) = label(\mu_{nT})\bullet\langle label(a)\rangle = label(\mu_n)\bullet\langle label(a)\rangle = label(\mu)$.

5. By construction in Step 2, $tail(\mu_T) = \langle\{\},\{\{\}\}\rangle$. The remaining vertices on $\mu_T$ are created in Step 3(b)(ii) and each is of the form $vertex^+(G, fore(v_t), BE)$. It follows that for all vertices $v$ on $\mu_T$, $v \in vertices(G)$.

6. By construction in Step 2, $tail(\mu_T) = \langle\{\},\{\{\}\}\rangle$.

7. By hypothesis, $\mu_{nT}$ satisfies Property 7. The sole remaining interior vertex of $\mu_T$ is $head(\mu_{nT})$, but we have already established that $head(\mu_{nT}) \neq \langle\{\{\}\},\{\}\rangle$. Thus for all interior vertices $v$ on $\mu_T$, $v \neq \langle\{\{\}\},\{\}\rangle$.

# REFERENCES

ACCELLERA 2004. *Property Specification Language Reference Manual, Version 1.1*. Accellera Organization (www.accellera.org), Napa, CA.

BALBES, R. AND DWINGER, P. 1974. *Distributive Lattices*. University of Missouri Press, Columbia, MO.

BEER, I., BEN-DAVID, S., EISNER, C., FISMAN, D., GRINGAUZE, A. AND RODEH, Y. 2001. The temporal logic Sugar. In *Proceedings of the 13th International Conference on Computer-Aided Verification* (*CAV* 2001), Paris, France, July 2001, Lecture Notes in Computer Science, Vol. 2102, G. BERRY, H. COMON AND A. FINKEL, Eds. Springer-Verlag, Berlin Heidelberg New York 363-367.

BIAŁYNICKI-BIRULA, A. AND RASIOWA, H. 1957. On the representation of quasi-Boolean algebras. *Bulletin de l'Academie Polonaise des Sciences*, 5, 259-261.

BIAŁYNICKI-BIRULA, A. 1957. Remarks on quasi-Boolean algebras. *Bulletin de l'Academie Polonaise des Sciences*, 5, 615-619.

BLAKE, A. 1937. *Canonical expressions in Boolean algebra*. PhD thesis, University of Chicago, 1937.

CIGNOLI, R. 1975. Injective De Morgan and Kleene algebras. *Proceedings of the American Mathematical Society*, Vol. 47, No. 2 (February 1975), 269-278.

CLARKE, E.M., EMERSON, E.A. AND SISTLA, A.P. 1986. Automatic verification of finite-state concurrent systems using temporal logic specifications. *ACM Transactions on Programming Languages and Systems* (*TOPLAS* 1986), Volume 8, Issue 2 (April 1986), 244-263.

CLARKE, E.M., GRUMBERG, O. AND PELED, D.A. 2000. *Model Checking*. The MIT Press, Cambridge, MA.

CLARKE, E.M., GRUMBERG, O., JHA S., LU Y. AND VEITH, H. 2001. Progress on the state explosion problem in model checking. In *Informatics: 10 Years Back, 10 Years Ahead*, Lecture Notes in Computer Science, Vol. 2000, R. WILHELM, Ed. Springer-Verlag, Berlin Heidelberg New York 176-194.

FIGALLO, A. AND MONTEIRO, L. 1981. The determinant system for the free De Morgan algebra over a finite ordered set. *Journal of Symbolic Logic*, Volume 46, (1), 185-185.

FURTEK, F.C. 1978. Constraints and compromise. In *Foundations of Secure Computation*, R.A. DEMILLO, D.P. DOBKIN, A.K. JONES AND R.J. LIPTON, Eds. Academic Press, New York, 189-204.

FURTEK, F.C. 1980. Specification and verification of real-time, distributed systems using the theory of constraints. In *Proceedings of the 5th Conference on Automated Deduction* (*CADE* 1980), Les Arcs, France, July 1980, Lecture Notes in Computer Science, Volume 87, W. BIBEL AND R. KOWALSKI Eds. Springer-Verlag, Berlin Heidelberg New York, 110-125.

FURTEK, F.C. 1982a. *The Theory of Constraints*. Report CSDL-P-1564, May 4, 1982, The Charles Stark Draper Laboratory, Cambridge, MA.

FURTEK, F.C. 1982b. Verifying ongoing behavior: a case study. In *Proceedings of the 6th Annual International Computer Software and Applications Conference* (*COMPSAC* 1982), Chicago, USA, November 1982. IEEE, 634-639.

FURTEK, F.C. 1984. A necessary and sufficient condition for a product relation to be total. *Journal of Combinatorial Theory, Series A*, Vol. 37, No. 3, November 1984, 320–326.

HOLT, A.W. 1968. *Information System Theory Project: Final Report*. Report RADC-TR-68-305, Applied Data Research, Inc., Princeton, NJ, February 1968. Available from the Department of Commerce Clearinghouse, Springfield, VA. as Report AD 676-972, September 1968.

HOLT, A.W. 1971. Introduction to Occurrence Systems. In *Associative Information Techniques*, L. Jacks, Ed. American Elsevier, New York, 175-203.

HOPCROFT, J.E., MOTWANI, R. AND ULLMAN J.D. 2006. *Introduction to Automata Theory, Languages, and Computation* (3$^{rd}$ *Edition*). Addison Wesley, Boston, USA.

KALMAN, J.A. 1958. Lattices with involution. *Transactions of the American Mathematical Society*, Vol. 87, No. 2 (Mar. 1958), pp. 485-491.

MCCLUSKEY, E.J. 1956. Minimization of Boolean functions. *Bell System Technical Journal*, Vol. 35, November 1956, 1417-1444.

MILLEN, J. 1978. Constraints and multilevel security. In *Foundations of Secure Computation*, R.A. DEMILLO, D.P. DOBKIN, A.K. JONES AND R.J. LIPTON, Eds. Academic Press, New York, 205-222.

OWRE, S., RUSHBY, J.M., SHANKAR, N. 1992. PVS: A prototype verification system. In *Proceedings of the 11th International Conference on Automated Deduction* (*CADE* 1992), Saratoga Springs, NY, June 1992, Lecture Notes in Artificial Intelligence, Volume 607, D. KAPUR, Ed. Springer-Verlag, Berlin Heidelberg New York, 748-752.

OWRE, S., RUSHBY, J., SHANKAR, N. AND STRINGER-CALVERT, D. 1998. PVS: An experience report. In *Applied Formal Methods - FM-Trends 98*, Boppard, Germany, October 1998, Lecture Notes in Computer Science, Volume 1641, D. HUTTER, W. STEPHAN, P. TRAVERSO and M. ULLMAN Eds. Springer-Verlag, Berlin Heidelberg New York, 338-345.

PETRI, C.A. 1962. *Kommunikation mit Automaten*. PhD Thesis, Institut für Instrumentelle Mathematik, Schriften des IIM Nr. 2, 1962, Second Edition, Bonn (In German). Also in Griffiss Air Force Base, Technical Report RADC-TR-65--377, Vol. 1, New York, 1966 (English translation).

PETRI, C.A. 1986. Concurrency Theory. In *Petri Nets: Central Models and Their Properties, Advances in Petri Nets 1986, Part I, Proceedings of an Advanced Course*, Bad Honnef, Germany, September 1986, Lecture Notes in Computer Science, Vol. 254, W. BRAUER, W. REISIG, G. ROZENBERG, Eds. Springer-Verlag, Berlin Heidelberg New York 4-24.

QUINE, W.V. 1952. The problem of simplifying truth functions. *American Mathematical Monthly*, Volume 59, Number 8 (1952), pp. 521-531.

REED, T.J. 1979. APL modeling of DeMorgan algebras. In *Proceedings of the International Conference on APL*, New York, New York, 1979. ACM Press, New York, NY, 302-305.

ROBINSON, J.A. 1965. A machine-oriented logic based on the resolution principle. *Journal of the ACM*, Volume 12, Issue 1 (January 1965), 23-41.

SANKAPPANAVAR, H.P. 1980. A characterization of principal congruences of De Morgan algebras and its applications. In *Mathematical Logic in Latin America*. North-Holland, Amsterdam, 341-349.

SIPSER, M. 2005. *Introduction to the Theory of Computation, Second Edition*. Thomson Course Technology, Boston, MA.

TISON, P. 1967. Generalized consensus theory and application to the minimization of Boolean functions. *IEEE Transactions on Electronic Computers*, Vol. EC-16, No. 4, 446-456.